\documentclass[twocolumn,english,superscriptaddress]{revtex4-2}
\usepackage{lmodern}
\usepackage{lmodern}

\usepackage[T1]{fontenc}
\usepackage[latin9]{inputenc}
\usepackage{babel}
\usepackage{float}
\usepackage{multirow}
\usepackage{amsmath}
\usepackage{amssymb}
\usepackage{graphicx}
\usepackage{color}
\usepackage[unicode=true,pdfusetitle,
 bookmarks=true,bookmarksnumbered=false,bookmarksopen=false,
 breaklinks=false,pdfborder={0 0 1},backref=false,colorlinks=false]
 {hyperref}

\makeatletter

\providecommand{\tabularnewline}{\\}
\newcommand{\lyxdot}{.}

\usepackage{babel}

\makeatother

\begin{document}
\title{A unified theory of the self-similar supersonic Marshak wave problem}
\author{Menahem Krief}
\email{menahem.krief@mail.huji.ac.il}

\affiliation{Racah Institute of Physics, The Hebrew University, 9190401 Jerusalem,
Israel}
\affiliation{Department of Aerospace and Mechanical Engineering, University of
Notre Dame, Fitzpatrick Hall, Notre Dame, IN 46556, USA}
\author{Ryan G. McClarren}
\affiliation{Department of Aerospace and Mechanical Engineering, University of
Notre Dame, Fitzpatrick Hall, Notre Dame, IN 46556, USA}
\begin{abstract}
We present a systematic study of the similarity solutions for the
Marshak wave problem, in the local thermodynamic equilibrium (LTE)
diffusion approximation and in the supersonic regime. Self-similar
solutions exist for a temporal power law surface temperature drive
and a material model with power law temperature dependent opacity
and energy density. The properties of the solutions in both linear
and nonlinear conduction regimes are studied as a function of the
temporal drive, opacity and energy density exponents. We show that
there exists a range of the temporal exponent for which the total
energy in the system decreases, and the solution has a local maxima.
For nonlinear conduction, we specify the conditions on the opacity
and energy density exponents under which the heat front is linear
or even flat, and does posses its common sharp character; this character is independent of the drive exponent. We specify
the values of the temporal exponents for which analytical solutions
exist and employ the Hammer-Rosen perturbation theory to obtain highly
accurate approximate solutions, which are parameterized using only
two numerically fitted quantities. The solutions are used to construct
a set of benchmarks for supersonic LTE radiative heat transfer, including
some with unusual and interesting properties such as local maxima
and non sharp fronts. The solutions are compared in detail to implicit
Monte-Carlo and discrete-ordinate transport simulations as well gray
diffusion simulations, showing a good agreement, which highlights
their usefulness as a verification test problem for radiative transfer
simulations. 
\end{abstract}
\maketitle

\section{Introduction}

Radiation hydrodynamics has a central role in the understating of
high energy density systems, such as laboratory astrophysics, inertial
confinement fusion and general astrophysical phenomena \cite{lindl2004physics,robey2001experimental,bailey2015higher,falize2011similarity,hurricane2014fuel,cohen2020key,heizler2021radiation}.
Solutions of the equations of radiation hydrodynamics are paramount
in the analysis, design and characterization of high energy density
experiments \cite{back2000diffusive,sigel1988x,lindl1995development,cohen2020key,heizler2021radiation}
and are frequently used in the verification of computer simulations
\cite{calder2002validating,krumholz2007equations,gittings2008rage,lowrie2007radiative,mcclarren2011benchmarks,bennett2021self,mcclarren2021two,mcclarren2011solutions,mcclarren2008analytic,kamm2008enhanced,rider2016robust,krief2021analytic,giron2021solutions,giron2023solutions,krief2023piston}.

The theory of Marshak waves was developed in the seminal work \cite{marshak1958effect},
and was further generalized by many authors \cite{petschek1960penetration,pert1977class,pakula1985self,kaiser1989x,hammer2003consistent,garnier2006self,smith2010solutions,lane2013new,shussman2015full,cohen2018modeling,hristov2018heat,krief2024self}.
It describes the dynamics which occurs due to an intense energy deposition
in a material, leading to a steep temperature gradient. In such circumstances,
radiative transfer plays a pivotal role in the description of energy
deposition, the subsequent thermalization of the material, and the
rapid emission and transport of radiative energy. At typical high
temperature scenarios, the radiative heat wave propagates faster than
the material speed of sound, giving rise to a supersonic Marshak wave
\cite{hammer2003consistent,garnier2006self,shussman2015full,malka2022supersonic},
for which the material motion is negligible. 
We note that in the last decades, numerous supersonic Marshak experiments have been carried out in high energy density facilities (see for example, Ref. \cite{cohen2020key} and references therein).
For optically thick systems,
local thermodynamic equilibrium (LTE) between the radiation field
and the heated material is reached very quickly and the diffusion
approximation of radiation transport is applicable. Under those not
uncommon circumstances, the Marshak wave problem is formulated mathematically
by the solutions of the planar LTE radiation diffusion equation, with
a boundary condition of a time dependent surface temperature drive
$T_{s}\left(t\right)$ and an initially cold homogeneous material.
Hammer and Rosen in their seminal work \cite{hammer2003consistent},
developed a perturbative method to obtain approximate solutions for
a general time dependent temperature drive, assuming power law temperature
dependence of the opacity $k\left(T\right)=k_{0}T^{-\alpha}$ and
total energy density $u\left(T\right)=u_{0}T^{\beta}$. For a surface
temperature with power law time dependence $T_{s}\left(t\right)=T_{0}t^{\tau}$,
the Marshak wave problem is self-similar \cite{pert1977class,hammer2003consistent,garnier2006self,smith2010solutions,shussman2015full,hristov2018heat}
and can be solved exactly using the method of dimensional analysis.
Solutions for a constant surface temperature and a diffusion coefficient
which is not a temperature power law, were discussed in Refs. \cite{fujita1952exact,fujita1952exactII,fujita1954exact,philip1960general,boyer1961some,tuck1976some,brutsaert1982some,king1990exact}.
Solutions for power law opacity and energy density, were studied in
Refs. \cite{marshak1958effect,petschek1960penetration,heaslet1961diffusion,kass1966numerical,castor2004radiation,mihalas1999foundations,lane2013new}
for a time independent surface temperature, and in Refs. \cite{pert1977class,garnier2006self,smith2010solutions,shussman2015full,hristov2018heat, malka2022supersonic}
for a general power law time dependence.

In this work we study in detail the self-similar solutions for surface
temperature, opacity and energy density with power law dependence to present a unified theory of these waves.
The behavior and characteristics of the solutions in both linear and
nonlinear conduction regimes are analyzed as a function of the surface
temperature drive, opacity and energy density exponents, $\tau$,
$\alpha$ and $\beta$. We discuss the monotonicity of the solutions
in different ranges of $\tau$, and relate it to the total energy
balance in the system. For nonlinear conduction, we state the conditions
on $\alpha$ and $\beta$ for which the heat front does not have the
familiar sharp character, and point out that the front can be linear
or even flat. We specify the values of $\tau$ for which closed form
analytical solutions exist. Classical solutions found by previous
authors are identified as special cases. We perform a detailed study
of the accuracy of the widely used Hammer-Rosen perturbation theory
\cite{hammer2003consistent}, as well as for the series expansion
method of Smith  \cite{smith2010solutions}, by comparing their results
to the exact self-similar Marshak wave solutions in a wide range of
the exponents $\tau,\alpha,\beta$. Finally, we use the similarity
solutions to construct a set of benchmarks for supersonic LTE radiative
heat transfer. These benchmarks are compared in detail to numerical
transport and diffusion computer simulations.

\section{Statement of the problem\label{sec:Statement-of-the}}

In supersonic radiation hydrodynamics flows, for which the hydrodynamic
motion is negligible in comparison to the radiation heat conduction,
the material density is constant in time, and the heat flow is supersonic.
Under these conditions, the LTE radiation transfer problem in planar
slab symmetry is given by:

\begin{equation}
\frac{\partial u}{\partial t}+\frac{\partial F}{\partial x}=0,\label{eq:main_eq}
\end{equation}
where $F$ is the radiation energy flux and 
\begin{equation}
u=u_{m}+aT^{4}\label{eq:utot}
\end{equation}
is the total (matter+radiation) energy density, with $u_{m}$ the
material energy density, $T$ the material temperature and $a=\frac{4\sigma}{c}$
the radiation constant with $c$ the speed of light and $\sigma$ the Stefan-Boltzmann constant. In the diffusion approximation of radiative
transfer, which is applicable for optically thick media, the radiation
energy flux obeys Fick's law: 
\begin{equation}
F=-D\frac{\partial}{\partial x}\left(aT^{4}\right),\label{eq:fick}
\end{equation}
with the radiation diffusion coefficient 

\begin{equation}
D=\frac{c}{3k},\label{eq:diffusion_coeff}
\end{equation}
where $k=\rho\kappa_{R}$ is the opacity,
defined as total (absorption+scattering) macroscopic transport cross
section with dimensions of inverse legnth,  $\kappa_{R}$ the Rosseland mean opacity and $\rho$
is the (time independent and spatially homogeneous) material mass
density.

In this work, we assume a material model with temperature power laws
for the opacity:

\begin{equation}
k\left(T\right)=k_{0}T^{-\alpha},\label{eq:ross_opac_powerlaw}
\end{equation}
and the total energy density equation of state:

\begin{equation}
u\left(T\right)=u_{0}T^{\beta}.\label{eq:eos}
\end{equation}
This approximation of a single power law for the two terms in Eq. \eqref{eq:utot} will be discussed below in Sec. \ref{subsec:Transport-setup}. We also note that the form \eqref{eq:ross_opac_powerlaw} is equivalent
to the common power law representation \cite{hammer2003consistent,garnier2006self,smith2010solutions,shussman2015full,heizler2016self,hristov2018heat,heizler2021radiation,krief2021analytic,malka2022supersonic,krief2024self}
of the Rosseland opacity $\kappa_{R}\left(T,\rho\right)=\frac{1}{g}T^{-\alpha}\rho^{\lambda}$,
with the coefficient $g=\rho^{\lambda+1}/k_{0}$. Similarly, the form
\eqref{eq:eos} is equivalent to the common power law representation
$u\left(T,\rho\right)=\mathcal{F}T^{\beta}\rho^{1-\mu}$, with $\mathcal{F}=u_{0}\rho^{\mu-1}$.

Using the opacity \eqref{eq:ross_opac_powerlaw}, the radiation energy
flux \eqref{eq:fick} can be written as:

\begin{equation}
F\left(x,t\right)=-u_{0}K\frac{\partial}{\partial x}T^{4+\alpha},\label{eq:FFF}
\end{equation}
and by plugging Eqs. \eqref{eq:eos}-\eqref{eq:FFF} into the radiation
diffusion equation \eqref{eq:main_eq}, a nonlinear diffusion equation
for the temperature is obtained:

\begin{equation}
\frac{\partial T^{\beta}}{\partial t}=K\frac{\partial^{2}}{\partial x^{2}}T^{4+\alpha},\label{eq:Tpde}
\end{equation}
where we have defined the dimensional constant:

\begin{equation}
K=\frac{4ac}{3\left(4+\alpha\right)u_{0}k_{0}}.
\end{equation}
The generalized self-similar Marshak problem is defined by the solution
of Eq. \eqref{eq:Tpde} with a temporal power law for the surface
temperature: 
\begin{equation}
T\left(x=0,t\right)=T_{0}t^{\tau},\label{eq:Tbc}
\end{equation}
which is applied on an initially cold medium, 
\begin{equation}
T\left(x,t=0\right)=0.\label{eq:initcond}
\end{equation}

The diffusion equation \eqref{eq:Tpde} can be alternatively written
as a canonical nonlinear diffusion equation \cite{pattle1959diffusion,ames1965similarity, kass1966numerical,king1990exact}, in terms of the total
energy density:

\begin{equation}
\frac{\partial u}{\partial t}=A\frac{\partial}{\partial x}\left(u^{n}\frac{\partial u}{\partial x}\right),\label{eq:uPDE}
\end{equation}
with the the nonlinearity index \cite{zeldovich1967physics,heaslet1961diffusion,pert1977class,kass1966numerical,hristov2018heat,krief2021analytic},
\begin{equation}
n=\frac{4+\alpha-\beta}{\beta},\label{eq:ndef}
\end{equation}
and $A=\left(n+1\right)u_{0}^{-n}K=\frac{4ac}{3\beta u_{0}^{n+1}k_{0}}$.

\section{Self-Similar solution\label{sec:Self-Similar-solution}}

\begin{figure}
\begin{centering}
\includegraphics[scale=0.6]{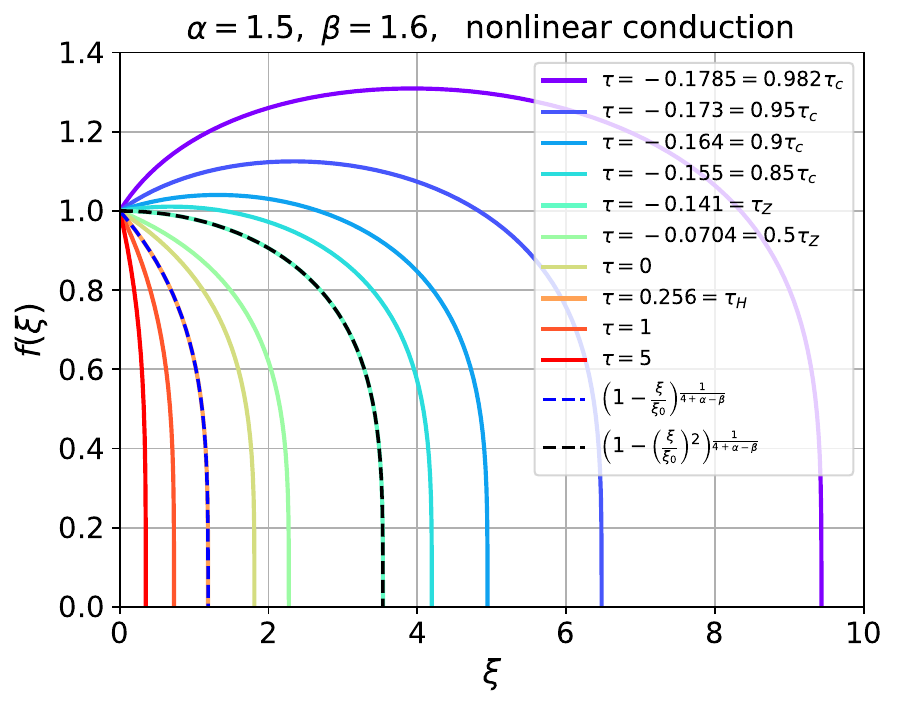} 
\par\end{centering}
\begin{centering}
\includegraphics[scale=0.6]{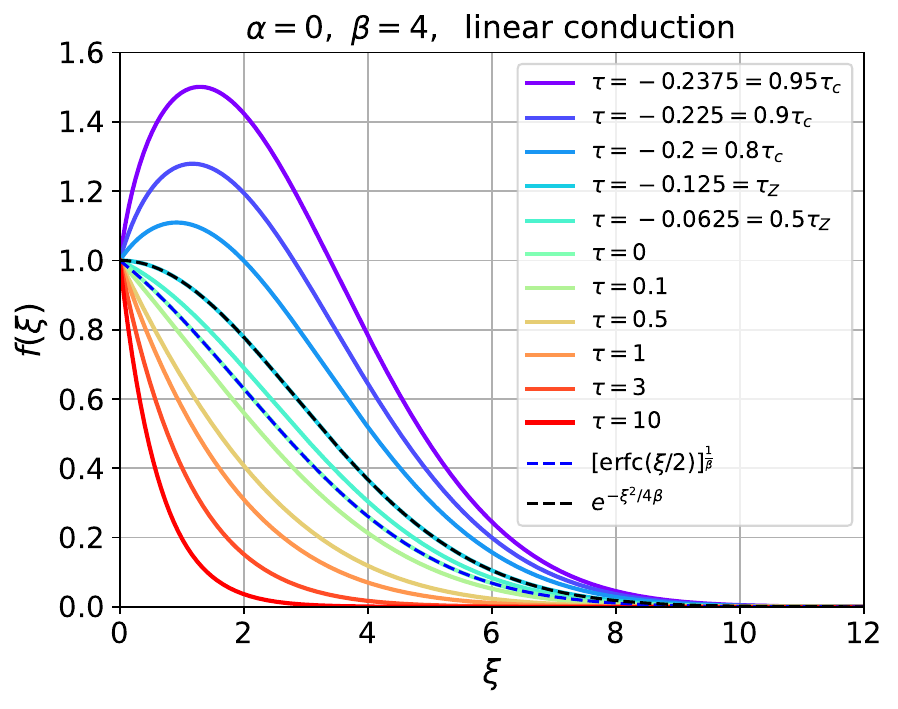} 
\par\end{centering}
\caption{The temperature similarity profiles (solutions of Eq. \eqref{eq:ode}),
for various values of the temporal exponent $\tau$. In the figures the value of $\tau$ decreases from left to right in the respective curves. The upper figure
shows the numerical solutions for the nonlinear Marshak wave problem
with $\alpha=1.5$, $\beta=1.6$ ($n=2.4375$). The analytical solutions
which are known for $\tau=\tau_{H}$ {[}Eqs. \eqref{eq:f_Hy_Exact}-\eqref{eq:xsi_0_hy}{]}
and $\tau=\tau_{Z}$ {[}Eqs. \eqref{eq:f_z}-\eqref{eq:xsi_0_Z}{]},
are shown for comparison. The lower figure shows the solutions for
the linear Marshak wave problem (given analytically in Eq. \eqref{eq:f_lin_sol}),
with $\alpha=0$, $\beta=4$ ($n=0$). The special solutions for $\tau=0$
{[}Eq. \eqref{eq:f_tau0_lin}{]} and for $\tau=\tau_{Z}$ {[}Eq. \eqref{eq:f_zel_lin}{]},
are also shown. It is evident that for both linear and nonlinear conduction,
the behavior of the solutions near the origin is in accordance with
Tab. \ref{tab:origin_behaviou}. \label{fig:profiles_shapes}}
\end{figure}

\begin{center}
\begin{table*}[t]
\begin{centering}
\begin{tabular}{|c|c||c|c|c|c|c|}
\cline{2-7} \cline{3-7} \cline{4-7} \cline{5-7} \cline{6-7} \cline{7-7} 
\multicolumn{1}{c|}{} & conduction  & $\tau_{c}<\tau<\tau_{Z}$  & $\tau=\tau_{Z}$  & $\tau_{Z}<\tau<\tau_{H}$  & $\tau=\tau_{H}$  & $\tau>\tau_{H}$\tabularnewline
\hline 
\multirow{3}{*}{$f\left(\xi\right)$} & \multirow{2}{*}{nonlinear} & \multirow{2}{*}{numerical solution} & analytic solution {[}Eq. \eqref{eq:f_z}{]}  & numerical  & analytic solution {[}Eq. \eqref{eq:f_Hy_Exact}{]}  & numerical \tabularnewline
 &  &  & $f\left(\xi\right)=\left(1-\xi^{2}/\xi_{0}^{2}\right)^{\nu}$  & solution  & $f\left(\xi\right)=\left(1-\xi/\xi_{0}\right)^{\nu}$  & solution\tabularnewline
\cline{2-7} \cline{3-7} \cline{4-7} \cline{5-7} \cline{6-7} \cline{7-7} 
 & linear  & \multicolumn{5}{c|}{analytic solution {[}Eq. \eqref{eq:f_lin_sol}{]}}\tabularnewline
\hline 
\multirow{2}{*}{heat front} & nonlinear  & \multicolumn{3}{c|}{decelerates} & constant speed, $x_{F}\left(t\right)\propto t$  & accelerates\tabularnewline
\cline{2-7} \cline{3-7} \cline{4-7} \cline{5-7} \cline{6-7} \cline{7-7} 
 & linear  & \multicolumn{5}{c|}{decelerates as $t^{0.5}$}\tabularnewline
\hline 
\multicolumn{2}{|c|||}{$f'\left(0\right)$} & $f'\left(0\right)>0$  & $f'\left(0\right)=0$  & \multicolumn{3}{c|}{$f'\left(0\right)<0$}\tabularnewline
\hline 
\multicolumn{2}{|c|||}{net surface flux} & outcoming  & none  & \multicolumn{3}{c|}{incoming}\tabularnewline
\multicolumn{2}{|c|||}{$F\left(x=0,t\right)$} & $<0$  & $=0$  & \multicolumn{3}{c|}{$>0$}\tabularnewline
\hline 
\multicolumn{2}{|c|||}{total energy} & energy exits the system  & total energy is constant  & \multicolumn{3}{c|}{energy enters the system}\tabularnewline
\hline 
\multicolumn{2}{|c|||}{monotonicity} & local maxima at $0<\xi<\xi_{0}$  & local maxima at $\xi=0$  & \multicolumn{3}{c|}{strictly monotonically decreasing}\tabularnewline
\hline 
\end{tabular}
\par\end{centering}
\caption{A summary of the behavior and properties of the solutions of the self-similar
Marshak wave problem in different ranges of the temporal exponent
$\tau$, which are described in the text. See also Fig. \ref{fig:profiles_shapes}. Note the special similarity exponents are  $\tau_H$ for the Henyey case, Eq.~\eqref{eq:tau_h}, $\tau_Z$ for the Zel'dovich case, Eq.~\eqref{eq:tau_Z}, and $\tau_c$ is the critical value from Eq.~\eqref{eq:tau_c}, below which no self-similar solution exists.
\label{tab:origin_behaviou}}
\end{table*}
\par\end{center}

The nonlinear diffusion equation \eqref{eq:Tpde} with the initial
and boundary conditions in Eqs. \eqref{eq:Tbc}-\eqref{eq:initcond},
can be solved using dimensional analysis \cite{buckingham1914physically,zeldovich1967physics,barenblatt1996scaling,shussman2015full,heizler2016self}.
This is performed in detail in Appendix \ref{sec:Dimensional-analysis}.
The result is a self-similar solution whose independent dimensionless
coordinate is: 
\begin{equation}
\xi=\frac{x}{t^{\delta}\left(KT_{0}^{4+\alpha-\beta}\right)^{\frac{1}{2}}},\label{eq:xsi_def}
\end{equation}
with the similarity exponent: 
\begin{equation}
\delta=\frac{1}{2}\left(1+\tau\left(4+\alpha-\beta\right)\right),\label{eq:similarity_exponent}
\end{equation}
and the solution is given in terms of a self-similar profile: 
\begin{equation}
T\left(x,t\right)=T_{0}t^{\tau}f\left(\xi\right).\label{eq:Tss}
\end{equation}
The similarity exponent can be written as $\delta=\frac{1}{2}\left(1+\tau/\tau_{H}\right)$,
where we have defined the temporal exponent 
\begin{equation}
\tau_{H}=\frac{1}{4+\alpha-\beta},\label{eq:tau_h}
\end{equation}
that is always positive, which sets the Marshak wave dynamics in space:
the propagation accelerates ($\delta>1$) for $\tau>\tau_{H}$, has
a constant speed ($\delta=1$) for $\tau=\tau_{H}$ and decelerates
($\delta<1$) for $\tau<\tau_{H}$. We see that for a constant surface
temperature drive ($\tau=0$), the similarity exponent does not depend
on $\alpha,\beta$ and is always $\delta=\frac{1}{2}$.

By plugging the self-similar form Eq. \eqref{eq:Tss} into the nonlinear
diffusion equation \eqref{eq:Tpde} and using the relations $\frac{\partial\xi}{\partial t}=-\frac{\delta\xi}{t}$
and $\frac{\partial\xi}{\partial x}=\frac{\xi}{x}$, all dimensional
quantities are factored out, and the following second order ordinary
differential equation (ODE) for the similarity profile is obtained
\cite{garnier2006self,shussman2015full,hristov2018heat}:

\begin{align}
f''\left(\xi\right) & =\frac{\beta\left[f\left(\xi\right)\right]^{\beta-\alpha-4}}{\left(4+\alpha\right)}\left(\tau f\left(\xi\right)-\delta f'\left(\xi\right)\xi\right)\nonumber \\
 & -\left(3+\alpha\right)\frac{f'^{2}\left(\xi\right)}{f\left(\xi\right)}.\label{eq:ode}
\end{align}
The surface temperature boundary condition {[}Eq. \eqref{eq:Tbc}{]},
is written in terms of the similarity profile as: 
\begin{equation}
f\left(0\right)=1.\label{eq:f0_ss_bc}
\end{equation}

The dimensionless problem defined by Eqs. \eqref{eq:ode}-\eqref{eq:f0_ss_bc}
is completely defined by the exponents $\alpha$, $\beta$, $\tau$.
Since the opacity is usually lower for a hotter material, we have
$\alpha\geq0$. Similarly, since the total energy density should be
an increasing function of temperature, we have $\beta>0$. Regarding
the temporal exponent $\tau$, it is evident from Eq. \eqref{eq:similarity_exponent}
that in order for heat to propagate outwards ($\delta>0$), we must
have $\tau>-\tau_{H}$, which is reported in Refs. \cite{garnier2006self,hristov2018heat}
as the minimal allowed value of $\tau$. However, as will be shown
below, the problem is only valid for $\tau>\tau_{c}>-\tau_{H}$, where
the correct minimal value is given by: 
\begin{equation}
\tau_{c}=-\frac{1}{4+\alpha}.\label{eq:tau_c}
\end{equation}

The solutions of the diffusion equation \eqref{eq:uPDE}, and as a
result, the solutions of Eq. \eqref{eq:Tpde} and the ODE Eq. \eqref{eq:ode}
take two different forms \cite{zeldovich1967physics,heaslet1961diffusion,pert1977class,kass1966numerical,barenblatt1996scaling,krief2021analytic}
depending on the nonlinearity index $n$ {[}Eq. \eqref{eq:ndef}{]},
as shown in Fig. \ref{fig:profiles_shapes}. If $n>0$ (that is, $\beta<4+\alpha$),
the problem is strictly nonlinear, and its solutions have a well defined
heat front, that is, there exists a finite heat front coordinate $\xi_{0}$,
for which $f\left(\xi\right)=0$ for $\xi\geq\xi_{0}$. These nonlinear
solutions are discussed below in Sec. \ref{sec:Non-linear-conduction}.
If $n=0$ (that is, $\beta=4+\alpha$), the problem is linear, as
the diffusion equation \eqref{eq:Tpde} and the ODE Eq.\eqref{eq:ode}
become linear equations in terms of $T^{\beta}$ and $f^{\beta}\left(\xi\right)$,
respectively, with solutions which decay to zero gradually as $\xi\rightarrow\infty$,
without a well defined front (that is, $\xi_{0}\rightarrow\infty$
when $n\rightarrow0$) \cite{heaslet1961diffusion,kass1966numerical,krief2021analytic}.
These linear solutions are discussed below in Sec. \ref{sec:Linear-conduction}.

Some general characteristics of the solutions in various ranges of
the temporal exponent $\tau$ are classified in Tab. \ref{tab:origin_behaviou}
and presented in Fig. \ref{fig:profiles_shapes}. Tab.~\ref{tab:origin_behaviou} is one of the major results of this study. It includes the specific
values of $\tau$ for which the ODE Eq. \eqref{eq:ode} can be solved
analytically, the ranges of $\tau$ for which the solution
is monotonically decreasing or has a maxima, and the total energy
in the system is constant, increasing or decreasing in time. These
results will be discussed below for both linear and nonlinear conduction
in sections \ref{sec:Linear-conduction}-\ref{sec:Non-linear-conduction}.

\subsection{The total energy in the system\label{subsec:The-total-energy}}

The total energy in the system at time $t$ is

\begin{equation}
E\left(t\right)=\int_{0}^{\infty}u\left(x,t\right)dx.
\end{equation}
Using the self-similar solution \eqref{eq:Tss} in the energy density
equation of state {[}Eq. \eqref{eq:eos}{]}, the total energy can
be written as the following temporal power law:

\begin{align}
E\left(t\right)= & u_{0}\left(KT_{0}^{4+\alpha+\beta}\right)^{\frac{1}{2}}t^{\frac{1}{2}\left(\tau\left(4+\alpha+\beta\right)+1\right)}\mathcal{E},\label{eq:Etotss}
\end{align}
where we have defined the dimensionless energy integral: 
\begin{equation}
\mathcal{E}=\int_{0}^{\infty}f^{\beta}\left(\xi\right)d\xi.\label{eq:Iint}
\end{equation}
Similarly, the radiation energy flux profile is obtained by using
Eq. \eqref{eq:Tss} in Eq. \eqref{eq:FFF}:

\begin{align}
F\left(x,t\right)= & -u_{0}\left(KT_{0}^{4+\alpha+\beta}\right)^{\frac{1}{2}}t^{\frac{1}{2}\left(\tau\left(4+\alpha+\beta\right)-1\right)}\nonumber \\
 & \times\left(4+\alpha\right)f^{3+\alpha}\left(\xi\right)f'\left(\xi\right).\label{eq:flux_selfsim}
\end{align}

It is evident from Eq. \eqref{eq:Etotss}, that for the following
temporal exponent:

\begin{equation}
\tau_{Z}=-\frac{1}{4+\alpha+\beta},\label{eq:tau_Z}
\end{equation}
which is always negative, the total energy is constant. This means
that for $\tau=\tau_{Z}$, the Marshak wave problem defined in Sec.
\ref{sec:Statement-of-the} in terms of a time dependent surface temperature
{[}Eq. \eqref{eq:Tbc}{]}, is identical to the instantaneous point
source problem that was studied by Zel'dovich, Barenblatt and others
\cite{zel1959propagation,zeldovich1967physics,barenblatt1996scaling,pattle1959diffusion,boyer1961some,ames1965similarity,lonngren1974field,tuck1976some,pert1977class,king1990exact,krief2021analytic,malka2022supersonic},
which is defined in terms of the initial energy which is deposited
at a point at $t=0$. Indeed, the time dependence of the temperature
at $x=0$ of the solution of the instantaneous point source problem
in planar symmetry, is given by $T_{0}t^{\tau_{Z}}$ (see for example
\cite[Eq.~(34)]{pert1977class} and 
\cite[Eqs.~(38)-(41)]{krief2021analytic}).
This equivalence will be used below for both linear and nonlinear
conduction.

The energy balance in the system as a function of $\tau$ results
from the relations $E\left(t\right)\sim t^{\left(\tau_{Z}-\tau\right)/2\tau_{Z}}$
and $dE/dt\sim F\left(x=0,t\right)\sim-f'\left(0\right)$, which hold
for both linear and nonlinear conduction. For $\tau=\tau_{Z}$, the
energy is constant and we must have $f'\left(0\right)=0$ (a local
maxima at $\xi=0$), as the net surface flux is zero (the incoming
and outcoming surface fluxes are equal). When $\tau>\tau_{Z}$ the
total energy in the system increases as the heat wave propagates,
due to the positive net surface flux, so we must have $f'\left(0\right)<0$
and a strictly monotonically decreasing solution, which is the familiar
scenario of the Marshak wave. On the other hand, when $\tau<\tau_{Z}$,
the energy is decreasing over time with a negative surface flux, so
that $f'\left(0\right)>0$, which means that the solution must have
a local maxima at some finite $\xi>0$. The existence of a local maxima
(as seen below in Figs. \ref{fig:profiles_shapes}, \ref{fig:heat_front_forms},
\ref{fig:origin_forms}-\ref{fig:origin_forms-eps013}), appears since
for $\tau<\tau_{Z}$, the drop in the energy density at the surface
(according to Eq. \eqref{eq:Tbc}) is faster than the rate of heat
propagation, causing the temperature at the origin to be lower than
the solution ahead. Indeed, this condition which is $\delta<-\beta\tau$
(see Eq. \eqref{eq:xsi_def}), is equivalent to $\tau<\tau_{Z}$.

This behavior of the solutions in different ranges of $\tau$ is evident
in Figs. \ref{fig:profiles_shapes}, \ref{fig:heat_front_forms},
\ref{fig:origin_forms}-\ref{fig:origin_forms-eps013}, \ref{fig:profiles_fit}
and \ref{fig:ftag0_err}.

\subsection{Marshak boundary condition\label{subsec:Marshak-boundary-condition}}

It is customary to define the Marshak wave problem in terms of a prescribed
incoming radiative flux \cite{pomraning1979non,bingjing1996benchmark,fleck1971implicit,larsen1988grey,olson2000diffusion,densmore2012hybrid,yee2017stable,mclean2022multi,steinberg2022multi,steinberg2023frequency,krief2024self,zhang2023fully,liu2023implicit,li2024unified},
rather than the surface temperature boundary condition {[}Eq. \eqref{eq:Tbc}{]}.
The latter boundary condition is more natural to apply under the diffusion
approximation, while the former is more natural to use in the solution
of the more elaborate radiation transport equation, which has the
angular surface flux as a boundary condition (see below in Sec. \ref{subsec:Transport-setup}).
The net surface flux, which is known from the analytical solution,
can be used to relate these two different boundary conditions.

In the LTE diffusion approximation, the incoming flux boundary condition,
also known as the Marshak (or Milne) boundary condition \cite{pomraning1979non,bingjing1996benchmark,olson2000diffusion,rosen2005fundamentals}
at $x=0$ is:

\begin{equation}
\frac{4}{c}F_{\text{inc}}\left(t\right)=aT^{4}\left(x=0,t\right)+\frac{2}{c}F\left(x=0,t\right),\label{eq:marsh_bc_def}
\end{equation}
where $F_{\text{inc}}\left(t\right)$ is a given time-dependent incoming
radiation energy flux at $x=0$. For a medium coupled to a heat bath
at temperature $T_{\text{bath}}\left(t\right)$, the incoming flux
is $F_{\text{inc}}\left(t\right)=\frac{ac}{4}T_{\text{bath}}^{4}\left(t\right)$.
Since the surface temperature is given by Eq. \eqref{eq:Tbc}, the
Marshak boundary condition \eqref{eq:marsh_bc_def} can be written
as:

\begin{equation}
T_{\text{bath}}\left(t\right)=\left(T_{s}^{4}\left(t\right)+\frac{2}{ac}F\left(x=0,t\right)\right)^{\frac{1}{4}},\label{eq:Tbath_marsh_bc_F}
\end{equation}
which relates the bath temperature, the surface temperature and the
net surface flux. Using the flux of the similarity solution from Eq.
\eqref{eq:flux_selfsim} in Eq. \eqref{eq:Tbath_marsh_bc_F}, gives
the following expression for the time dependent bath temperature:
\begin{equation}
T_{\text{bath}}\left(t\right)=\left(1+Bt^{\frac{1}{2}\left(\tau\left(\alpha+\beta-4\right)-1\right)}\right)^{\frac{1}{4}}T_{0}t^{\tau},\label{eq:Tbath}
\end{equation}
where we defined the bath constant:

\begin{align}
B & =-\frac{2u_{0}}{ac}\left(KT_{0}^{\alpha+\beta-4}\right)^{\frac{1}{2}}\left(4+\alpha\right)f'\left(0\right)\nonumber \\
 & =-\left(\frac{16\left(4+\alpha\right)u_{0}}{3ack_{0}T_{0}^{4-\alpha-\beta}}\right)^{\frac{1}{2}}f'\left(0\right)\label{eq:Bbath}
\end{align}
It is evident that only for $\tau=\frac{1}{\alpha+\beta-4}$, the
bath temperature is given by a temporal power law, which has the same
temporal power $\tau$ of the surface temperature. We see that $T_{\text{bath}}\left(t\right)>T_{s}\left(t\right)$
for $\tau>\tau_{Z}$, $T_{\text{bath}}\left(t\right)=T_{s}\left(t\right)$
for $\tau=\tau_{Z}$, and $T_{\text{bath}}\left(t\right)<T_{s}\left(t\right)$
for $\tau<\tau_{Z}$. It is also evident that $B$ decreases as $k_{0}^{-1/2}$,
so that $T_{\text{bath}}\left(t\right)\approx T_{s}\left(t\right)$
for opaque problems.

The results above agree with several previous works on on supersonic
\cite{rosen2005fundamentals,cohen2018modeling,cohen2020key} and subsonic
\cite{heizler2021radiation} LTE Marshak waves as well as non-LTE
supersonic Marshak waves \cite{krief2024self}.

\section{Linear conduction\label{sec:Linear-conduction}}

As discussed above, when

\begin{equation}
\beta=4+\alpha,\label{eq:betalin}
\end{equation}
that is, $n=0$, Eq. \eqref{eq:uPDE} is linear, and the diffusion
equation \eqref{eq:Tpde} becomes a linear equation in terms of the
total energy density $u$ which is proportional to $T^{\beta}$. In
this case, the ODE \eqref{eq:ode} becomes a linear ODE in terms of
the total energy density similarity profile, $f^{\beta}$: 
\begin{equation}
\left[f^{\beta}\left(\xi\right)\right]''+\frac{1}{2}\xi\left[f^{\beta}\left(\xi\right)\right]'-\beta\tau f^{\beta}\left(\xi\right)=0.
\end{equation}
Under the boundary condition in Eq. \eqref{eq:f0_ss_bc} and assuming
$f\left(\xi\rightarrow\infty\right)=0$, this equation has the solution:
\begin{equation}
f^{\beta}\left(\xi\right)=\frac{\Gamma\left(1+\beta\tau\right)}{\sqrt{\pi}}e^{-\xi^{2}/4}U\left(\frac{1}{2}+\beta\tau,\frac{1}{2},\frac{\xi^{2}}{4}\right),\label{eq:f_lin_sol}
\end{equation}
where $\Gamma$ denotes the Gamma function and $U$ is the Tricomi
confluent hypergeometric function of the second kind (also known as
the Kummer $U$ function), which is related to the Hermite polynomial
of fractional order by $H_{\nu}\left(x\right)=2^{\nu}U\left(-\frac{\nu}{2},\frac{1}{2},x^{2}\right)$.
The $U$ function is available in most scientific scripting languages
such as python scipy (as scipy.special.hyperu \cite{virtanen2020scipy}),
Matlab and Mathematica. The solution in Eq. \eqref{eq:f_lin_sol}
is an exact analytic solution of the Marshak wave problem for a general
$\tau$, and was derived in Ref. \cite{carslaw_jaeger1959} for integer
$\beta\tau$ and in Refs. \cite{pert1977class,kot2019parabolic} for
general $\tau$. It is evident that this solution breaks down at $\tau\leq-\frac{1}{\beta}=\tau_{c}$.
As expected, the linear solution does not have a well defined heat
front, and it decays according to: 
\begin{align}
\lim_{\xi\rightarrow\infty} & e^{-\xi^{2}/4}U\left(\frac{1}{2}+\beta\tau,\frac{1}{2},\frac{\xi^{2}}{4}\right)\nonumber \\
 & =e^{-\xi^{2}/4}\xi^{-\beta\tau-\frac{1}{2}}\left(1+\mathcal{O}\left(\frac{1}{\xi}\right)\right).\label{eq:fxsi_lin_asymp}
\end{align}
The derivative, which gives the radiation energy flux {[}Eq. \eqref{eq:flux_selfsim}{]},
can be calculated using the relation: 
\begin{equation}
f'\left(\xi\right)=-\frac{\xi}{2\beta}f\left(\xi\right)\frac{U\left(\frac{1}{2}+\beta\tau,\frac{3}{2},\frac{\xi^{2}}{4}\right)}{U\left(\frac{1}{2}+\beta\tau,\frac{1}{2},\frac{\xi^{2}}{4}\right)},
\end{equation}
and the derivative at the origin, which determines the bath temperature
{[}Eqs. \eqref{eq:Tbath}-\eqref{eq:Bbath}{]}, is given by: 
\begin{equation}
f'\left(0\right)=\begin{cases}
-\frac{\Gamma\left(1+\beta\tau\right)}{\beta\Gamma\left(\frac{1}{2}+\beta\tau\right)}, & \beta\tau\neq-\frac{1}{2}\\
0, & \beta\tau=-\frac{1}{2}
\end{cases}\label{eq:f'0_linear}
\end{equation}
and switches from a positive to a negative value at $\tau=\tau_{Z}=-\frac{1}{2\beta}$.
This means that in the range $-\frac{1}{\beta}<\tau<-\frac{1}{2\beta}$,
the solution is not monotonic and has a local maxima, the surface
flux is negative, and that the total energy in the system is decreasing
in time, as was discussed in Sec. \ref{subsec:The-total-energy}.
The solutions are shown in Fig. \ref{fig:profiles_shapes} for various
values of $\tau$.

The energy integral of the solution in Eq. \eqref{eq:f_lin_sol} can
be calculated analytically, and is given by: 
\begin{align}
\mathcal{E}=\int_{0}^{\infty}f^{\beta}\left(\xi\right)d\xi & =\frac{\Gamma\left(1+\beta\tau\right)}{\Gamma\left(\frac{3}{2}+\beta\tau\right)}.\label{eq:Eint_linear_anal}
\end{align}
As discussed in Sec. \ref{subsec:The-total-energy} above, for $\tau=\tau_{Z}=-\frac{1}{2\beta}$,
the surface temperature is decreasing in such a way that the total
energy (Eq. \eqref{eq:Etotss}) is time independent, and the solution
should be identical to the Green's function of the (linear) diffusion
equation, which is the solution of the instantaneous point source
problem \cite{zel1959propagation,zeldovich1967physics,barenblatt1996scaling,krief2021analytic}.
Indeed, since $U\left(0,\frac{1}{2},z\right)=1$, in this case Eq.
\eqref{eq:f_lin_sol} is reduced to the well known Gaussian Green's
function: 
\begin{align}
f^{\beta}\left(\xi\right) & =e^{-\frac{\xi^{2}}{4}}.\label{eq:f_zel_lin}
\end{align}
The analogous solution for nonlinear conduction is described below
in Sec. \ref{subsec:The-Zel'dovich-Barenblatt-analyt}.

For $\tau=0$, using the identity $U\left(\frac{1}{2},\frac{1}{2},z\right)=\sqrt{\pi}e^{z}\text{erfc}\left(\sqrt{z}\right)$,
we find:

\begin{equation}
f^{\beta}\left(\xi\right)=\text{erfc}\left(\frac{\xi}{2}\right),\label{eq:f_tau0_lin}
\end{equation}
where $\text{erfc}$ is the complementary error function. This result
is in agreement with Refs. \cite{jost1952diffusion,carslaw_jaeger1959,philip1960general,heaslet1961diffusion,kass1966numerical,kot2018parabolic}
that give the linear conduction solution for $\tau=0$.

Finally, we note that for linear conduction we have $\tau_{H}\rightarrow\infty$,
and the similarity exponent does not depend on $\alpha,\beta,\tau$
and is always $\delta=\frac{1}{2}$, that is, heat propagates as $t^{1/2}$
for any value of $\tau$ (in contrast to nonlinear conduction).

\section{Nonlinear conduction\label{sec:Non-linear-conduction}}

\begin{table*}[t]
\centering{}%
\begin{tabular}{|c|c|c|c|c|}
\cline{2-5} \cline{3-5} \cline{4-5} \cline{5-5} 
\multicolumn{1}{c|}{$ $} & $0\leq\beta<3+\alpha$  & $\beta=3+\alpha$  & $3+\alpha<\beta<4+\alpha$  & $\beta=4+\alpha$\tabularnewline
\hline 
$f'\left(\xi_{0}\right)$  & $-\infty$  & finite  & 0  & \tabularnewline
\hline 
$\nu$  & $0<\nu<1$  & $\nu=1$  & $\nu>1$  & $\nu=\infty$\tabularnewline
\hline 
front  & sharp  & linear  & flat  & -\tabularnewline
\hline 
conduction  & \multicolumn{3}{c|}{nonlinear} & linear\tabularnewline
\hline 
\end{tabular}\caption{Behavior of the heat front in different ranges of the energy density
temperature exponent $\beta$ (see also Figs. \ref{fig:mesh_ny}-\ref{fig:heat_front_forms_and_asymp}).\label{tab:front_behaviour}}
\end{table*}

\begin{figure}
\begin{centering}
\includegraphics[scale=0.65]{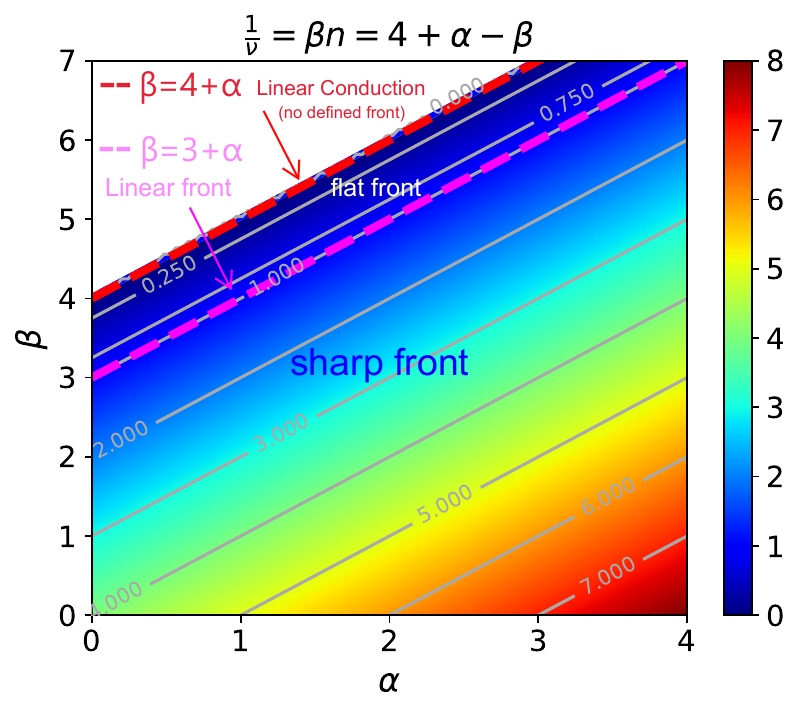} 
\par\end{centering}
\caption{A mapping of the various heat propagation modes as a function of opacity
and energy density exponents $\alpha$, $\beta$ (see also Tab. \ref{tab:front_behaviour}).
The color represents the inverse of the front exponent $\nu$ which is a measures the heat
front steepness (see Eqs. \eqref{eq:asymp_front}-\eqref{eq:nudef}). The dashed magenta
line represents the points $\beta=3+\alpha$, for which the front
is linear ($\nu=1$). The points below this line ($\beta<3+\alpha$)
result in sharp heat fronts ($\nu<1$), while the point above it ($3+\alpha<\beta<4+\alpha$),
result in flat heat fronts ($\nu>1$). The dashed red line represents
the points $\beta=4+\alpha$, for which heat is conducted linearly
with no defined heat front ($\nu\rightarrow\infty$), for which the
solution decays indefinitely, according to Eq. \eqref{eq:fxsi_lin_asymp}.
\label{fig:mesh_ny}}
\end{figure}

\begin{figure*}[t]
\begin{centering}
\includegraphics[scale=0.4]{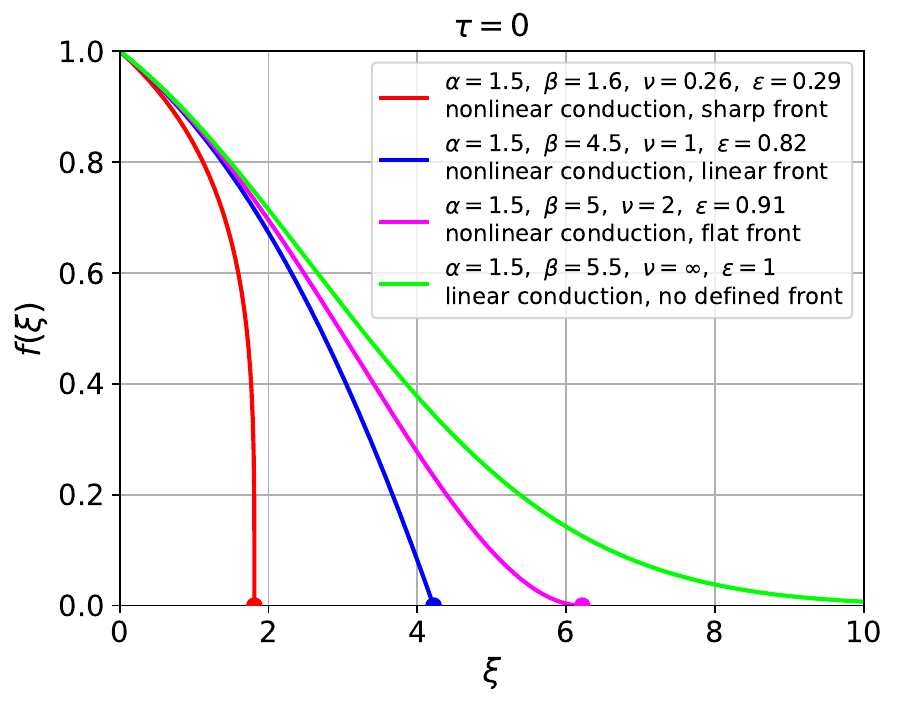}\includegraphics[scale=0.4]{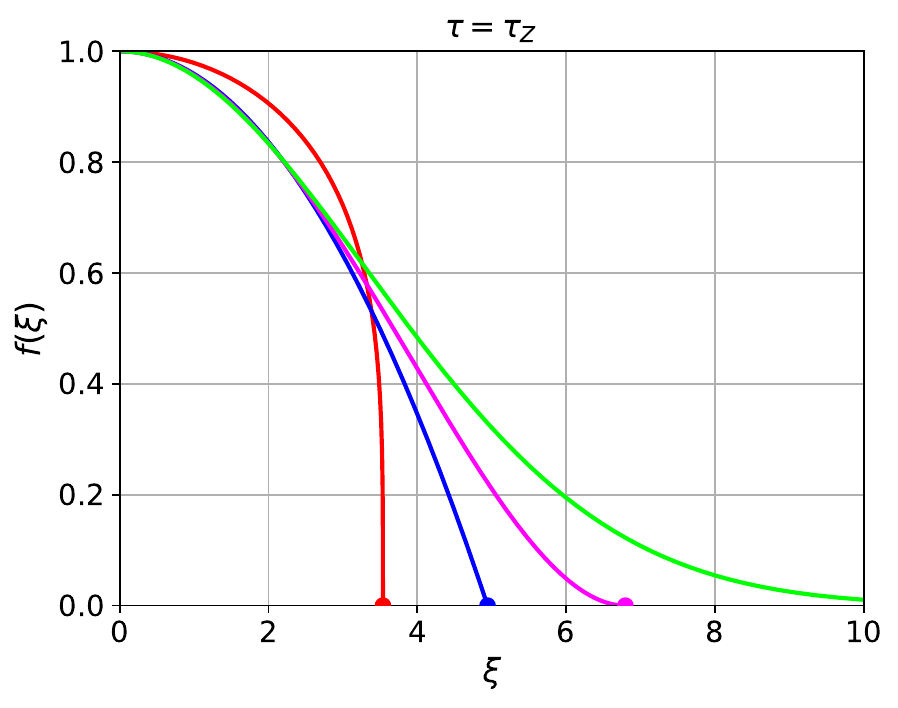}\includegraphics[scale=0.4]{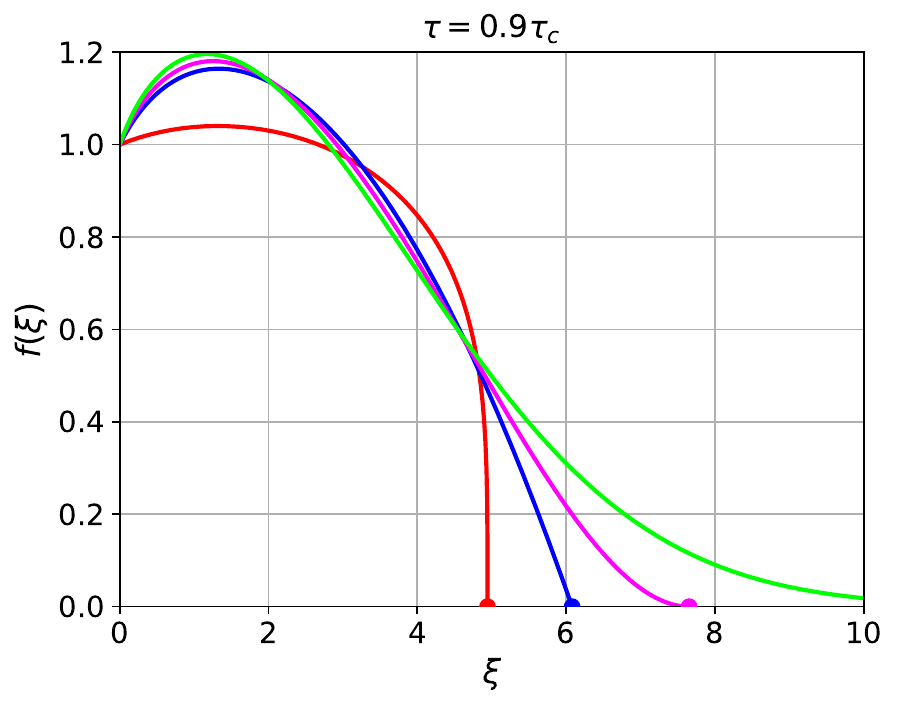} 
\par\end{centering}
\caption{Various forms of the heat front structure, as outlined in Tab. \ref{tab:front_behaviour}
and in Fig. \ref{fig:mesh_ny}. The similarity profiles are shown
for $\alpha=1.5$ and varying $\beta$: a nonlinear sharp front $\beta=1.6<3+\alpha$
(in red); a linear front $\beta=4.5=3+\alpha$ (in blue); a flat (parabolic)
front $\beta=5>3+\alpha$ (in magenta); and the linear conduction
solution (undefined front) $\beta=4+\alpha=5.5$ (in green). The profiles
are shown for $\tau=0$ (left figure), $\tau=\tau_{Z}$ (middle figure)
and $\tau=0.9\tau_{c}$ (right figure). The values of the front exponent
$\nu$ {[}Eq. \eqref{eq:nudef}{]} and steepness $\epsilon$ {[}Eq.
\eqref{eq:eps_HR}{]} are listed in the legend. The front locations
$\xi_{0}$ are marked with a circle. The behavior of the solutions
for the different values of $\tau$ agree with Tab. \ref{tab:origin_behaviou}.
\label{fig:heat_front_forms}}
\end{figure*}

\begin{figure*}[t]
\begin{centering}
\includegraphics[scale=0.38]{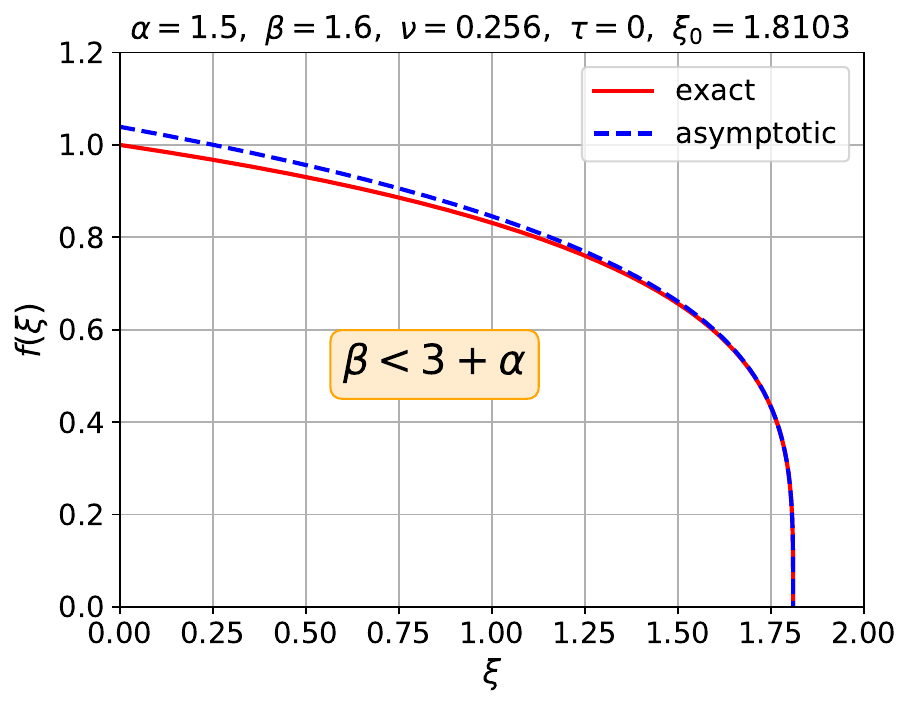}\includegraphics[scale=0.38]{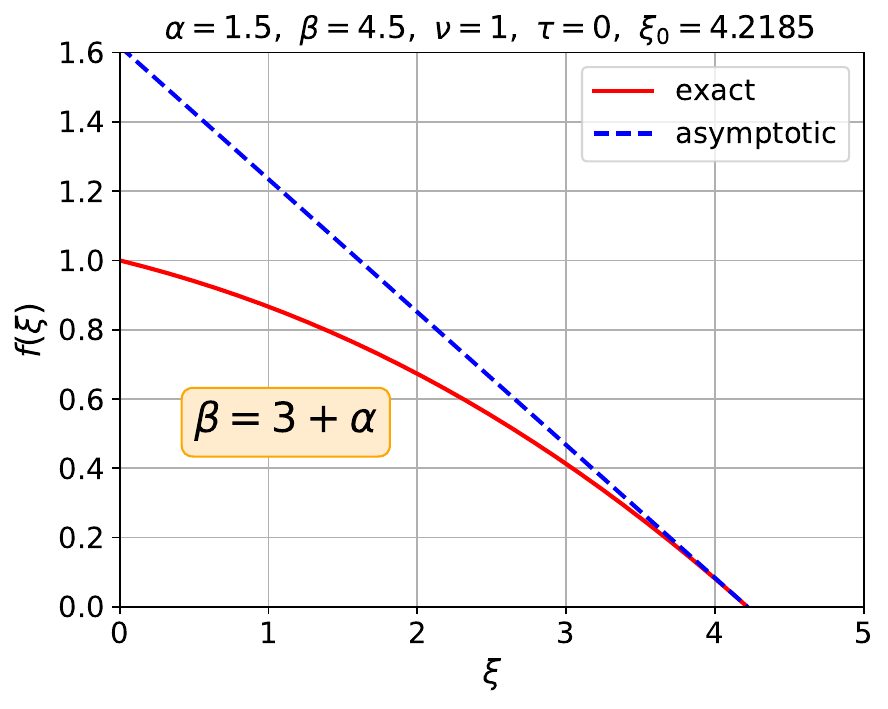}\includegraphics[scale=0.38]{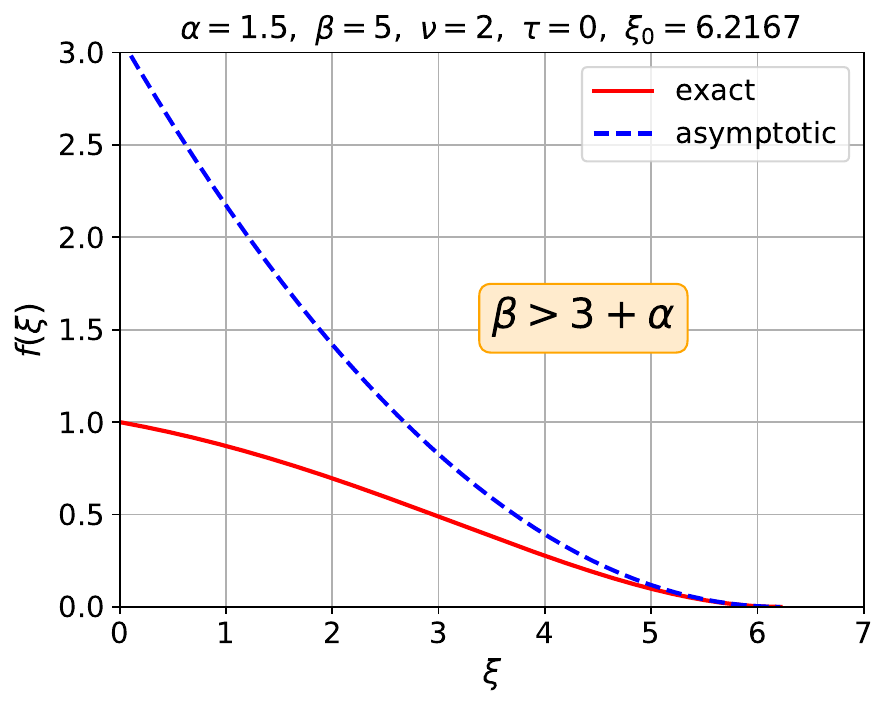} 
\par\end{centering}
\begin{centering}
\includegraphics[scale=0.38]{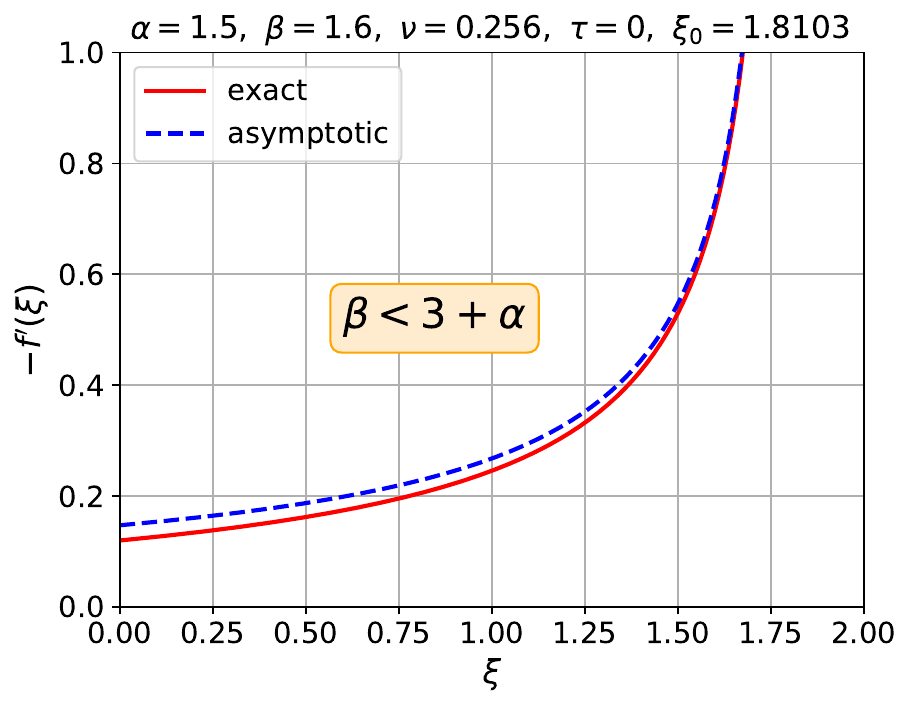}\includegraphics[scale=0.38]{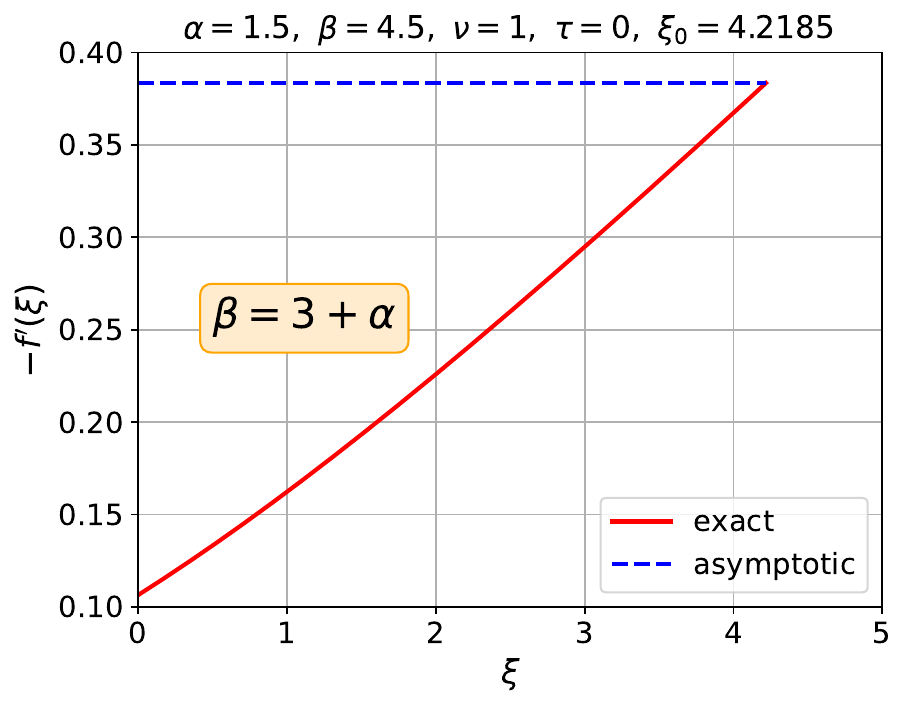}\includegraphics[scale=0.38]{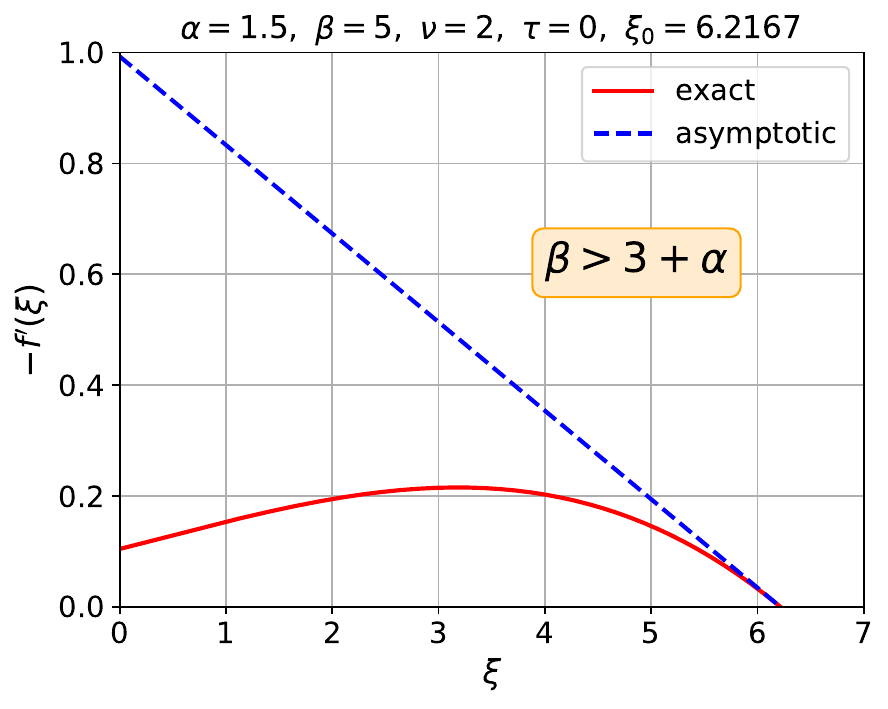} 
\par\end{centering}
\caption{A comparison of nonlinear solution profiles (upper figures) and derivatives
(lower figures), to the near-front asymptotic approximation {[}Eqs.
\eqref{eq:asymp_front}-\eqref{eq:f0_asymp_front}{]}, for $\tau=0$.
These solutions were also shown in the leftmost pane of Fig. \ref{fig:heat_front_forms}.
\label{fig:heat_front_forms_and_asymp}}
\end{figure*}

For nonlinear conduction, $n>0$ (that is, $\beta<4+\alpha$), the
solution has a well defined heat front, that is, we are looking for
solutions which must obey $f\left(\xi\right)=0$ for $\xi\geq\xi_{0}$,
where the similarity coordinate at the heat front, $\xi_{0}$, is
finite. According to Eq. \eqref{eq:xsi_def}, the front position propagates
in time according to: 
\begin{equation}
x_{F}\left(t\right)=\xi_{0}t^{\delta}\left(KT_{0}^{4+\alpha-\beta}\right)^{\frac{1}{2}}.\label{eq:xheat}
\end{equation}
Near the front $\xi\approx\xi_{0}$, the solution has the following
asymptotic form:

\begin{equation}
f\left(\xi\right)=f_{0}\left(1-\frac{\xi}{\xi_{0}}\right)^{\nu},\label{eq:asymp_front}
\end{equation}
with $\nu\geq0$. By plugging this ansatz into Eq. \eqref{eq:ode},
one finds the front exponent: 
\begin{equation}
\nu=\tau_{H}=\frac{1}{4+\alpha-\beta}=\frac{1}{\beta n},\label{eq:nudef}
\end{equation}

\begin{align}
f_{0} & =\left[\frac{\delta\xi_{0}^{2}}{\left(4+\alpha\right)\nu}\right]^{\nu}.\label{eq:f0_asymp_front}
\end{align}
It is evident that the wave front is steeper for larger values of
the nonlinearity index $n$ (smaller $\nu$). The exponent given in
Eq. \eqref{eq:nudef} defines three different wave fronts, since the
derivative at the front is

\begin{equation}
\lim_{\xi\rightarrow\xi_{0}^{-}}f'\left(\xi\right)=\begin{cases}
-\infty & ,\ \beta<3+\alpha\\
-\nu f_{0} & ,\ \beta=3+\alpha\\
0 & ,\ \beta>3+\alpha
\end{cases}
\end{equation}
In the common case where $\beta<3+\alpha$, we have $0<\nu<1$, resulting
in $f'\left(\xi_{0}\right)=-\infty$, and the solution has the familiar
steep heat front. If $\beta>3+\alpha$, we have $\nu>1$ so that $f'\left(\xi_{0}\right)=0$,
and the solution has a flat heat front. For the special value $\beta=3+\alpha$,
we have $\nu=1$ so that $f'\left(\xi_{0}\right)$ is finite, and
the solution has a linear heat front. This is summarized in table
\ref{tab:front_behaviour}. The various front types are mapped as
a function of $\alpha,\beta$ in Fig. \ref{fig:mesh_ny}. It is interesting
to note that the front shape does not depend on the temperature drive
exponent $\tau$. The similarity profiles for these different cases are shown in Figure \ref{fig:origin_forms}.

Figure \ref{fig:heat_front_forms_and_asymp} shows
a comparison between the exact (numerical) solution of Eq. \eqref{eq:ode}
to the asymptotic near-front solution {[}Eq. \eqref{eq:asymp_front}{]},
for steep, linear and flat heat fronts. The expected behavior of the
solution $f\left(\xi\right)$ and its derivative $f'\left(\xi\right)$
near the front is evident. Other comparisons are shown for various
values of $\tau$ in Figs. \ref{fig:origin_forms}-\ref{fig:origin_forms-eps013},
showing a great agreement with the exact solution near the front.
At the origin, the asymptotic form has $f'\left(0\right)=-\nu f_{0}\xi_{0}^{\nu-1}$
which does not depend on $\tau$, and is always finite and negative,
as opposed to the exact solution, which, as discussed above can have
a positive or zero slope at the origin. This disagreement near the
origin is evident in Figs. \ref{fig:origin_forms}-\ref{fig:origin_forms-eps013}.

We note that the discussion above is consistent with the numerical
results in Refs. \cite{heaslet1961diffusion,kass1966numerical,hristov2016integral},
which examine nonlinear diffusion for a constant boundary temperature
($\tau=0$). In addition, we note that Eqs. \eqref{eq:nudef}-\eqref{eq:f0_asymp_front}
are consistent with Refs. \cite{heaslet1961diffusion,garnier2006self}.

\subsection{Numerical solution\label{subsec:Numerical-solution}}

\begin{figure*}[t]
\begin{centering}
\includegraphics[scale=0.38]{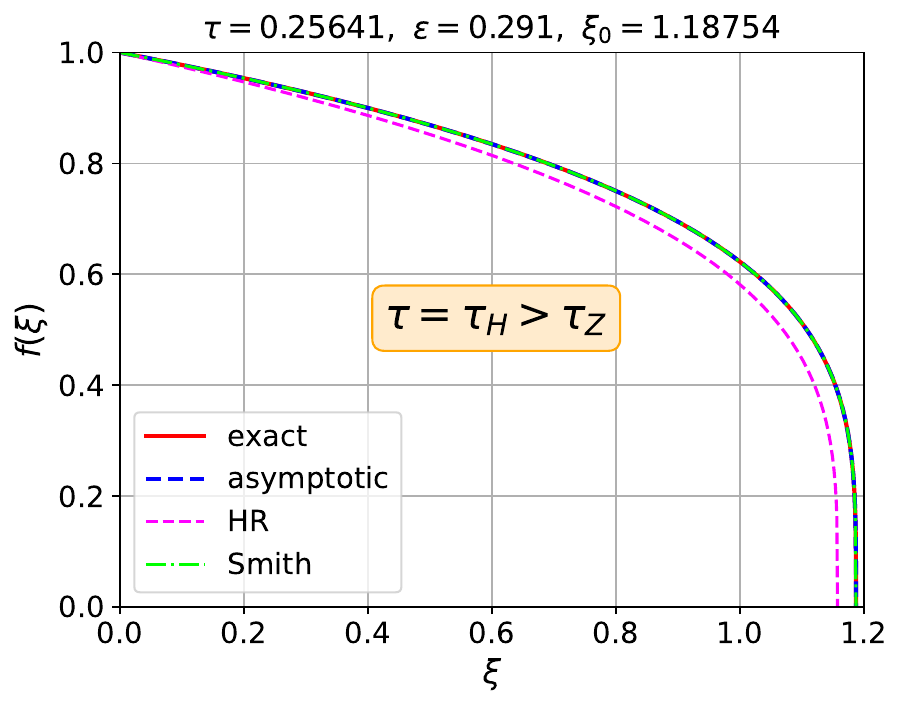}\includegraphics[scale=0.38]{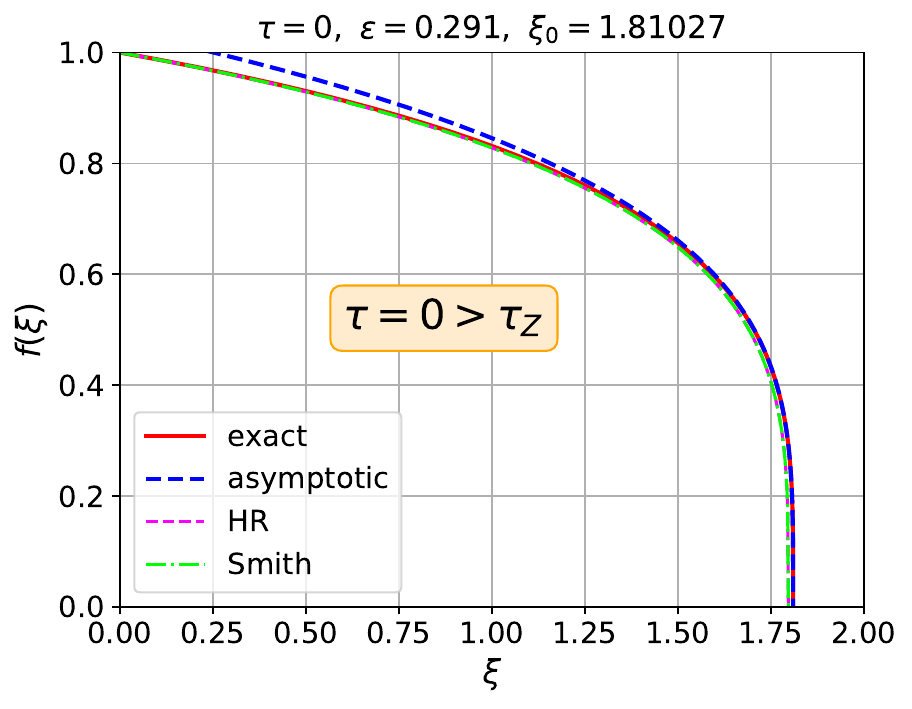}\includegraphics[scale=0.38]{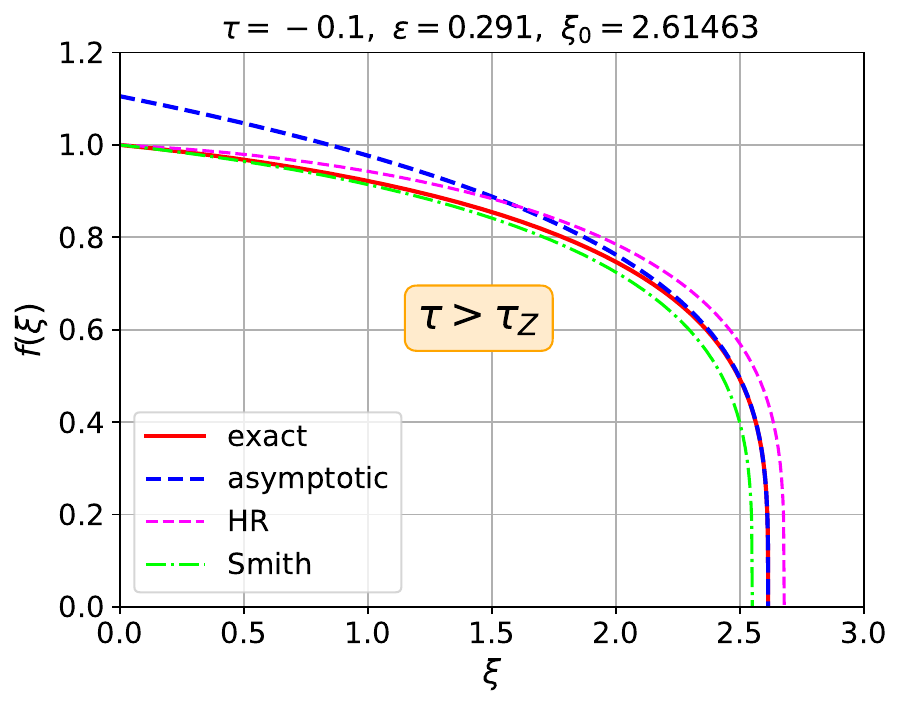} 
\par\end{centering}
\begin{centering}
\includegraphics[scale=0.38]{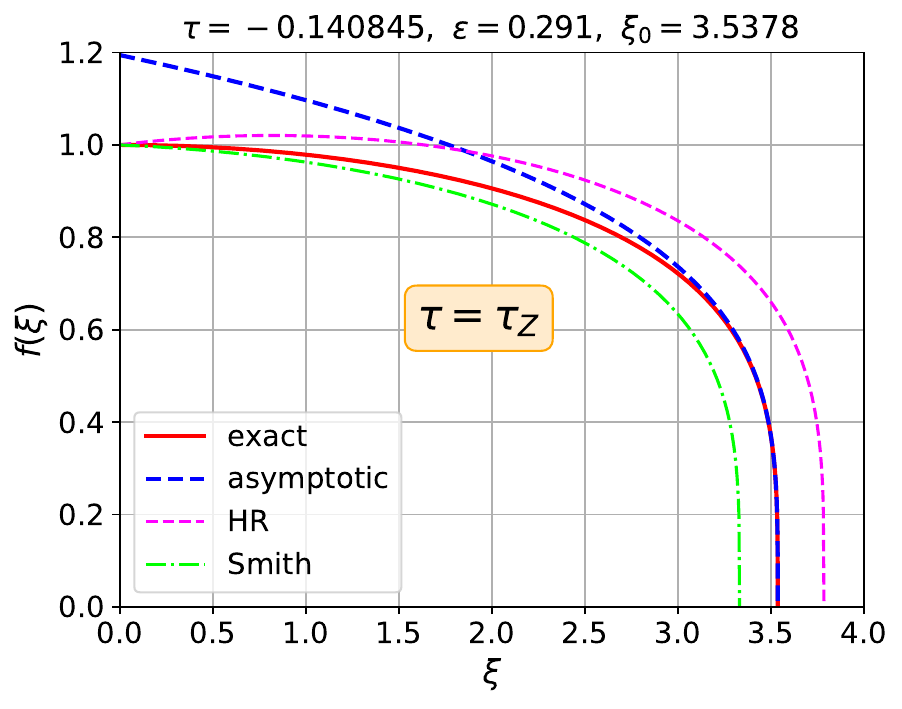}\includegraphics[scale=0.38]{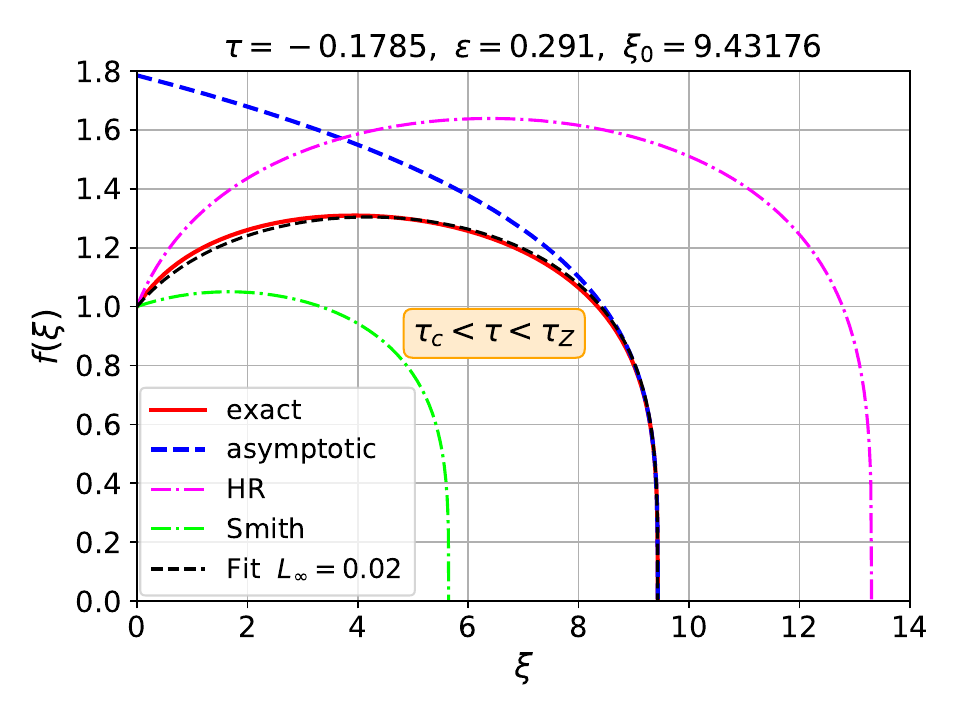}\includegraphics[scale=0.38]{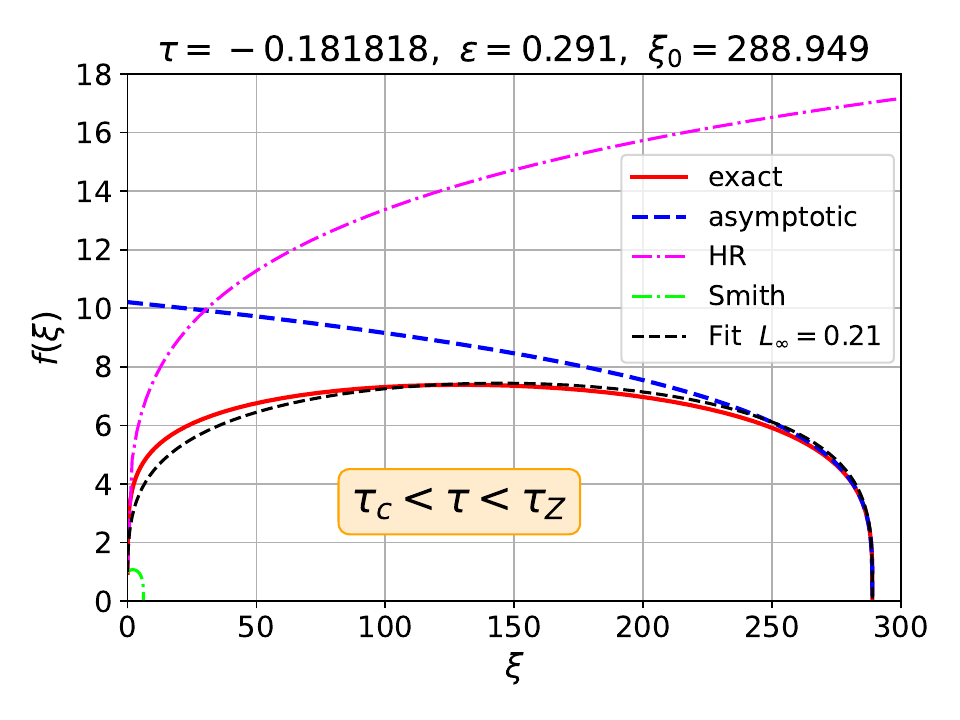} 
\par\end{centering}
\caption{Various forms of the temperature similarity profile $f\left(\xi\right)$
for various values of $\tau$ (as listed in the titles). Shown are
the exact profile (in red) which is obtained by a numerical solution
of the ODE Eq. \eqref{eq:ode} (red line), the near-front asymptotic
form (Eq. \eqref{eq:asymp_front}, blue line), the Hammer-Rosen approximation
(Eqs. \eqref{eq:f_hr},\eqref{eq:xsi0_HR},\eqref{eq:R_HR}, magenta
line) and the Smith approximation (Eqs. \eqref{eq:f_hr},\eqref{eq:XSI0_SM},\eqref{eq:R_SMITH},
green line). The approximated form Eq. \eqref{eq:f_hr} with fitted
parameters, are shown (in black) only for the bottom middle and right
figures, where the difference with respect to the exact solution is
visible (the $L_{\infty}$ (maximal) error norm is listed in the legend).
All profiles shown are for $\alpha=1.5$, $\beta=1.6$, for which
$\epsilon\approx0.29$, $\tau_{c}=-\frac{2}{11}=-0.181818...$, $\tau_{H}=\frac{10}{39}\approx0.2564$
and $\tau_{Z}=-\frac{10}{71}\approx-0.1408$. As discussed in the
text (see also table \ref{tab:origin_behaviou}), it is evident that
for $\tau>\tau_{Z}$, the solutions have a finite negative slope at
$\xi=0$ (upper figures) which represents a positive net incoming
flux, for $\tau=\tau_{Z}$ the solution has a zero slope at the origin
(lower left plot) which represents a zero net surface flux, while
for $\tau_{c}<\tau<\tau_{Z}$ the solution has a positive slope near
the origin, which leads to a non-monotonic behavior and the existence
of a local maxima (lower middle and right figures), with a net negative
surface flux.\label{fig:origin_forms}}
\end{figure*}

\begin{figure*}[t]
\begin{centering}
\includegraphics[scale=0.38]{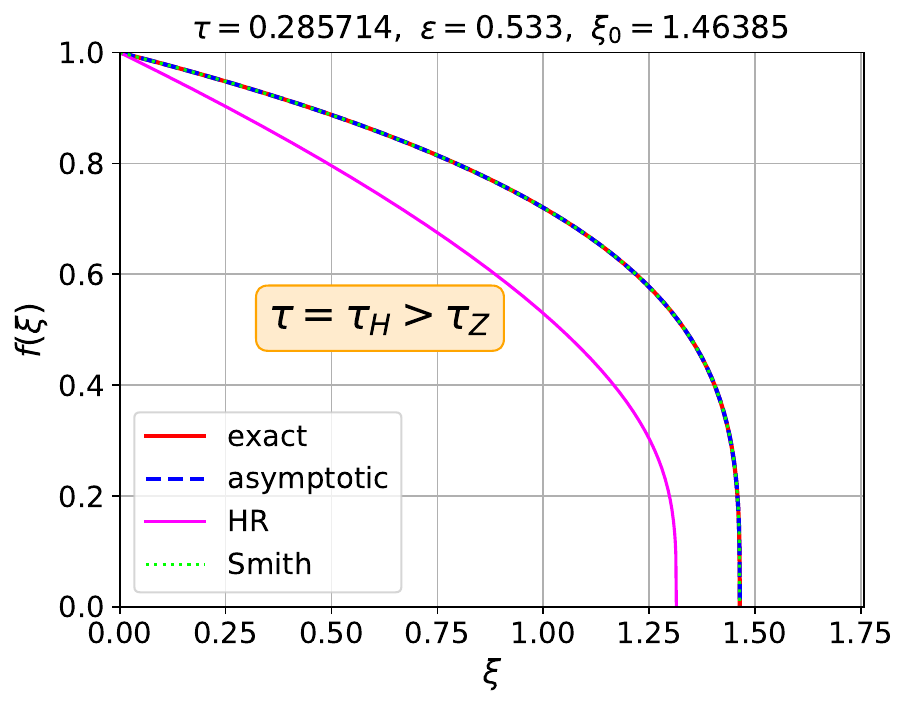}\includegraphics[scale=0.38]{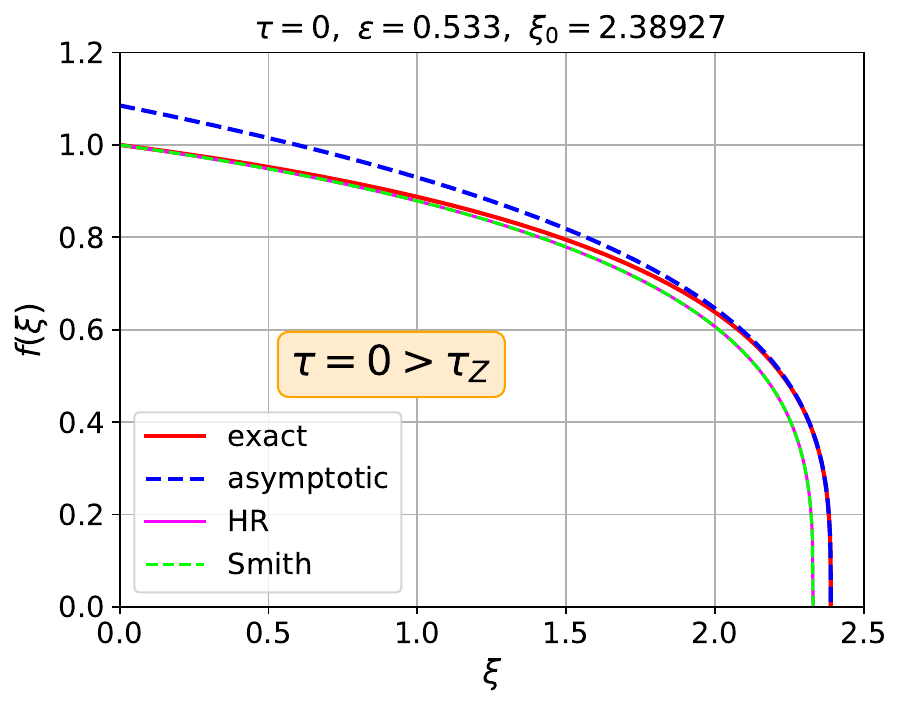}\includegraphics[scale=0.38]{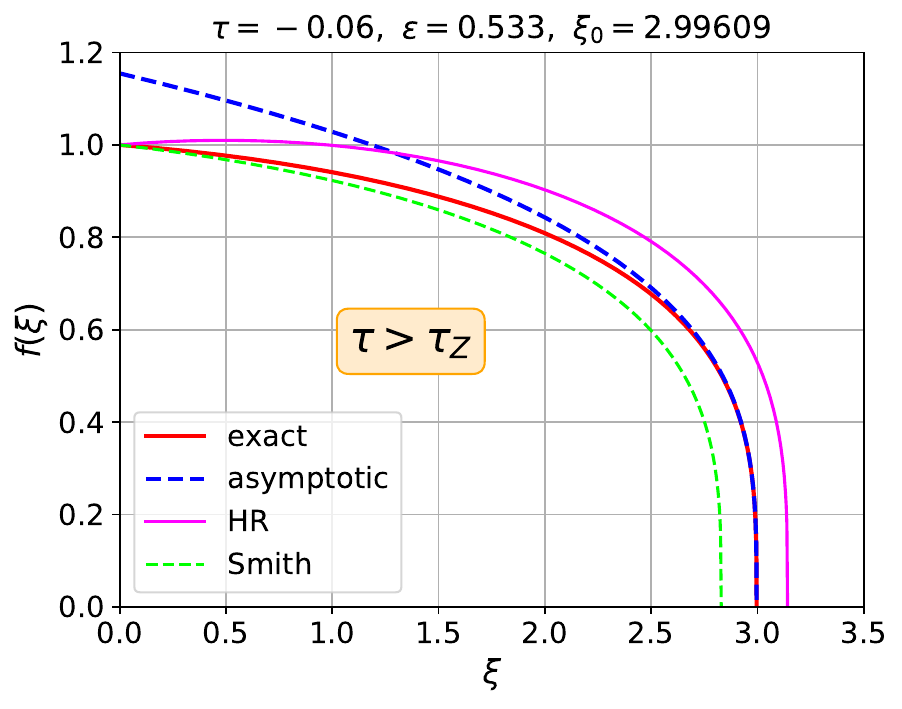} 
\par\end{centering}
\begin{centering}
\includegraphics[scale=0.38]{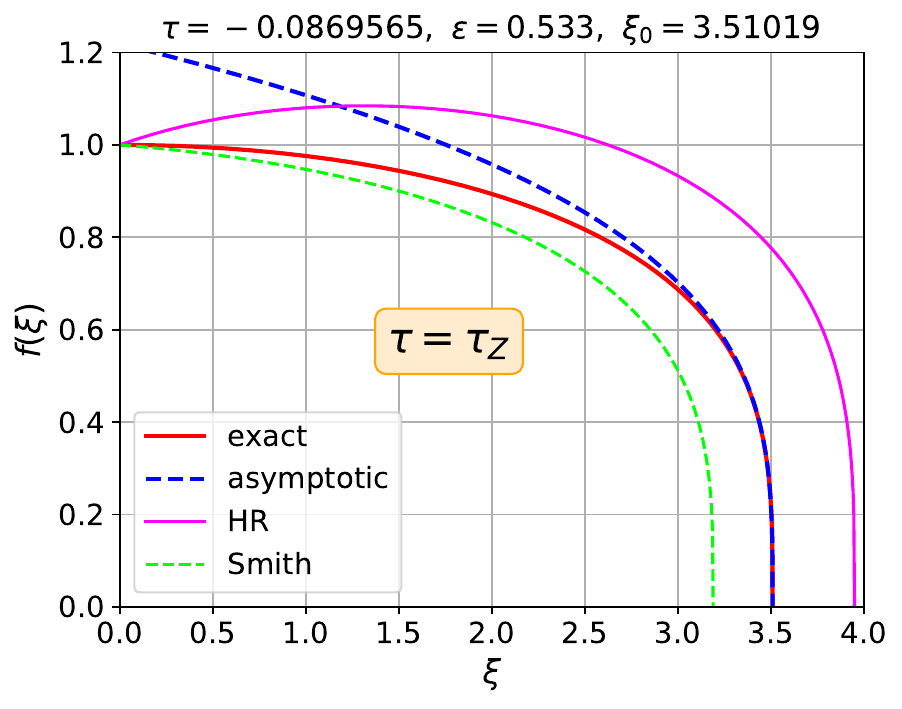}\includegraphics[scale=0.38]{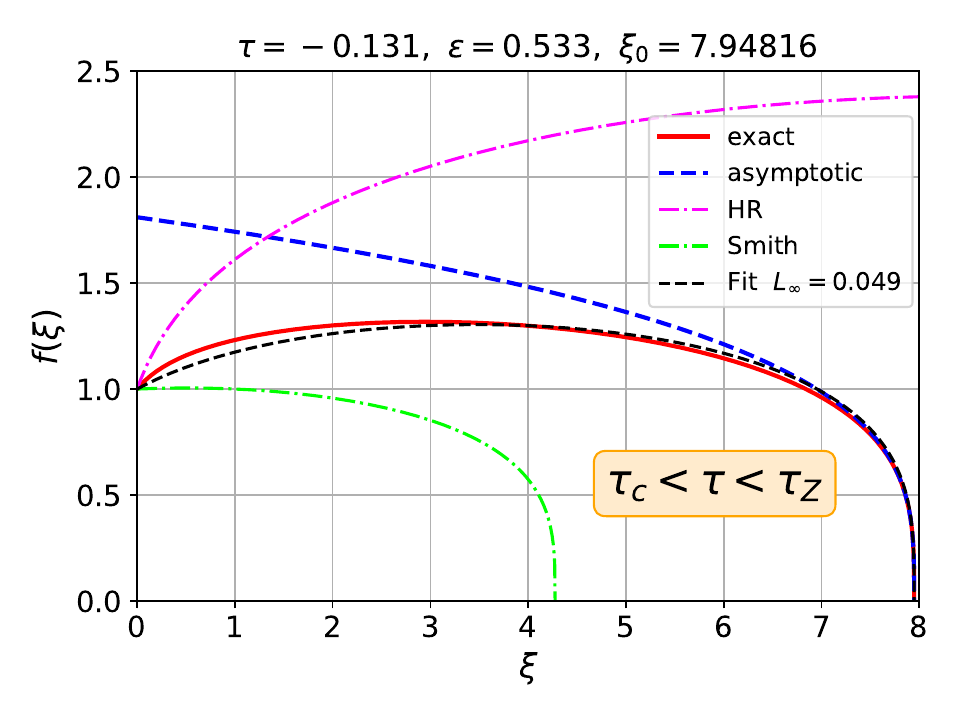}\includegraphics[scale=0.38]{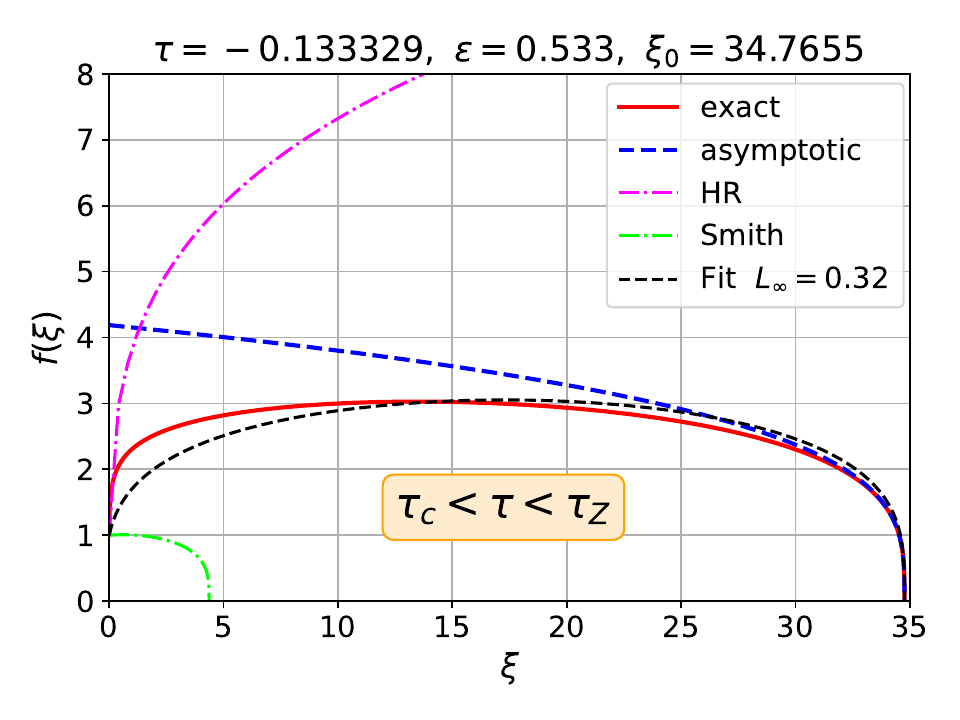} 
\par\end{centering}
\caption{Same as Fig. \ref{fig:origin_forms}, but for $\alpha=3.5$, $\beta=4$.
For these parameters we have $\epsilon\approx0.53$, $\tau_{c}=-\frac{2}{15}=-0.133333...$,
$\tau_{H}=\frac{2}{7}\approx0.2857$ and $\tau_{Z}=-\frac{2}{23}\approx-0.08696$.
The larger value of $\epsilon$ relative to Fig. \ref{fig:origin_forms},
results in a less steep front and larger error for the HR, Smith and
fit approximations.\label{fig:origin_forms-eps05}}
\end{figure*}

\begin{figure*}[t]
\begin{centering}
\includegraphics[scale=0.38]{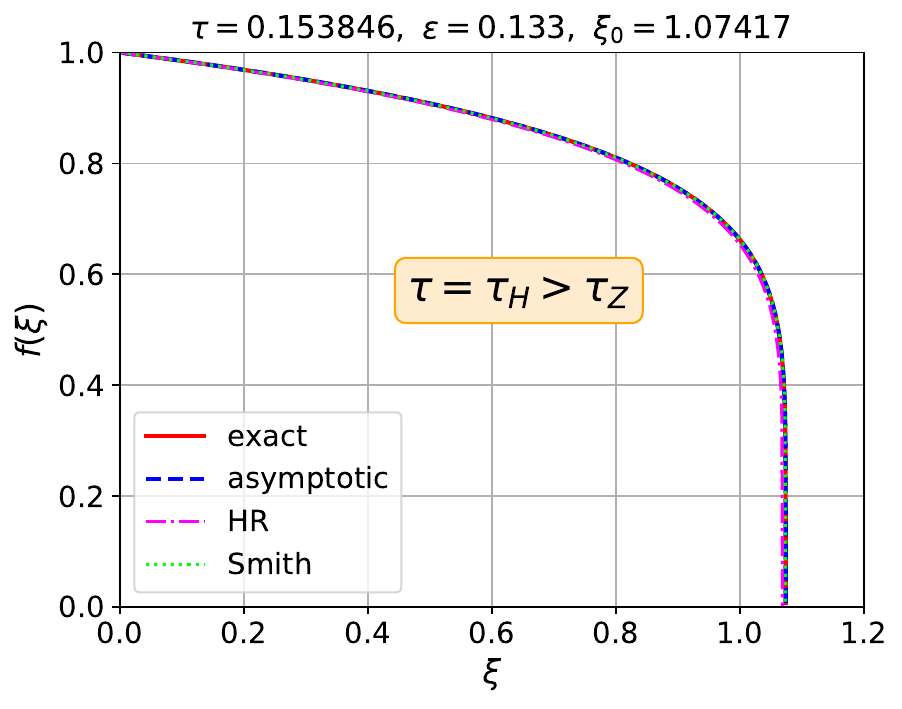}\includegraphics[scale=0.38]{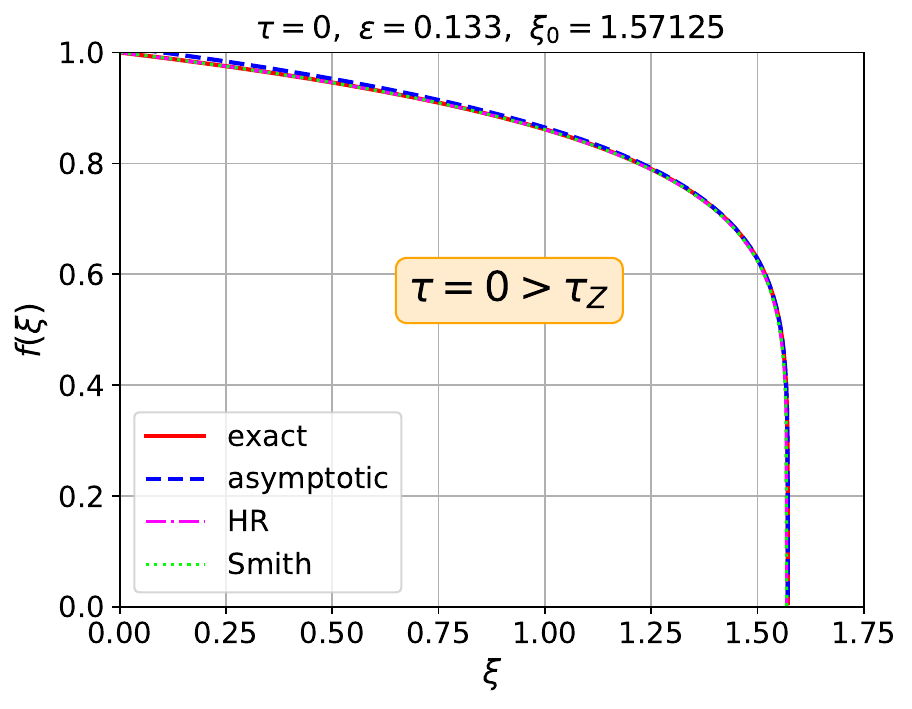}\includegraphics[scale=0.38]{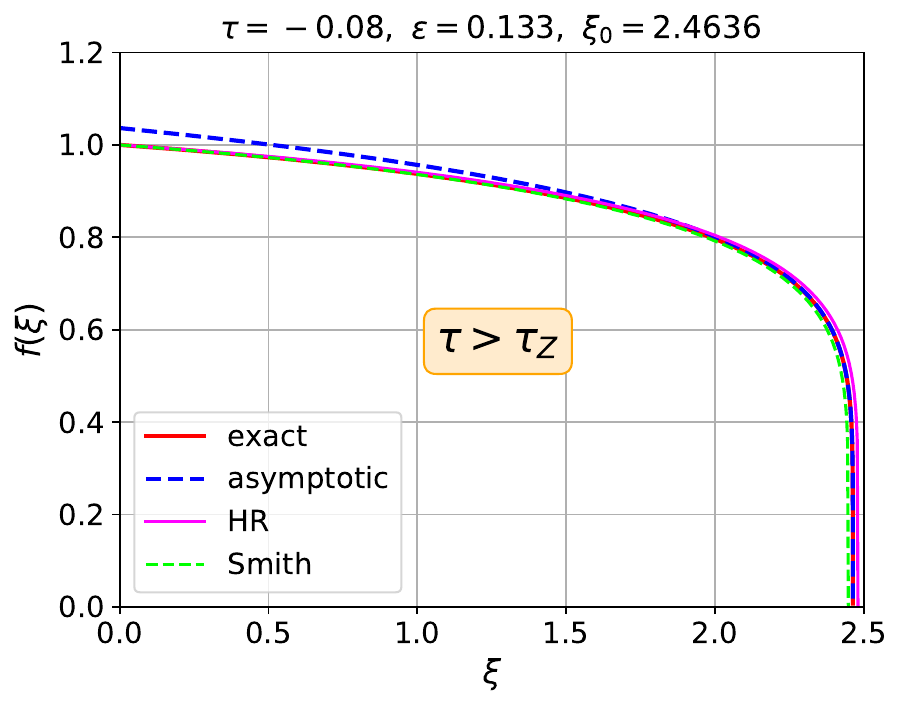} 
\par\end{centering}
\begin{centering}
\includegraphics[scale=0.38]{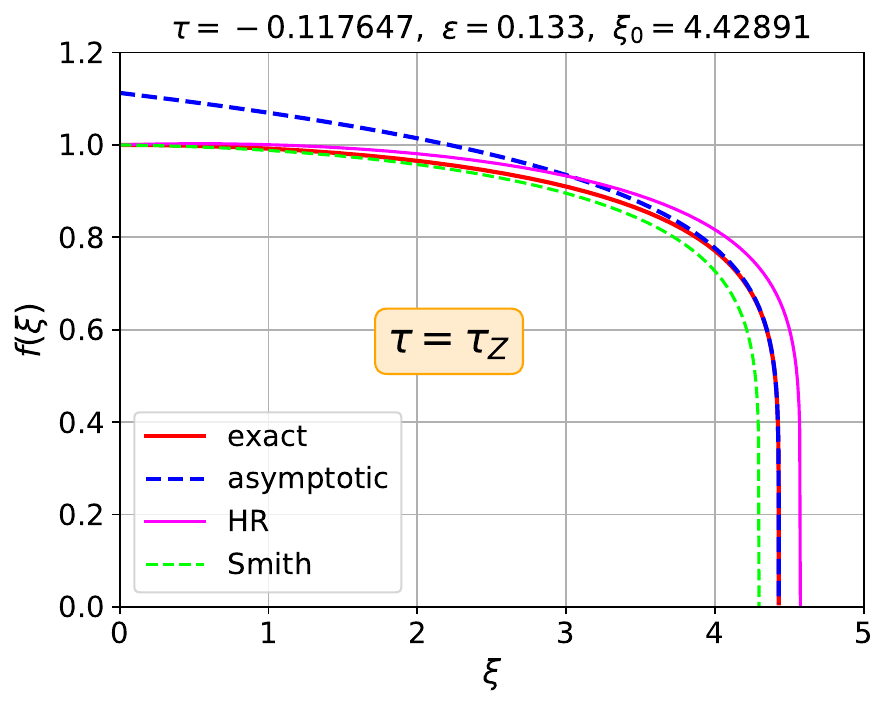}\includegraphics[scale=0.38]{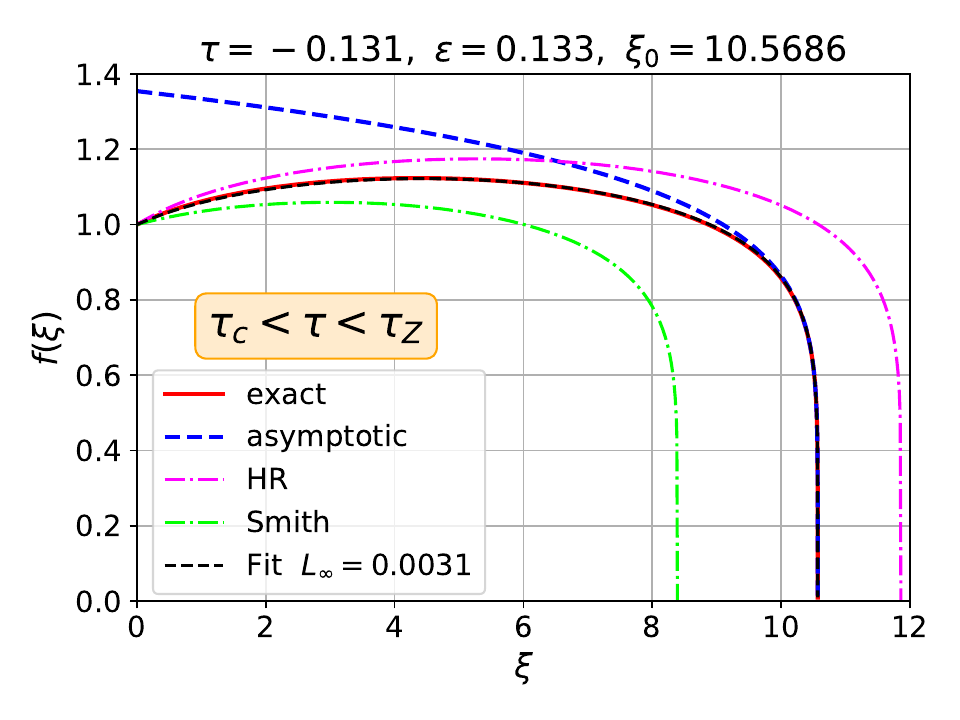}\includegraphics[scale=0.38]{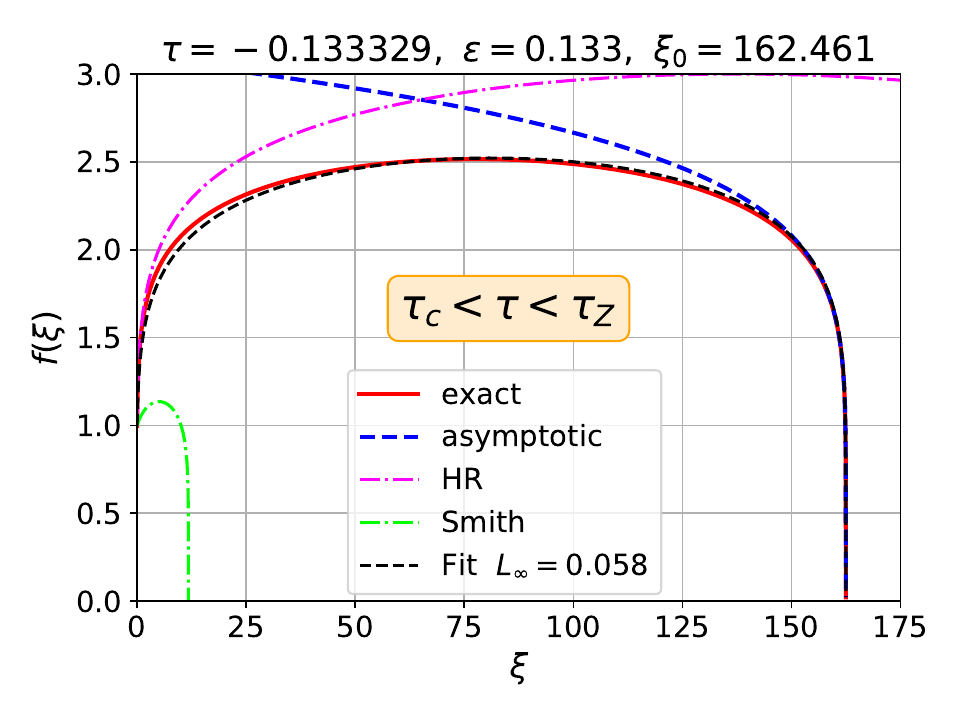} 
\par\end{centering}
\caption{Same as Fig. \ref{fig:origin_forms}, but for $\alpha=3.5$, $\beta=1$.
For these parameters we have the same $\tau_{c}=-\frac{2}{15}=-0.133333...$
as in Fig. \ref{fig:origin_forms-eps05}, and $\epsilon\approx0.13$,
$\tau_{H}=\frac{2}{13}\approx0.1538$, $\tau_{Z}=-\frac{2}{17}\approx-0.1176$.
The smaller value of $\epsilon$ relative to Fig. \ref{fig:origin_forms},
results a steeper front and in a much smaller error for the HR, Smith
and fit approximations.\label{fig:origin_forms-eps013}}
\end{figure*}

It will be shown below that there are only two special values of the
temperature drive exponent $\tau$ for which the ODE Eq. \eqref{eq:ode}
has an exact analytic solution for the similarity temperature profile
$f\left(\xi\right)$. For a general $\tau$, $f\left(\xi\right)$
is obtained by a numerical integration of the ODE \eqref{eq:ode}
(see also Refs. \cite{marshak1958effect,petschek1960penetration,heaslet1961diffusion,castor2004radiation,mihalas1999foundations,garnier2006self,nelson2009semi,lane2013new,shussman2015full,krief2024self}).
This is in contrast to the linear conduction case (see Sec. \ref{sec:Linear-conduction}),
which has a closed form analytic solution for a general $\tau$.

The value of $\xi_{0}$ is obtained by iterations of a ``shooting
method'', which is applied on the numerical solution of the ODE Eq.
\eqref{eq:ode}, which is integrated inwards from a trial $\xi_{0}$
to $\xi=0$. The iterative process adjusts the trial $\xi_{0}$ until
the result obeys the boundary condition Eq. \eqref{eq:f0_ss_bc} at
$\xi=0$. Moreover, since the derivative $f'$ diverges at the front,
in practice, given a trial $\xi_{0}$, the integration actually starts
from $\bar{\xi}_{0}=\xi_{0}\left(1-\varepsilon\right)$ with $\varepsilon$
a small number (we take $\varepsilon=10^{-12}$). The near-front asymptotic
form {[}Eq. \eqref{eq:asymp_front}{]} evaluated at $\xi=\bar{\xi}_{0}$
(for the current trial $\xi_{0}$), is used to obtain the initial
values $f\left(\bar{\xi}_{0}\right)$ and $f'\left(\bar{\xi}_{0}\right)$
which are then given to a standard numerical ODE solver. Numerical
results for the similarity profiles in various cases are shown in
Figs. \ref{fig:profiles_shapes},\ref{fig:heat_front_forms}-\ref{fig:origin_forms-eps013}.

\subsection{The Henyey exact analytic solution\label{subsec:The-Henyey-analytic}}

For the temporal exponent $\tau=\tau_{H}$ {[}Eq. \eqref{eq:tau_h}{]},
the ODE Eq. \eqref{eq:ode} has an exact analytic solution, given
by:

\begin{equation}
f\left(\xi\right)=\left(1-\frac{\xi}{\xi_{0}}\right)^{\frac{1}{4+\alpha-\beta}},\label{eq:f_Hy_Exact}
\end{equation}
where: 
\begin{equation}
\xi_{0}=\sqrt{\frac{4+\alpha}{4+\alpha-\beta}}.\label{eq:xsi_0_hy}
\end{equation}
This result is known as the Henyey Marshak wave (see Sec. 5.2 in Ref.
\cite{pert1977class}, Sec. II-B of Ref. \cite{hammer2003consistent},
Appendix A of Ref. \cite{malka2022supersonic}, Sec. III.D in Ref.
\cite{krief2024self} and Refs. \cite{polyanin2003handbook,rosen2005fundamentals,smith2010solutions,cohen2018modeling}).
For $\tau=\tau_{H}$, the exact solution and the asymptotic near-front
approximation {[}Eq. \eqref{eq:asymp_front}{]}, are identical for
all $\xi$ (as seen in the top left pane of Figs. \ref{fig:origin_forms}-\ref{fig:origin_forms-eps013}).
In addition, the similarity exponent in this case is $\delta=1$,
that is, the front propagates at constant speed $x_{F}\propto t$.

The bath constant {[}Eq. \eqref{eq:Bbath}{]} can be obtained from
the derivative at the origin, 
\begin{equation}
f'\left(0\right)=-\frac{1}{\sqrt{\left(4+\alpha-\beta\right)\left(4+\alpha\right)}},\label{eq:FTAG_HY}
\end{equation}
which is always negative, since $\tau_{H}>\tau_{Z}$ (as discussed
in Sec. \ref{subsec:The-total-energy}). The energy integral {[}Eq.
\eqref{eq:Iint}{]} is given by 
\begin{equation}
\mathcal{E}=\frac{1}{\xi_{0}},\label{eq:eint_hy_Exact}
\end{equation}
and the total energy as a function of time is 
\begin{equation}
E\left(t\right)=u_{0}\left(KT_{0}^{4+\alpha+\beta}\right)^{\frac{1}{2}}\mathcal{E}t^{\frac{4+\alpha}{4+\alpha-\beta}}.
\end{equation}

\subsection{The Zel'dovich-Barenblatt exact analytic solution - the nonlinear
instantaneous point source problem\label{subsec:The-Zel'dovich-Barenblatt-analyt}}

As discussed in Sec. \ref{subsec:The-total-energy}, for $\tau=\tau_{Z}$
{[}Eq. \eqref{eq:tau_Z}{]}, the surface temperature is reduced in
such a way that the total energy is constant in time, and the solution
to the nonlinear diffusion problem with a prescribed surface temperature
drive {[}Eq. \eqref{eq:Tbc}{]}, should be identical to the solution
of the nonlinear instantaneous point source problem \cite{zeldovich1967physics,pert1977class,barenblatt1996scaling,krief2021analytic},
where a finite amount of energy is deposited at at the origin at $t=0$.
Indeed, for $\tau=\tau_{Z}$ the ODE Eq. \eqref{eq:ode} has the exact
analytic solution:

\begin{equation}
f\left(\xi\right)=\left(1-\left(\frac{\xi}{\xi_{0}}\right)^{2}\right)^{\frac{1}{4+\alpha-\beta}},\label{eq:f_z}
\end{equation}
where: 
\begin{equation}
\xi_{0}=\sqrt{\frac{2\left(4+\alpha+\beta\right)\left(4+\alpha\right)}{\left(4+\alpha-\beta\right)\beta}}.\label{eq:xsi_0_Z}
\end{equation}
This solution is in agreement with the nonlinear instantaneous point
source problem (i.e. Eq. 30 in Ref. \cite{pert1977class} and Eq.
42 in Ref. \cite{krief2021analytic}). Due to the energy preserving
nature of this solution, it has a maxima at the origin, a zero net
surface flux, and a zero bath constant which results in a bath temperature
{[}Eq. \eqref{eq:Bbath}{]} that is equal to the surface temperature.
The (time independent) total energy is 
\begin{equation}
E=u_{0}\left(KT_{0}^{4+\alpha+\beta}\right)^{\frac{1}{2}}\mathcal{E},\label{eq:Ezld}
\end{equation}
where for the solution \eqref{eq:f_z}, the energy integral {[}Eq.
\eqref{eq:Iint}{]} can be calculated analytically: 
\begin{equation}
\mathcal{E}=\frac{\xi_{0}}{2}\mathcal{B}\left(\frac{1}{2},\frac{4+\alpha}{4+\alpha-\beta}\right),
\end{equation}
where the Beta function is defined in terms of the Gamma function:

\begin{equation}
\mathcal{B}\left(x,y\right)=\frac{\Gamma\left(x\right)\Gamma\left(y\right)}{\Gamma\left(x+y\right)}.
\end{equation}

Finally, we note that looking at the linear conduction limit of the
solution \eqref{eq:f_z}, that is, $4+\alpha\rightarrow\beta$ and
$\nu\rightarrow\infty$, we have $\xi_{0}\approx\sqrt{4\beta\nu}\rightarrow\infty$
and 
\begin{equation}
\lim_{\nu\rightarrow\infty}f\left(\xi\right)=\lim_{\nu\rightarrow\infty}\left(1-\frac{\xi^{2}}{4\beta\nu}\right)^{\nu}=e^{-\frac{\xi^{2}}{4\beta}},
\end{equation}
which agrees with the linear solution {[}Eq. \eqref{eq:f_zel_lin}{]}
for $\tau=\tau_{Z}$.

\subsection{The Hammer-Rosen, Smith and fitted approximate solutions\label{subsec:The-Hammer-Rosen-approximate}}

\begin{figure}
\begin{centering}
\includegraphics[scale=0.55]{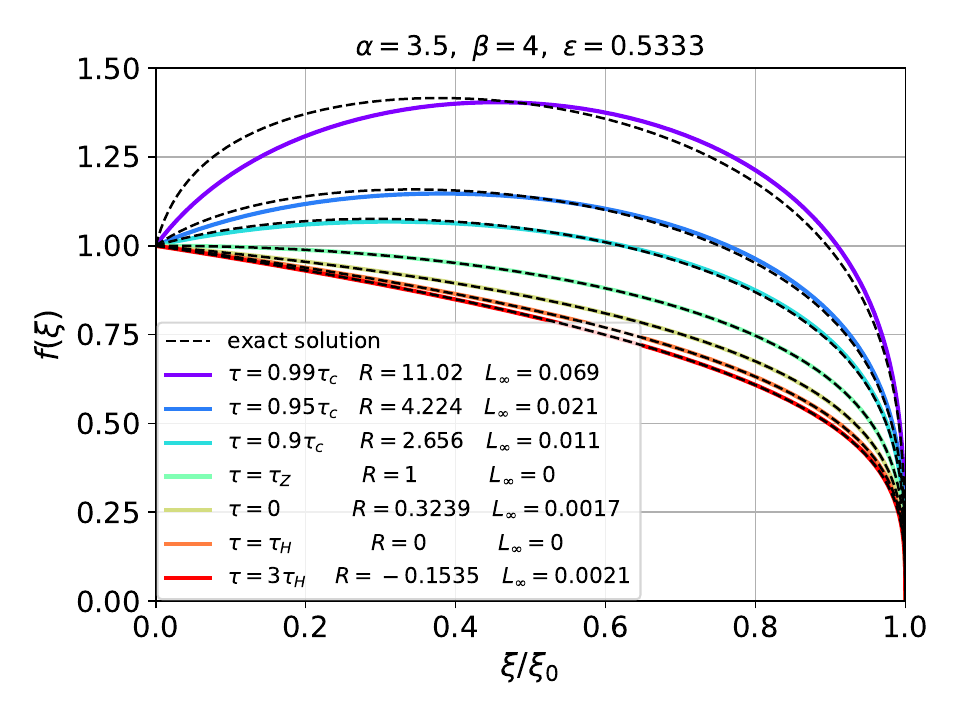} 
\par\end{centering}
\begin{centering}
\includegraphics[scale=0.55]{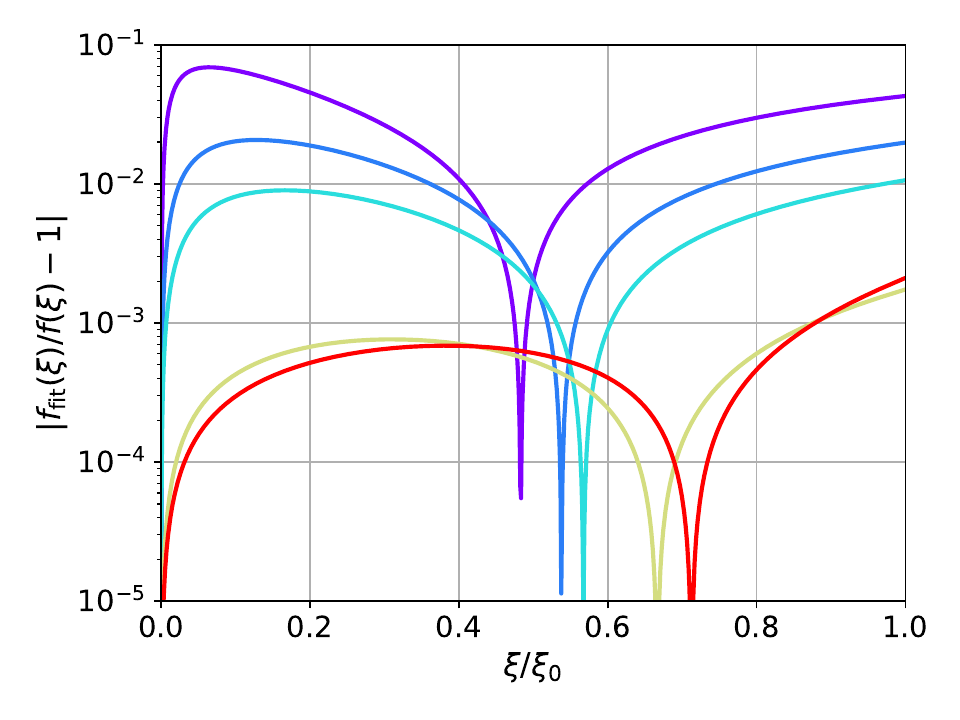} 
\par\end{centering}
\caption{A demonstration of the numerical fit approximate solution. The upper
figure shows a comparison between the exact solution $f\left(\xi\right)$
(black dashed lines) and the approximate profile form Eq. \eqref{eq:f_hr}
(colorful lines), using a value of $R$ (listed in the legend) that
is calculated by a fit to the exact numerical solution. The results
are shown for various values of $\tau$ and for $\alpha=3.5$, $\beta=4$,
for which $\epsilon\approx0.53$ (see also Fig. \ref{fig:origin_forms-eps05}).
The lower figure shows the pointwise relative error as a function
of $\xi/\xi_{0}$ (not shown are the results for $\tau=\tau_{H}$
and $\tau=\tau_{Z}$, for which the fit error is identically zero,
since the exact solution has the form of Eq. \eqref{eq:f_hr}). The
resulting maximal fit errors ($L_{\infty}$) are also listed in the
legend. \label{fig:profiles_fit}}
\end{figure}

\begin{figure*}[t]
\begin{centering}
\includegraphics[scale=0.38]{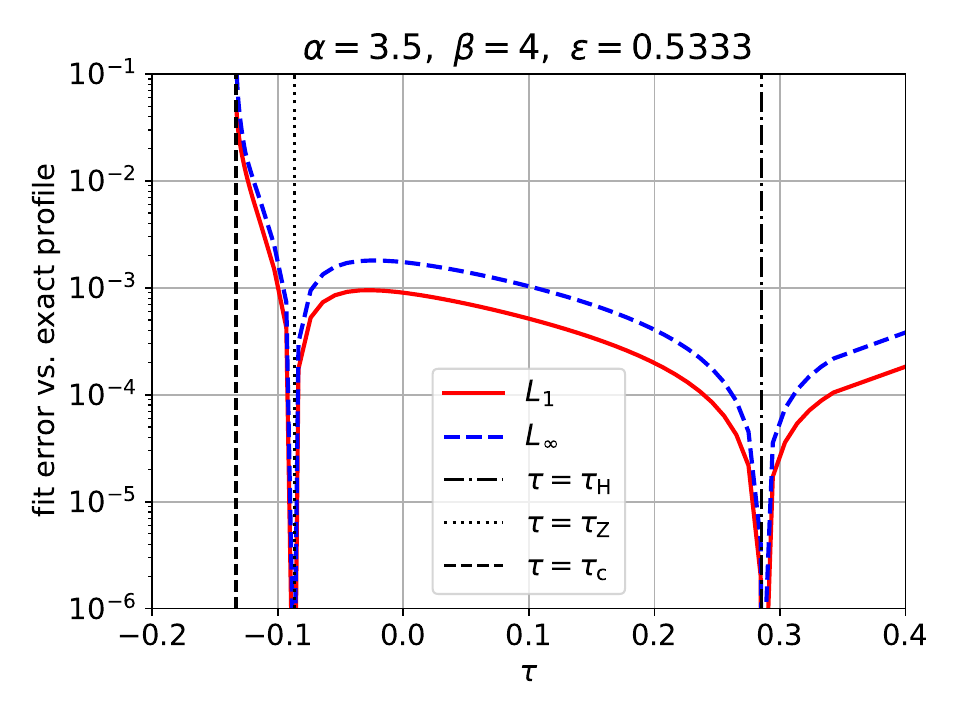}\includegraphics[scale=0.38]{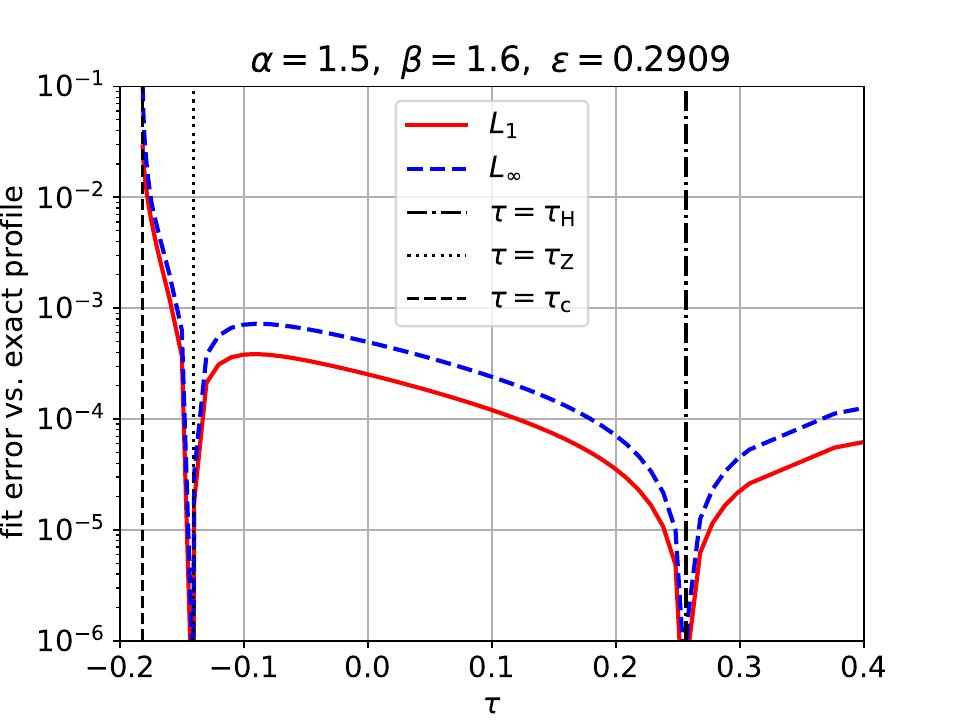}\includegraphics[scale=0.38]{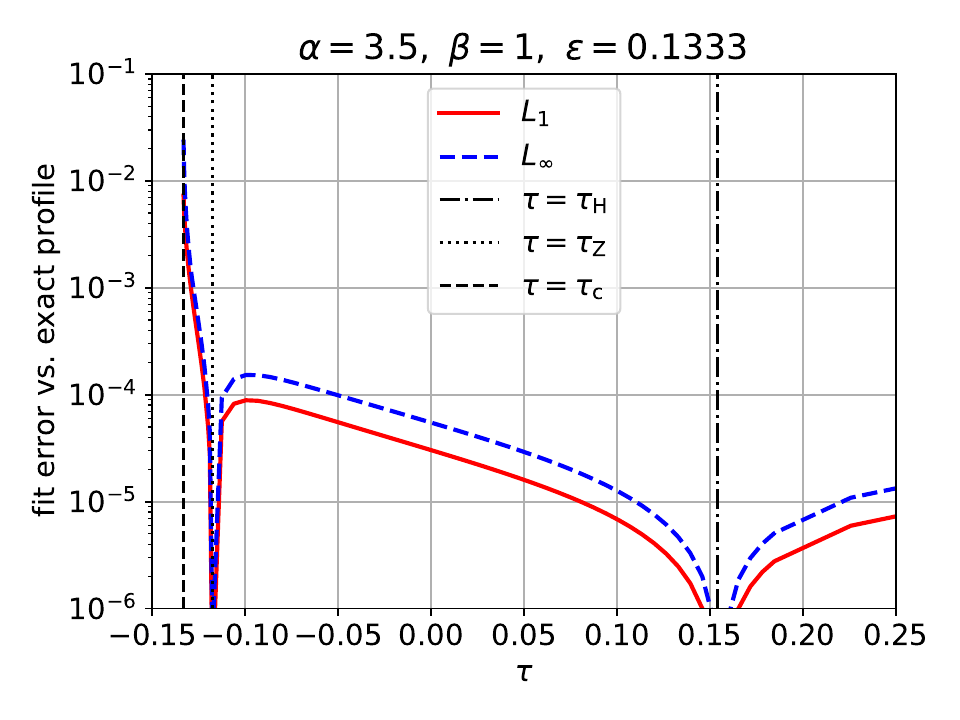} 
\par\end{centering}
\caption{The $L_{1}$ (average) and $L_{\infty}$ (maximal) errors of the approximate
fitted profile {[}Eq. \eqref{eq:f_hr}{]} and the exact solution (see
also Fig. \ref{fig:profiles_fit}), as a function of $\tau$. The
results were calculated for the three choices of $\alpha,\beta$ that
were shown in Figs. \ref{fig:origin_forms}-\ref{fig:origin_forms-eps013},
as listed in the titles. Also shown are vertical lines for $\tau=\tau_{H}$,
$\tau=\tau_{Z}$, for which the exact solution has the Form Eq. \eqref{eq:f_hr}
(so that the errors are zero), as well as $\tau=\tau_{c}$ which is
the minimal possible value of $\tau$. \label{fig:fit_err}}
\end{figure*}

\begin{figure*}[t]
\begin{centering}
\includegraphics[scale=0.38]{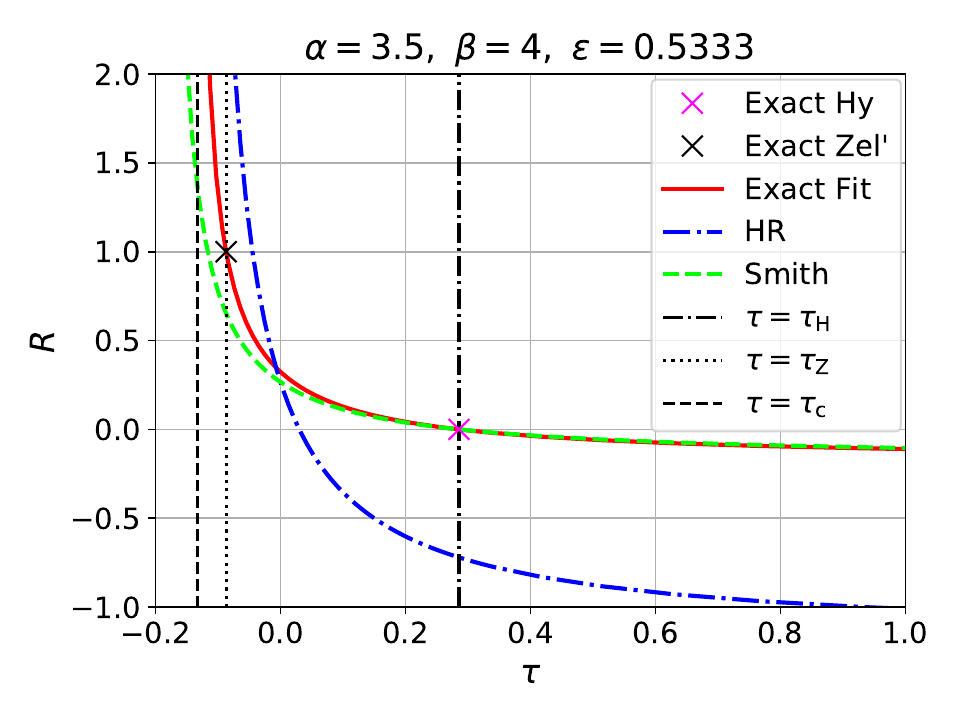}\includegraphics[scale=0.38]{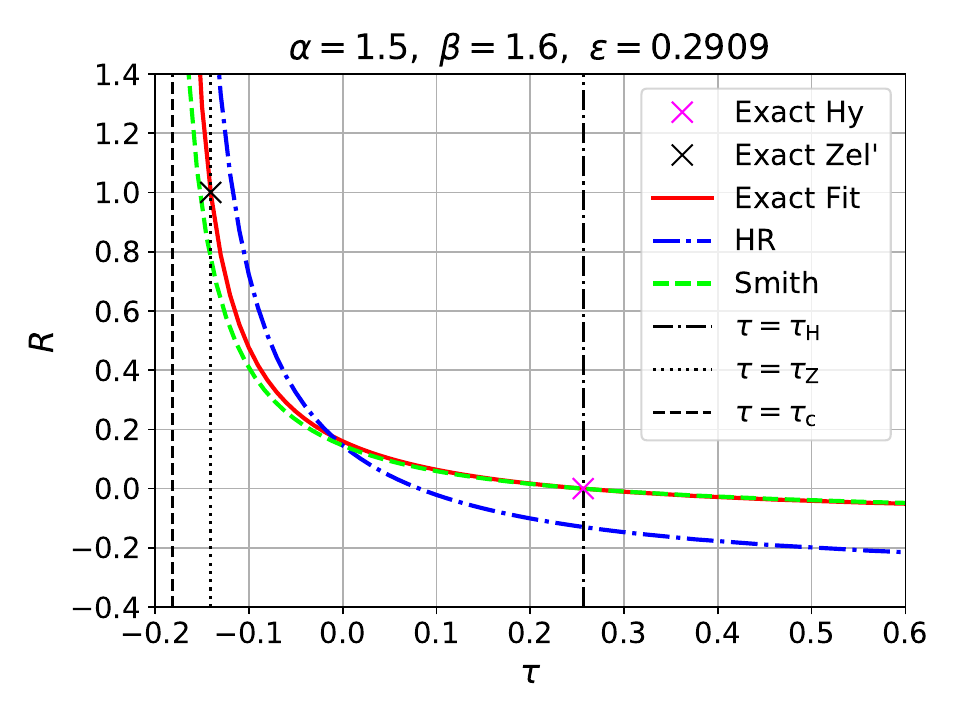}\includegraphics[scale=0.38]{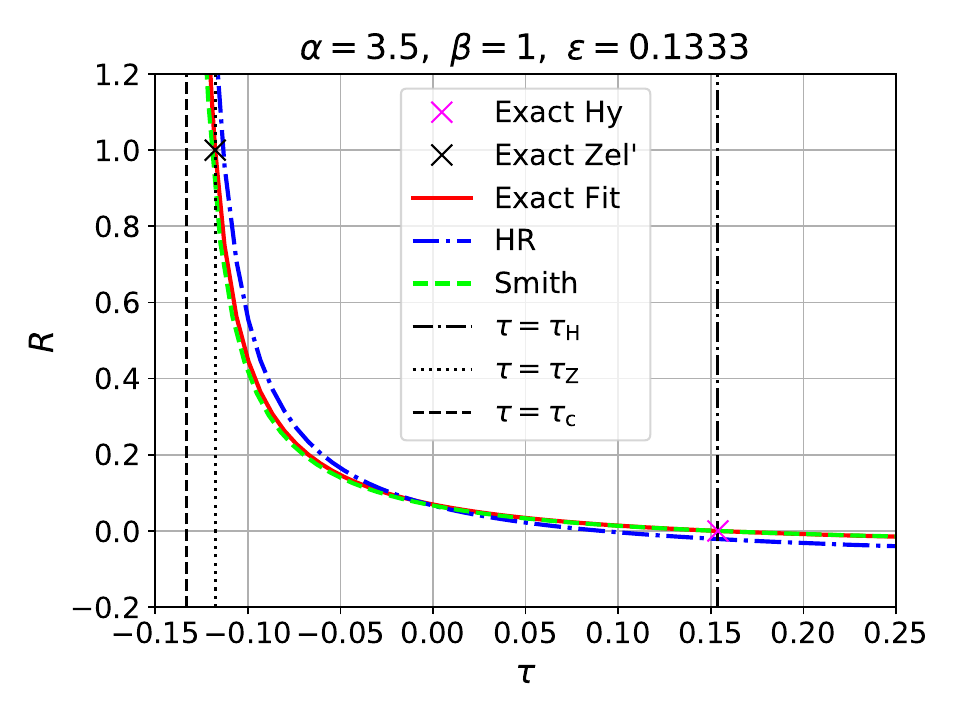} 
\par\end{centering}
\caption{The parameter $R$ {[}see Eq. \eqref{eq:f_hr}{]}, as a function of
$\tau$, as calculated from the Hammer-Rosen (HR) theory (Eq. \eqref{eq:R_HR},
blue line), from the Smith theory (Eq. \eqref{eq:R_SMITH}, green
line) and from a fit to the exact (numerical) similarity profile (red
line). The results were calculated for the three choices of $\alpha,\beta$
that were shown in Figs. \ref{fig:origin_forms}-\ref{fig:origin_forms-eps013},
\ref{fig:fit_err}, as listed in the titles. As discussed in the text,
it is evident that the exact Henyey solution ($R=0$, pink 'x' marker)
coincides with the Smith and fit results, while the exact Zel'dovich-Barenblatt
solution ($R=1$, black 'x' marker) only agrees with the fit result.
Also shown are vertical lines for $\tau=\tau_{H}$, $\tau=\tau_{Z}$,
as well as $\tau=\tau_{c}$ which is the minimal possible value of
$\tau$, where it is evident that the exact and HR solutions break
down. \label{fig:Rfit}}
\end{figure*}

\begin{figure*}[t]
\begin{centering}
\includegraphics[scale=0.38]{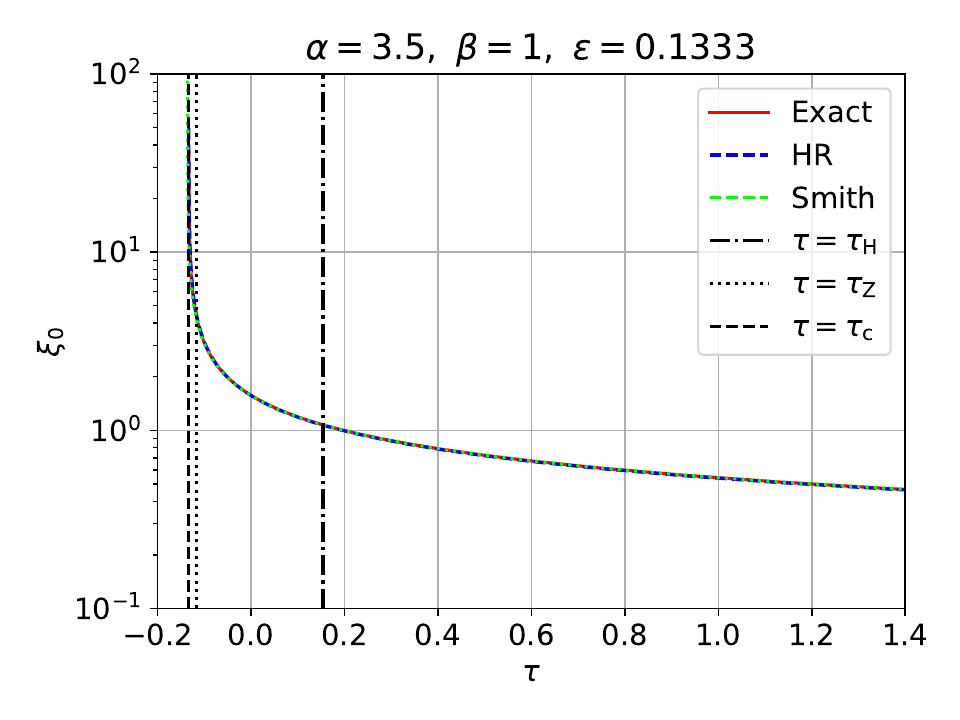}\includegraphics[scale=0.38]{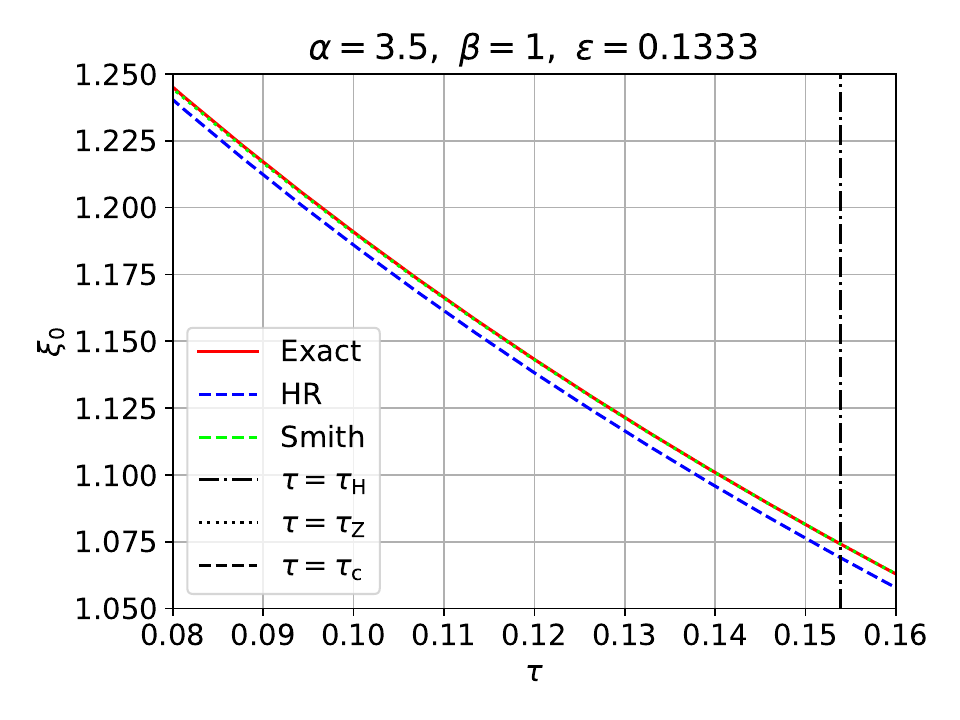}\includegraphics[scale=0.38]{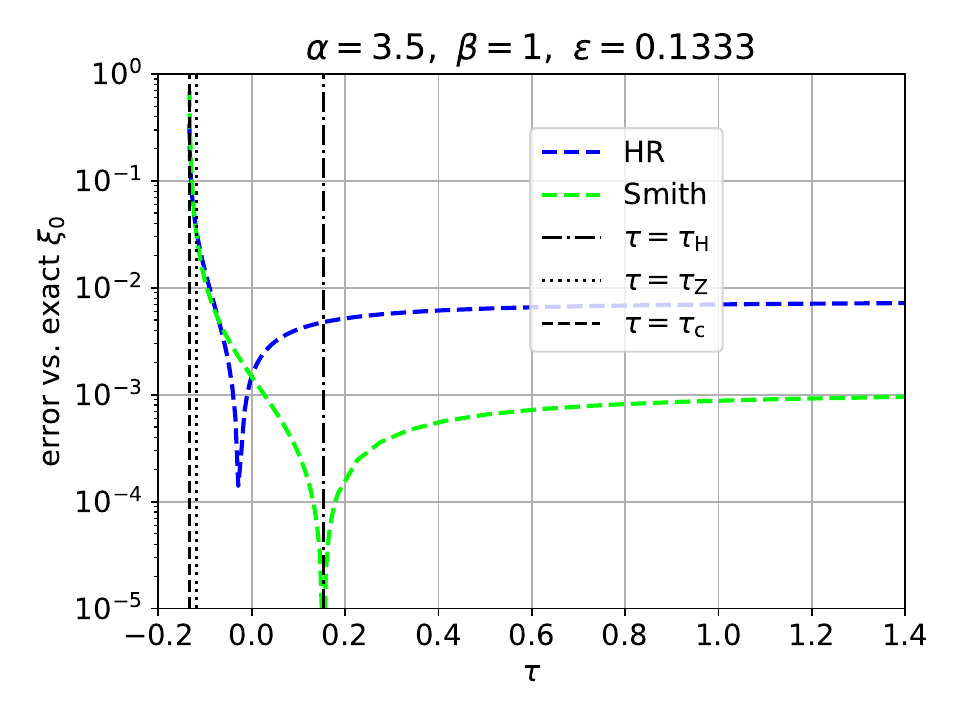} 
\par\end{centering}
\begin{centering}
\includegraphics[scale=0.38]{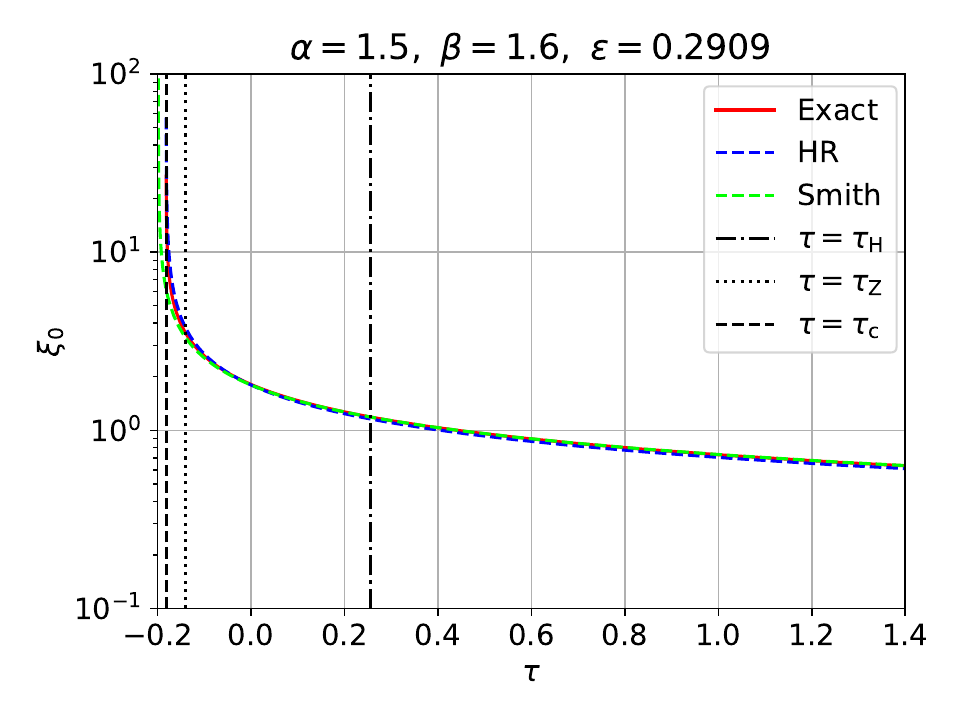}\includegraphics[scale=0.38]{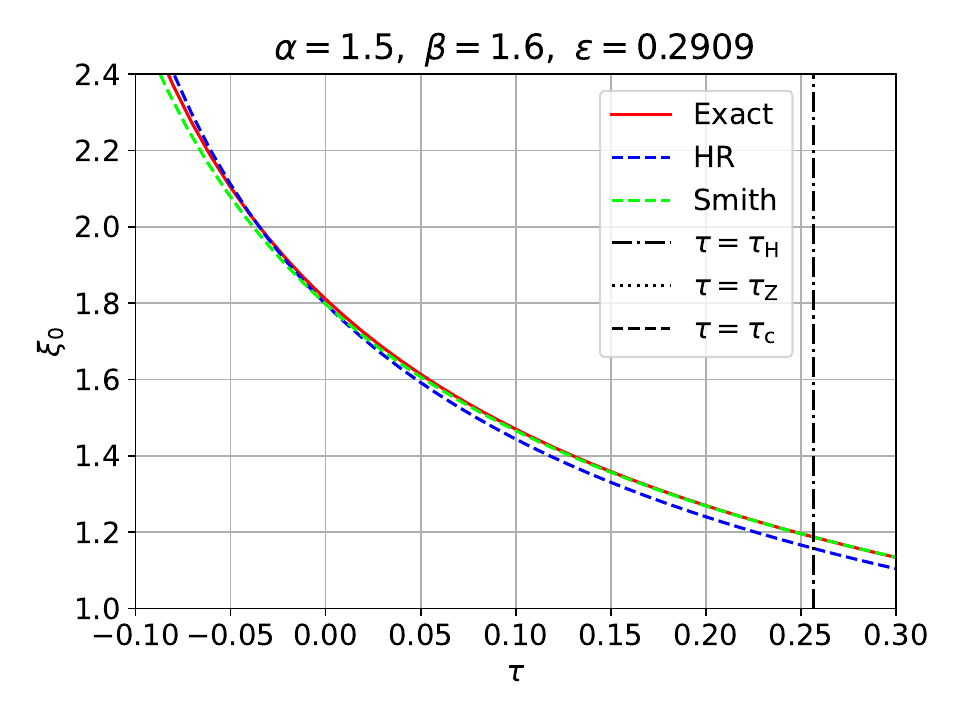}\includegraphics[scale=0.38]{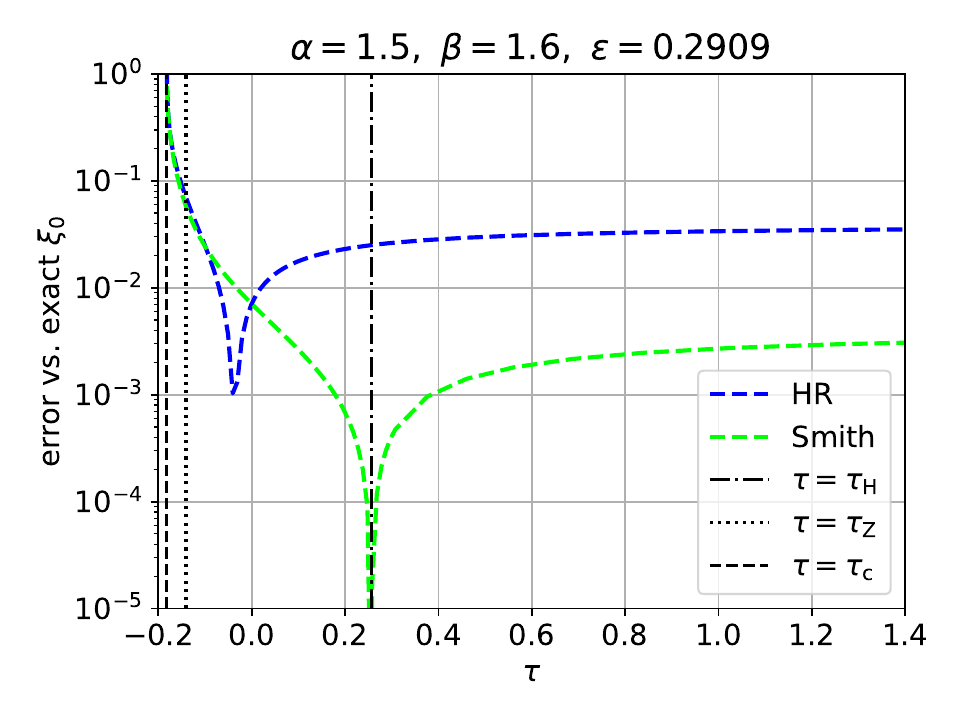} 
\par\end{centering}
\begin{centering}
\includegraphics[scale=0.38]{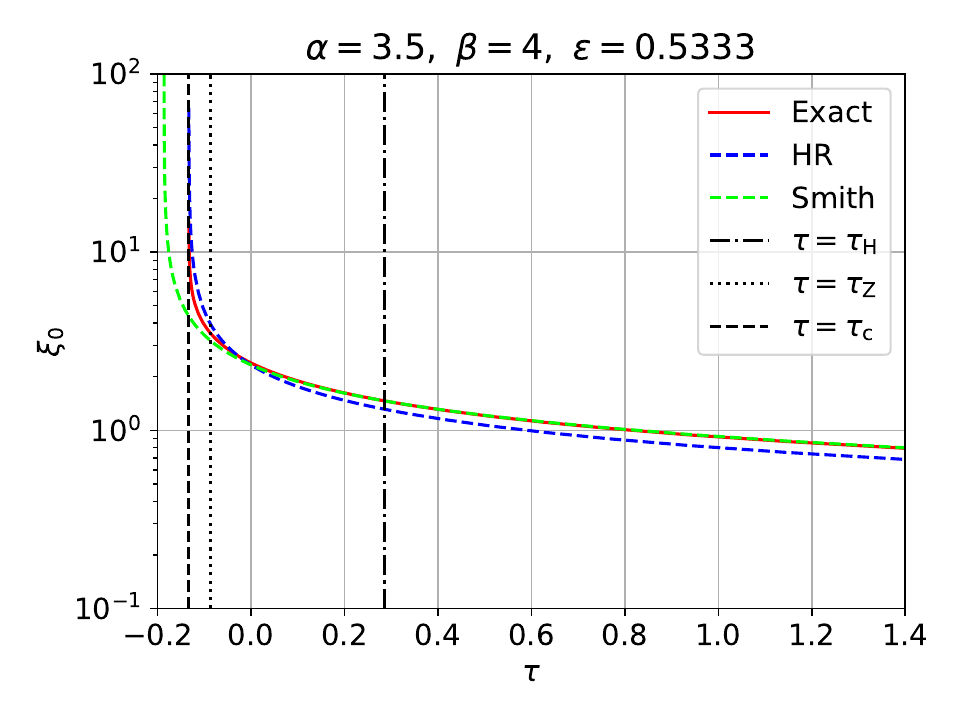}\includegraphics[scale=0.38]{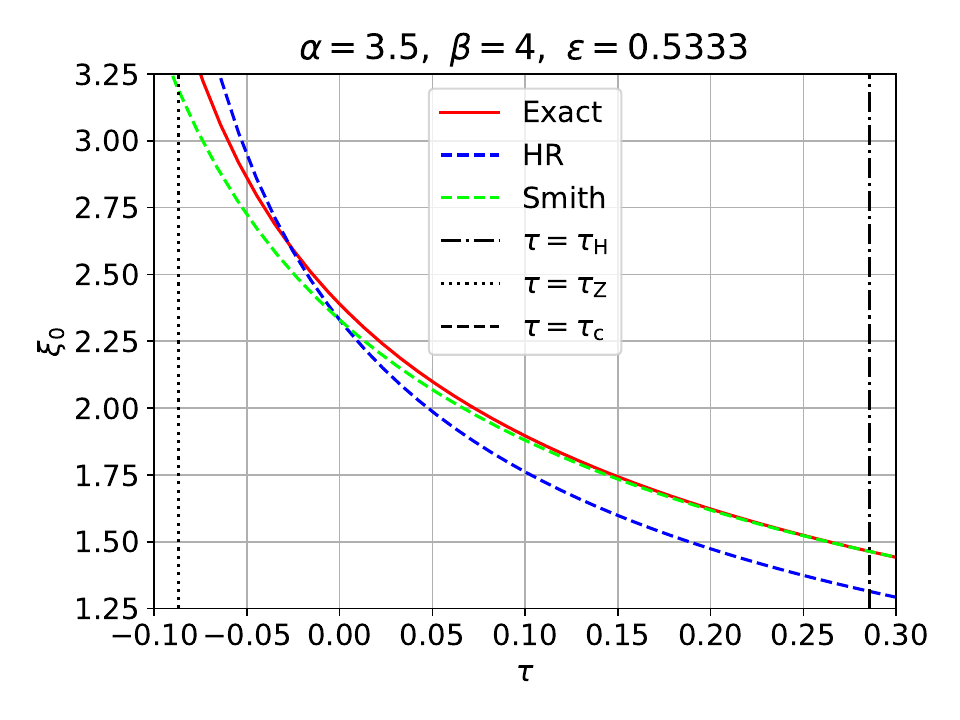}\includegraphics[scale=0.38]{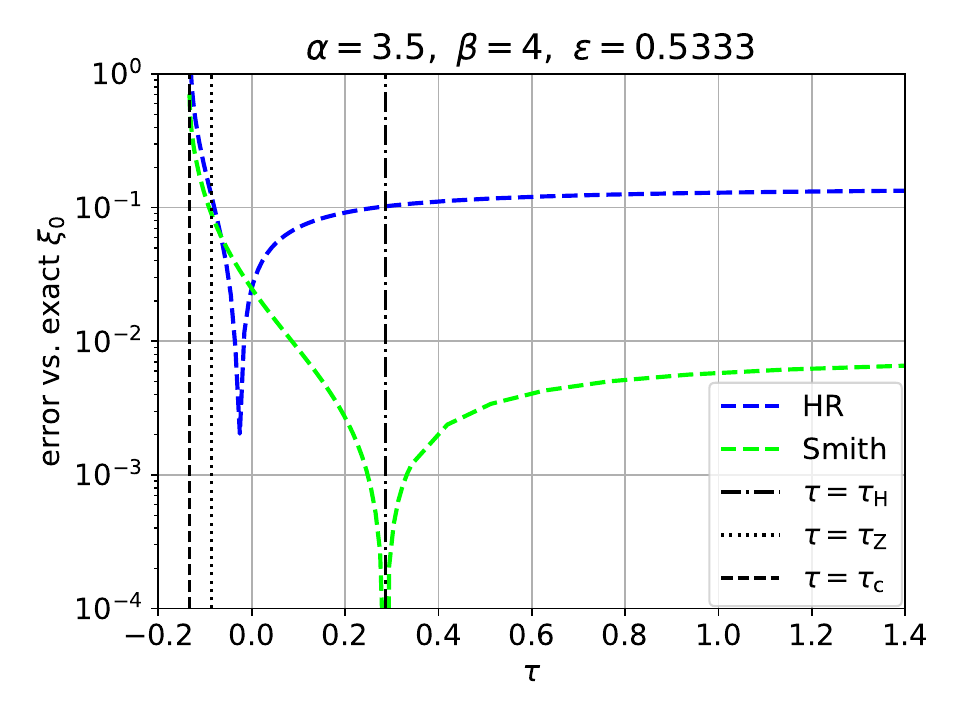} 
\par\end{centering}
\caption{A comparison between the exact value (obtained numerically, as described
in Sec. \ref{subsec:Numerical-solution}), of the heat front coordinate
$\xi_{0}$ (red lines) and the approximate Hammer-Rosen (Eq. \eqref{eq:xsi0_HR},
blue lines) and Smith (Eq. \eqref{eq:XSI0_SM}, green lines) results,
as a function of $\tau$. Each row represents a specific choice of
$\alpha$ and $\beta$, yielding different values of $\epsilon$,
as listed in the titles (some profiles for these cases were shown
in Figs. \ref{fig:origin_forms},\ref{fig:origin_forms-eps05},\ref{fig:origin_forms-eps013},
respectively). The leftmost plots show the comparison in a wide range
of $\tau$. The middle plots show a closed up view, which demonstrates
the differences between the approximations. The rightmost plots show
the relative error between the exact and approximate results. Vertical
lines are shown at the special values $\tau=\tau_{H}$, for which
the Smith and exact results coincide with the analytic Henyey solution
(see Sec. \ref{subsec:The-Henyey-analytic}); $\tau=\tau_{Z}$, for
which the exact solution coincides with the analytic Zel'dovich-Barenblatt
solution (see Sec. \ref{subsec:The-Zel'dovich-Barenblatt-analyt});
$\tau=\tau_{c}$ which is the minimal value of $\tau$, for which
the exact and Hammer-Rosen solutions break down (while the Smith solution
breaks down at a smaller $\tau$). It is evident that for $\tau>0$,
the Smith results are more accurate than HR by about an order of magnitude,
and that the overall accuracy of the both approximations decreases
with $\epsilon$. \label{fig:xsi0_err}}
\end{figure*}

\begin{figure*}[t]
\begin{centering}
\includegraphics[scale=0.38]{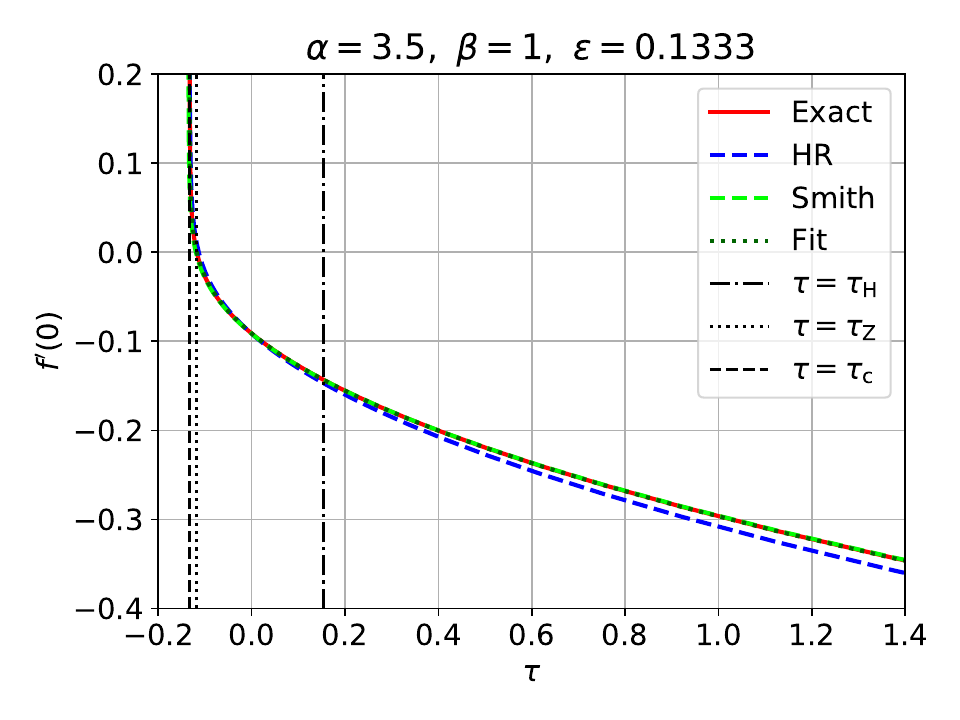}\includegraphics[scale=0.38]{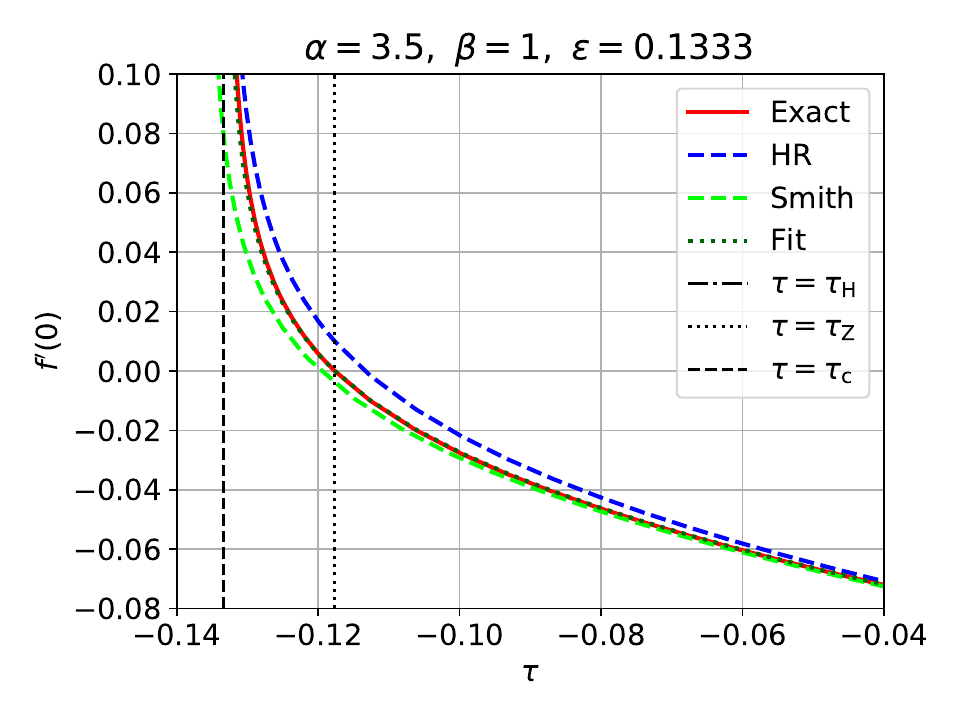}\includegraphics[scale=0.38]{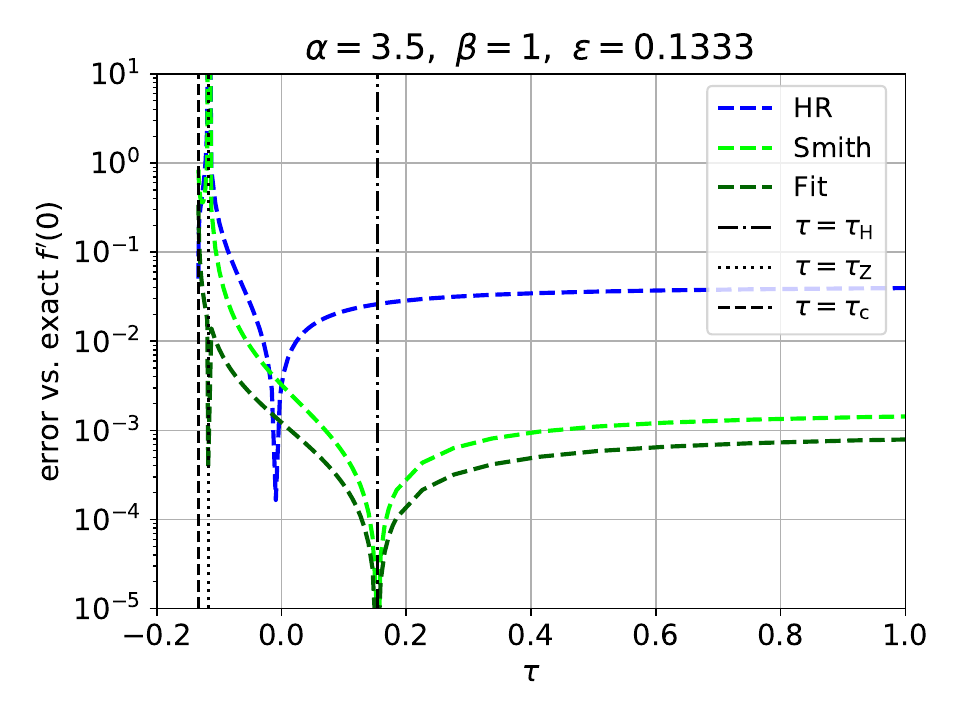} 
\par\end{centering}
\begin{centering}
\includegraphics[scale=0.38]{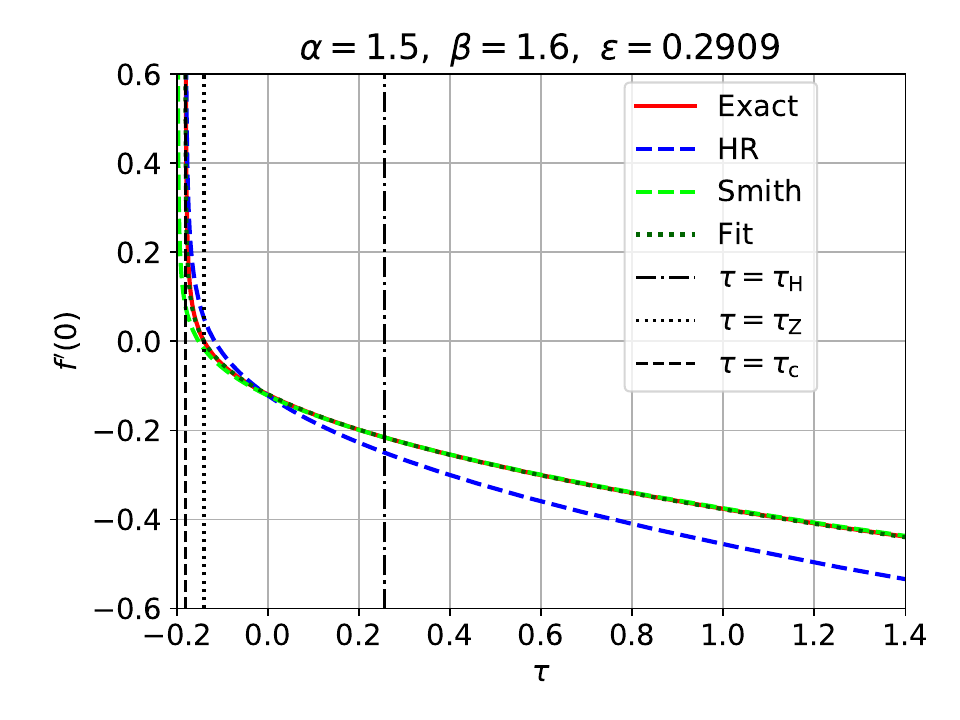}\includegraphics[scale=0.38]{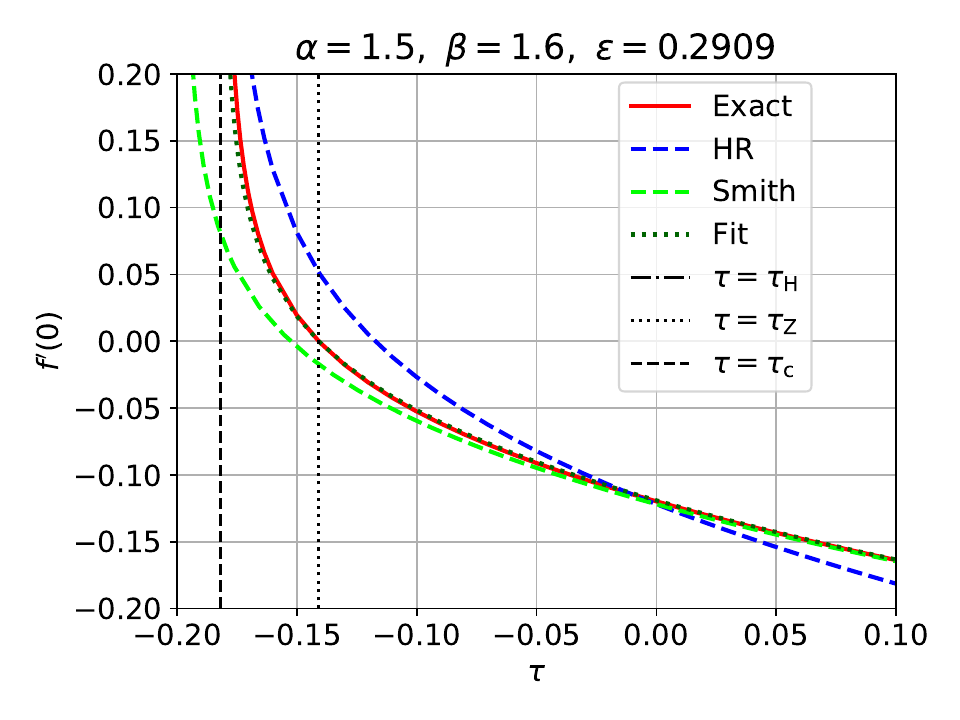}\includegraphics[scale=0.38]{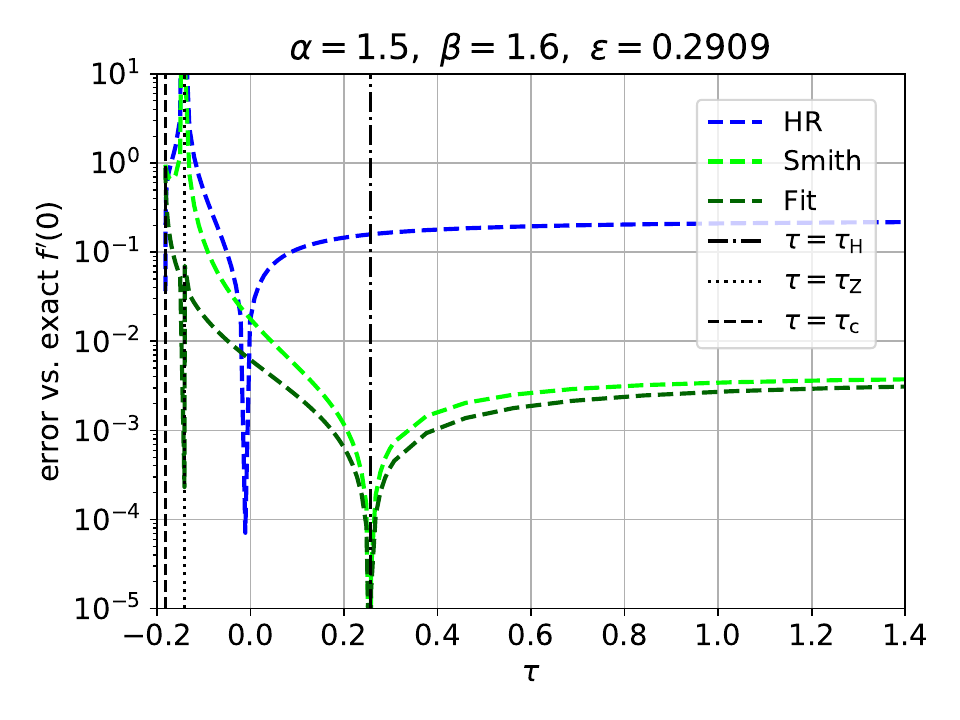} 
\par\end{centering}
\begin{centering}
\includegraphics[scale=0.38]{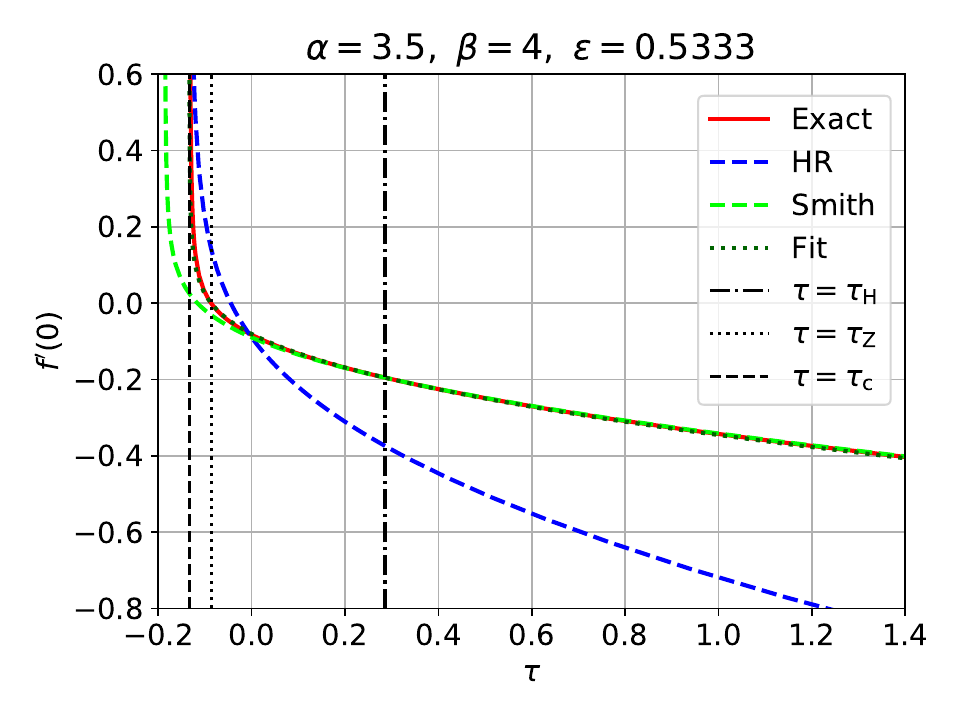}\includegraphics[scale=0.38]{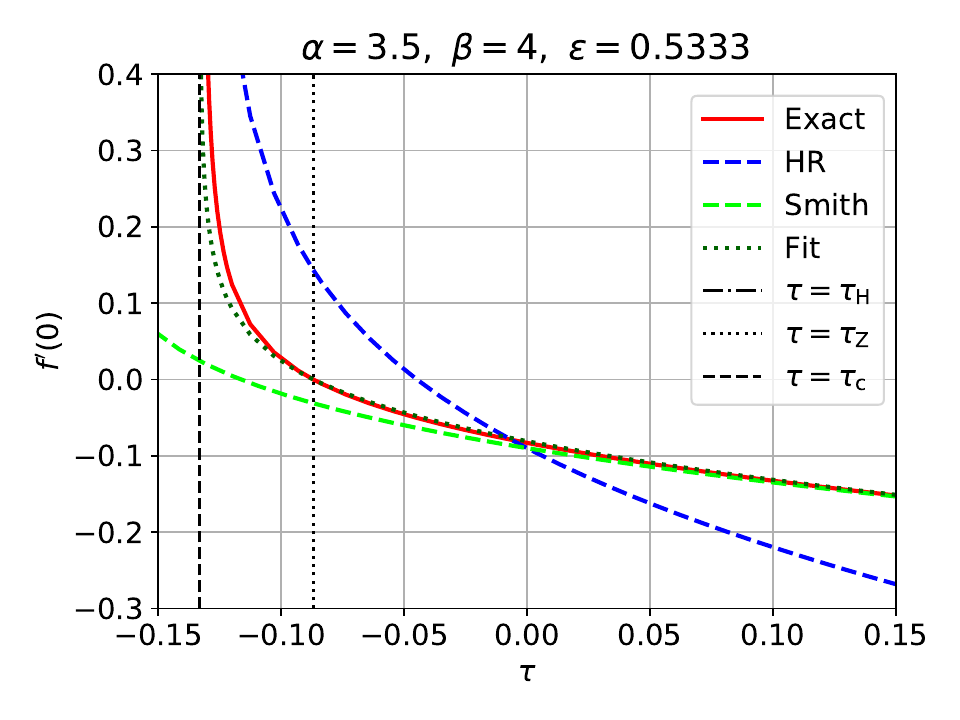}\includegraphics[scale=0.38]{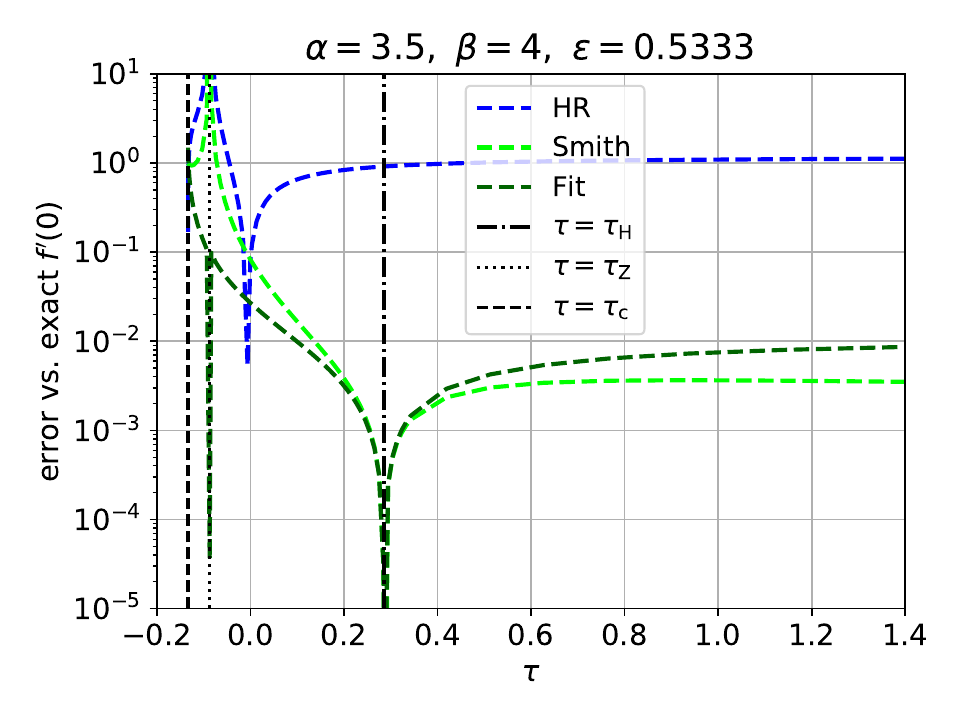} 
\par\end{centering}
\caption{Same as Fig. \ref{fig:xsi0_err}, but for $f'\left(0\right)$ which
determines the surface flux (see Eq. \eqref{eq:flux_selfsim}). The
results obtained by the fitted approximate profile {[}Eq. \eqref{eq:f_hr}{]}
are also shown (dotted green line). The approximate Hammer-Rosen,
Smith and fit results are obtained using Eq. \eqref{eq:ftag0_HR_SMITH}.
It is evident that only for the exact and fit results the derivative
correctly changes its sign precisely at $\tau=\tau_{Z}$, as the solution
passes between the positive net surface flux ($\tau>\tau_{Z}$), zero
surface flux (constant energy, $\tau=\tau_{Z}$) and negative surface
flux ($\tau<\tau_{Z}$) scenarios (see Tab. \ref{tab:origin_behaviou}).
It is evident that \label{fig:ftag0_err}}
\end{figure*}

\begin{figure*}[t]
\begin{centering}
\includegraphics[scale=0.38]{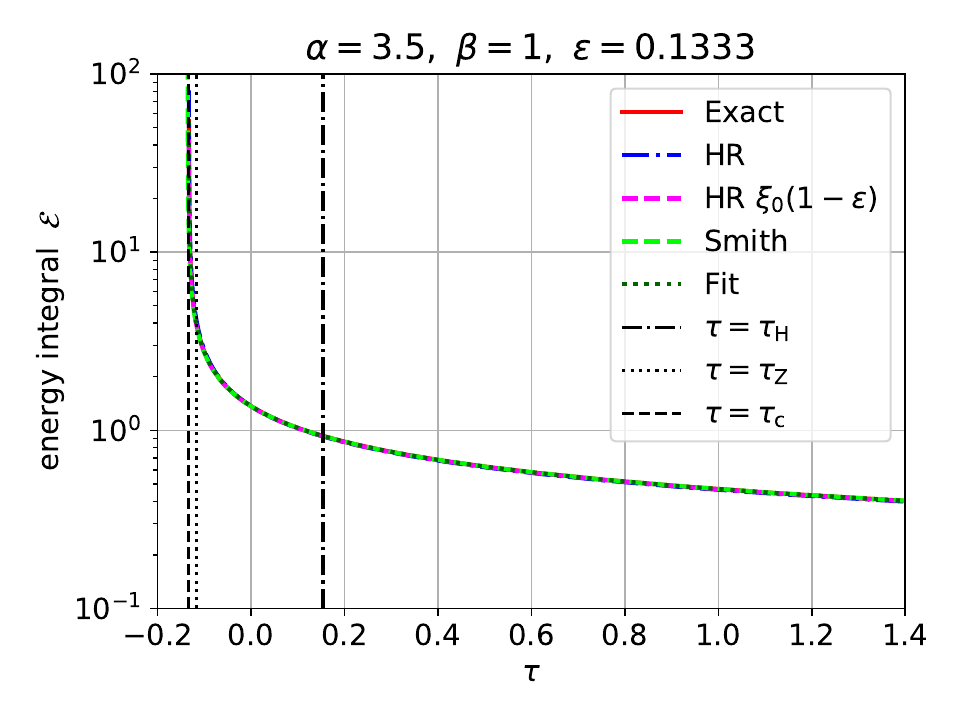}\includegraphics[scale=0.38]{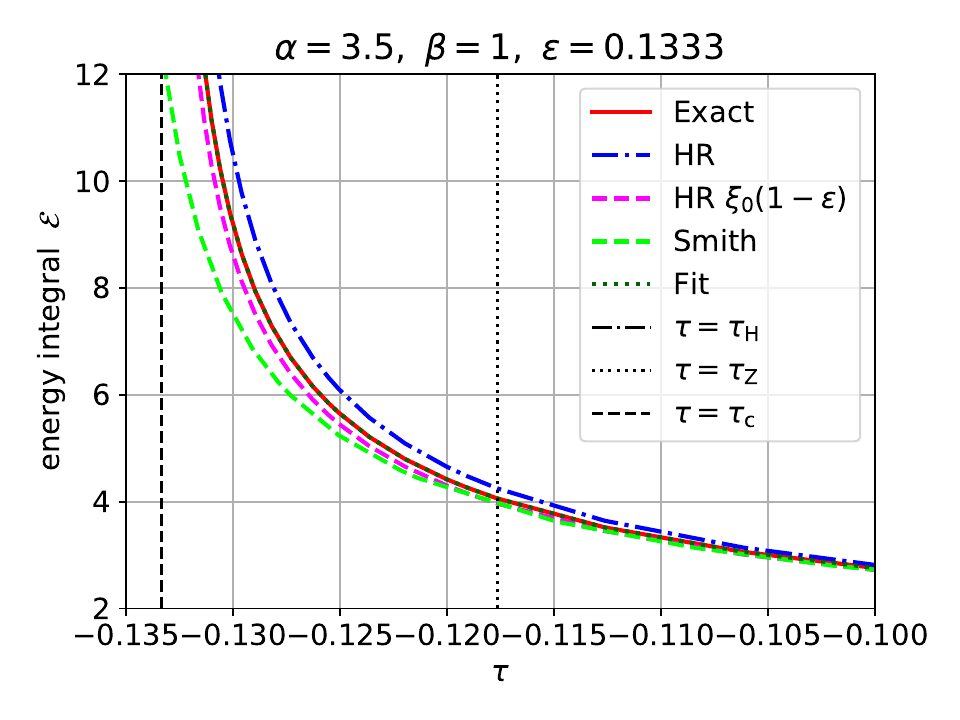}\includegraphics[scale=0.38]{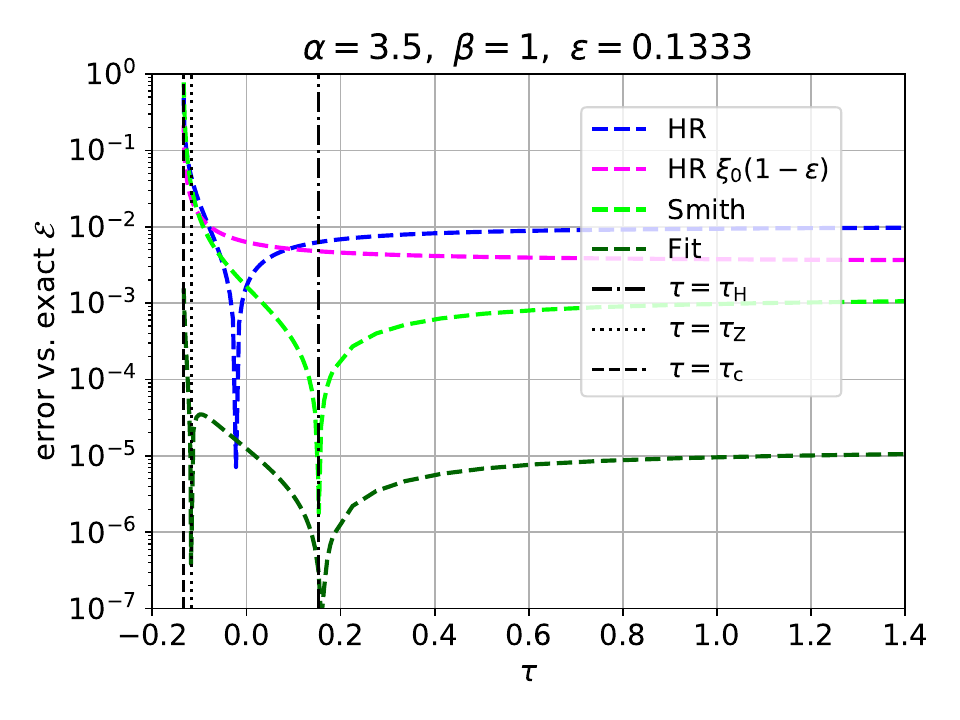} 
\par\end{centering}
\begin{centering}
\includegraphics[scale=0.38]{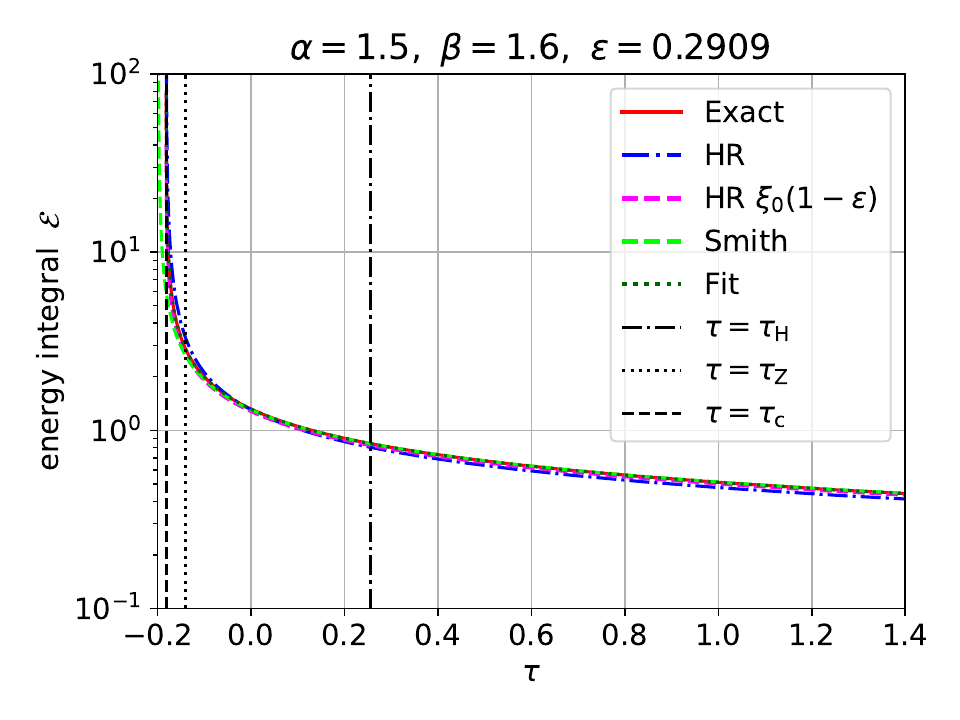}\includegraphics[scale=0.38]{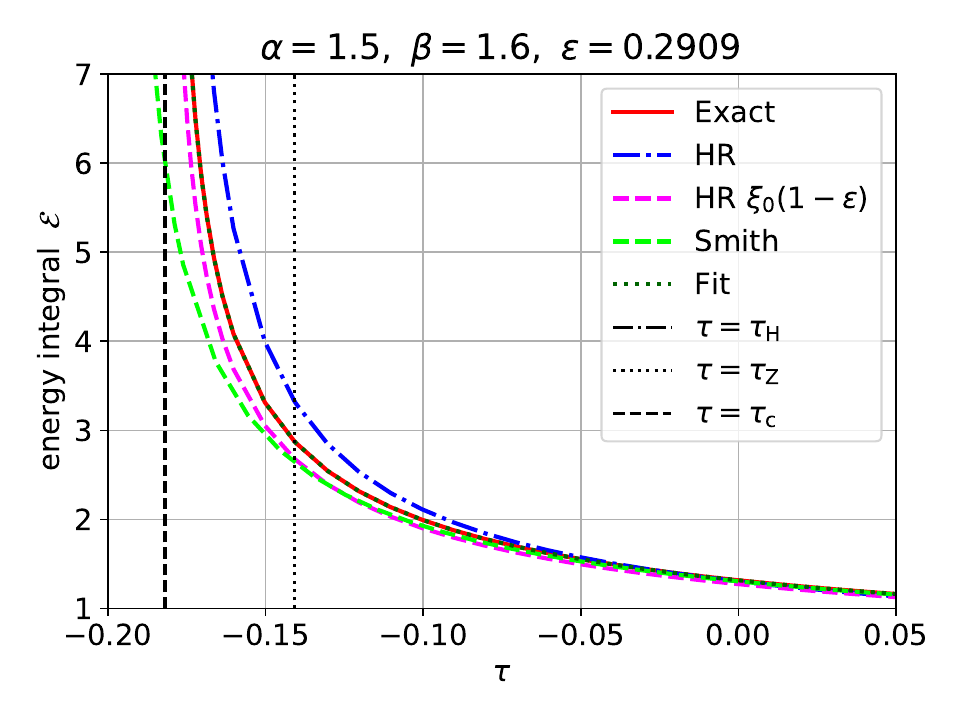}\includegraphics[scale=0.38]{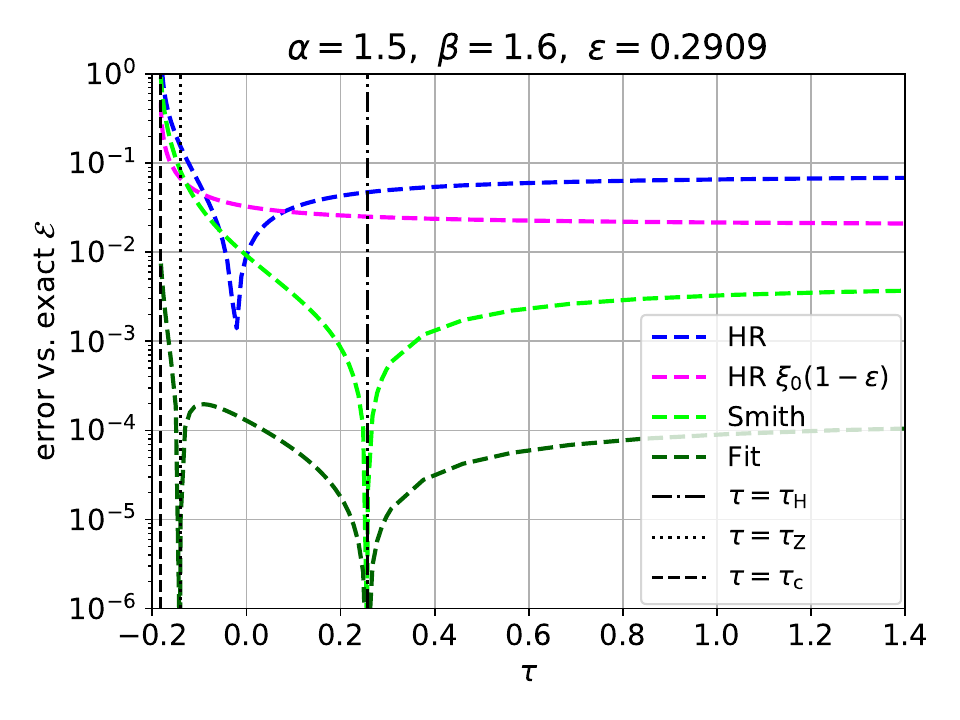} 
\par\end{centering}
\begin{centering}
\includegraphics[scale=0.38]{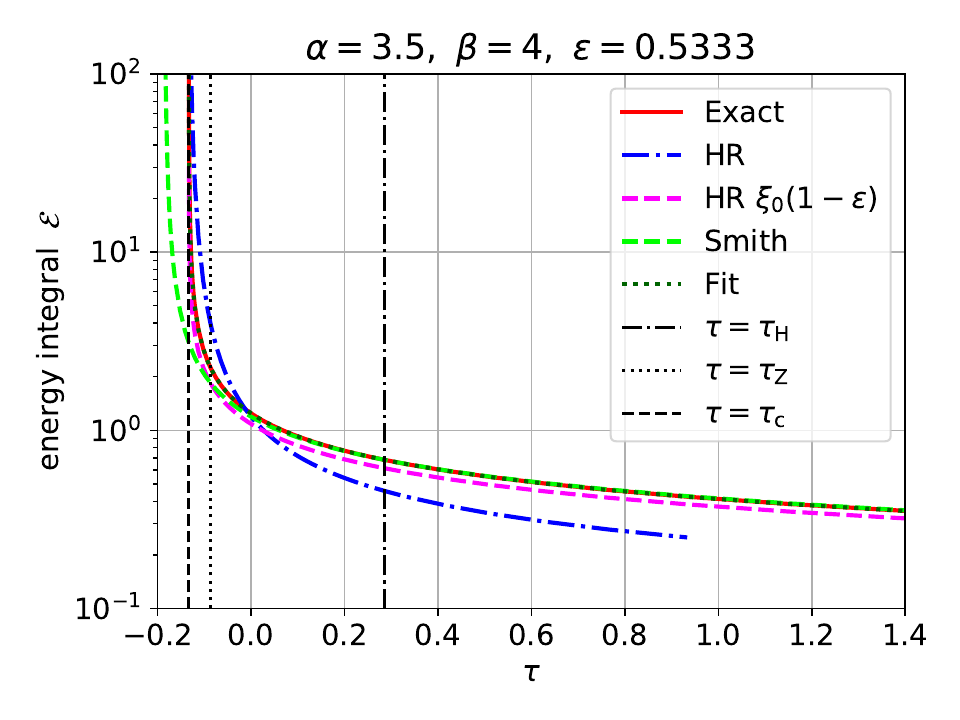}\includegraphics[scale=0.38]{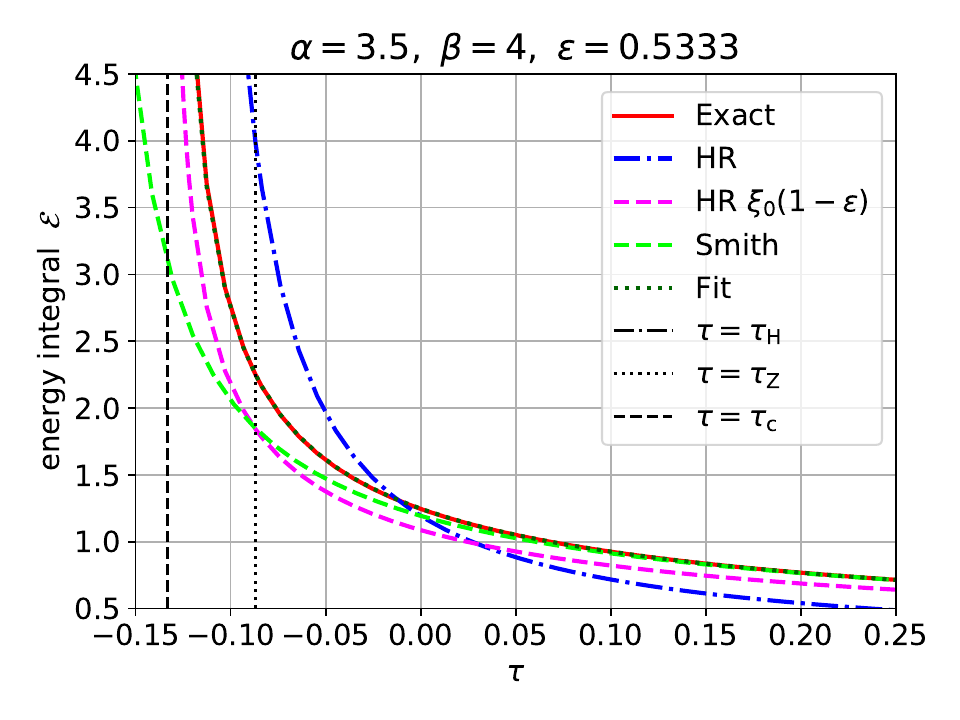}\includegraphics[scale=0.38]{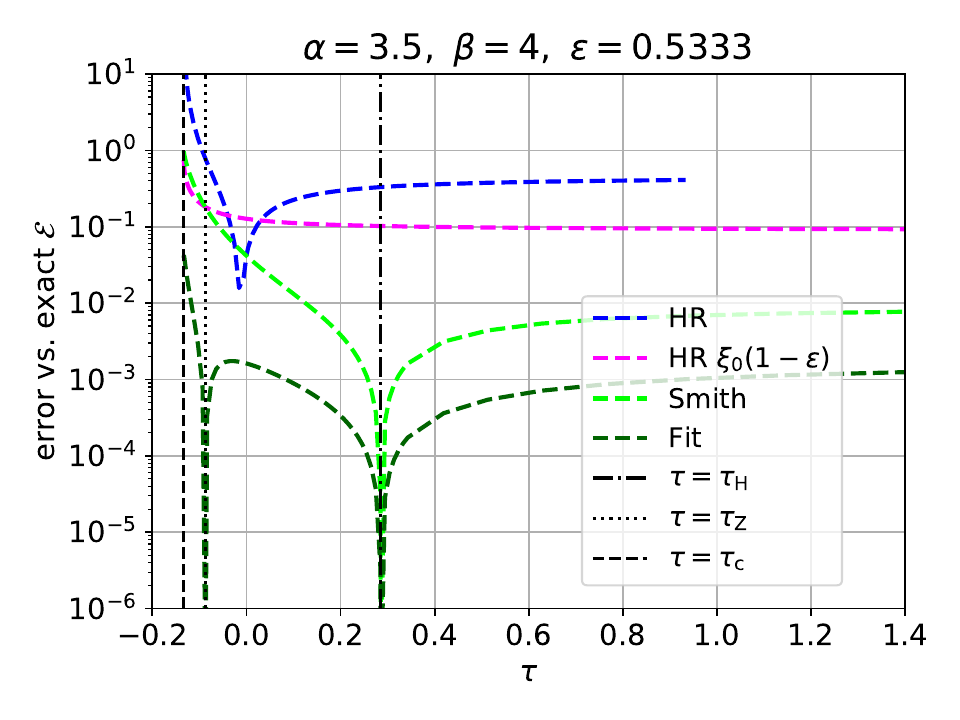} 
\par\end{centering}
\caption{Same as Fig. \ref{fig:ftag0_err}, but for the dimensionless energy
integral $\mathcal{E}$, which determines the total energy in the
system (see Eqs. \eqref{eq:Etotss}-\eqref{eq:Iint}). The HR, Smith
and fit results are obtained by an exact integration of the approximate
profile {[}Eqs. \eqref{eq:f_hr},\eqref{eq:energy_integral_HR_SMITH}{]}.
Also shown (magenta line) is the result of Eq. \eqref{eq:E_HR_INTEGRAL_apprx},
which does not use the exact integration. As discussed in the text,
for $\epsilon\approx0.533$, the HR profile becomes invalid for $\tau\protect\geq\frac{14}{15}$
.\label{fig:E_err}}
\end{figure*}

In Ref. \cite{hammer2003consistent} Hammer and Rosen (HR) introduced
a perturbative approach which results in an approximate solution of
the diffusion equation \eqref{eq:Tpde}, for a general time dependent
surface temperature $T\left(x=0,t\right)=T_{S}\left(t\right)$. The
perturbation expansion is employed through the parameter:

\begin{equation}
\epsilon=\frac{\beta}{4+\alpha},\label{eq:eps_HR}
\end{equation}
which is small for steep heat fronts, since the front exponent {[}see
Eqs. \eqref{eq:asymp_front}-\eqref{eq:nudef}{]} is: 
\begin{equation}
\nu=\frac{\epsilon}{\beta\left(1-\epsilon\right)}.
\end{equation}
In Ref. \cite{smith2010solutions}, Smith provides a more accurate
analysis based on a series expansion for which the velocity of the
heat front serves as a boundary condition, rather than the surface
temperature.

An application of either the HR or Smith approaches to the specific
case of a power law surface temperature $T_{S}\left(t\right)=T_{0}t^{\tau}$,
{[}Eq. \eqref{eq:Tbc}{]}, gives the following approximate form for
the temperature profile (see Eq. 37 in Ref. \cite{hammer2003consistent},
and Eq. 23 in Ref. \cite{smith2010solutions}):

\begin{equation}
f\left(\xi\right)=\left[\left(1-\frac{\xi}{\xi_{0}}\right)\left(1+R\frac{\xi}{\xi_{0}}\right)\right]^{\frac{1}{4+\alpha-\beta}},\label{eq:f_hr}
\end{equation}
where $R$ and $\xi_{0}$ are some functions of $\alpha$, $\beta$
and $\tau$. This form Eq. \eqref{eq:f_hr} is in fact a generalization
of the exact Henyey (for which $R=0$, see Sec. \ref{subsec:The-Henyey-analytic})
and Zel'dovich-Barenblatt (for which $R=1$, see Sec. \ref{subsec:The-Zel'dovich-Barenblatt-analyt})
solutions, which now applies to a general temperature drive exponent
$\tau$. We note that since $f\left(\xi\right)>0$ for $\xi<\xi_{0}$,
we must have $R\geq-1$ in Eq. \eqref{eq:f_hr}.

Given the values of $\xi_{0}$ and $R$, we can calculate some properties
of the approximate profile in Eq. \eqref{eq:f_hr}. The derivative
at the origin, which determines the net surface flux and the bath
temperature (see Sec. \ref{subsec:Marshak-boundary-condition}), is
is given by: 
\begin{equation}
f'\left(0\right)=\frac{\epsilon}{1-\epsilon}\frac{R-1}{\beta\xi_{0}}.\label{eq:ftag0_HR_SMITH}
\end{equation}
The energy integral {[}Eq. \eqref{eq:Iint}{]} can be calculated analytically:
\begin{equation}
\mathcal{E}={}_{2}F_{1}\left(1,-\frac{\epsilon}{1-\epsilon};\frac{2-\epsilon}{1-\epsilon};-R\right)\xi_{0}\left(1-\epsilon\right),\label{eq:energy_integral_HR_SMITH}
\end{equation}
where $_{2}F_{1}$ is the Gaussian hypergeometric function.

We discuss three choices for the calculation of the parameters $\xi_{0}$
and $R$ in Eq. \eqref{eq:f_hr}.

\subsubsection*{The Hammer-Rosen approximation}

The first choice is to employ the HR perturbation theory for a power
law surface temperature drive, which gives the following approximate
expressions for $\xi_{0}$ and $R$ (see Eqs. 35,37 in Ref. \cite{hammer2003consistent}
and the discussion following Eq. 3 in Ref. \cite{malka2022supersonic}):
\begin{equation}
\xi_{0,\text{HR}}^{2}=\frac{\left(2+\epsilon\right)\epsilon}{\left(1-\epsilon\right)\left(\epsilon+\tau\beta\right)}=\frac{2+\epsilon}{\left(1-\epsilon\right)\left(1+\tau\left(4+\alpha\right)\right)},\label{eq:xsi0_HR}
\end{equation}
\begin{equation}
R_{\text{HR}}=\frac{\epsilon}{2}-\frac{\beta\tau}{2}\xi_{0,\text{HR}}^{2}=\frac{\epsilon\left(\beta\tau\left(2\epsilon+1\right)+\left(\epsilon-1\right)\epsilon\right)}{2\left(\epsilon-1\right)\left(\beta\tau+\epsilon\right)}.\label{eq:R_HR}
\end{equation}
We note that the HR results correctly diverge at the value $\tau=\tau_{c}$,
in agreement with the exact solution. On the other hand, while the
exact solution is valid for any $\tau>\tau_{c}$, the HR results become
invalid for $\tau\geq\frac{\left(1-\epsilon\right)\epsilon}{\beta\left(2\epsilon-1\right)}$
(since $R\leq-1$), which are positive for $\epsilon>\frac{1}{2}$.

\subsubsection*{The Smith approximation}

The second choice is based on formulation of Smith, for which we find
(by employing Eqs. 27,30,33 in Ref. \cite{smith2010solutions}, for
a power law surface temperature) the following expressions for $\xi_{0}$
and $R$ : 
\begin{align}
\xi_{0,\text{Sm}}^{2} & =\frac{\left(2+\epsilon\right)\epsilon}{\left(1-\epsilon\right)\left(\epsilon+\beta\tau\left(1-\epsilon^{2}\right)\right)}\label{eq:XSI0_SM}\\
 & =\frac{2+\epsilon}{\left(1-\epsilon\right)\left(1+\tau\left(4+\alpha-\beta\right)\left(1+\epsilon\right)\right)},\nonumber 
\end{align}
\begin{align}
R_{\text{Sm}} & =\frac{1}{2}\left(\epsilon-\beta\tau\left(1-\epsilon\right)^{2}\xi_{0,\text{Sm}}^{2}\right)\label{eq:R_SMITH}\\
 & =\frac{\epsilon}{2}\left(1-\frac{\beta\tau\left(2+\epsilon\right)\left(1-\epsilon\right)}{\epsilon+\beta\tau\left(1-\epsilon^{2}\right)}\right).\nonumber 
\end{align}
It is evident that for $\tau=\tau_{H}$ the Smith solution coincides
with the exact Henyey solution (for which $R=0$ and $\xi_{0}$ is
given by Eq. \eqref{eq:xsi_0_hy}), while, as discussed in Ref. \cite{hammer2003consistent},
the agreement of the HR results {[}Eqs. \eqref{eq:xsi0_HR}-\eqref{eq:R_HR}{]}
with the Henyey solution is not exact but only accurate through first
order in $\epsilon$. We also note that for $\tau=0$ the Smith and
HR expressions coincide, while for $\tau=\tau_{Z}$ they differ, and
also do not match the exact Zel'dovich-Barenblatt solution (for which
$R=1$ and $\xi_{0}$ is given by Eq. \eqref{eq:xsi_0_Z}).

\subsubsection*{The least squares fit approximation}
The third choice is to obtain the parameters $\xi_{0}$ and $R$ numerically,
by finding the best fit of the form Eq. \eqref{eq:f_hr} to the exact
numerical solution. This approach is different than the HR and Smith
formulations above, which give simple, closed form but approximate
expressions for the parameters $\xi_{0}$ and $R$ to be used in the
approximate profile in Eq. \eqref{eq:f_hr}. Since we know that the
exact solution coincides with the form in Eq. \eqref{eq:f_hr} for
$\tau=\tau_{H}$ and $\tau=\tau_{Z}$, then if the exact profile $f\left(\xi\right)$
for a general $\tau$ does not differ significantly from this approximate
form, using the exact value of $\xi_{0}$ and calculating $R$ by
fitting Eq. \eqref{eq:f_hr} to the exact numerical solution, should
result in a reasonably accurate solution based on just two numerically
calculated parameters: the exact $\xi_{0}$ and a fitted parameter
$R$. Here we use a standard nonlinear least squares fitting method for the calculation of $R$. Fig. \ref{fig:profiles_fit} shows a comparison between the
exact solution and the approximated fit by Eq. \eqref{eq:f_hr}, for
$\epsilon\approx0.53$ and various values of $\tau$. In Fig. \ref{fig:fit_err}
we show the error between the exact numerical profile and the fit
approximation, as a function of $\tau$. We plot the $L_{1}$ and
$L_{\infty}$ error norms, which are, respectively, the average and
maximal relative error between the profiles, at each value of $\tau$.
It is evident that, as expected, for $\tau=\tau_{H}$ and $\tau=\tau_{Z}$
the error diminishes, since in these cases the exact solution has
the form \eqref{eq:f_hr}. It is remarkable that for $\tau\gtrsim\tau_{Z}$
the fitted profiles are in a very good agreement with the exact solution
with an error smaller than $10^{-3}$, even for a quite large value
of $\epsilon\approx0.53$. In the range $\frac{1}{2}\left(\tau_{Z}+\tau_{c}\right)\lesssim\tau\lesssim\tau_{Z}$,
where the exact solution and the form \eqref{eq:f_hr} have a local
maxima (as also seen in the bottom middle and right panes of Figs.
\ref{fig:origin_forms}-\ref{fig:origin_forms-eps013}), the fit solution
error remains reasonable, about $10^{-2}$. When $\tau$ approaches
$\tau_{c}$, the fit approximation becomes inaccurate, as $R\rightarrow\infty$.
It is also clear that the overall fit error decreases with $\epsilon$,
which means that the form \eqref{eq:f_hr} is a better approximation
for smaller $\epsilon$, as expected, since it is based on the HR
perturbation theory. We conclude that the parametrization of the exact
solution by Eq. \eqref{eq:f_hr}, which is much more simple to use
than a full numerical solution for $f\left(\xi\right)$, is highly
accurate in a wide range of $\tau$ and even for intermediate values
of $\epsilon$.

In Figs. \ref{fig:origin_forms}-\ref{fig:origin_forms-eps013}, we
compare the similarity profiles resulting from the exact numerical
solution, the asymptotic near front approximation {[}Eq. \eqref{eq:asymp_front}{]}
and the approximate form in Eq. \eqref{eq:f_hr} using the HR, Smith
and numerically fitted values for $\xi_{0}$ and $R$. The comparisons
are shown for various values of $\tau$ and for three choices of $\alpha,\beta$
resulting in $\epsilon\approx0.13,\ 0.29$ and $0.53$. In Fig. \ref{fig:Rfit}
we plot the values of $R$ as a function of $\tau$, as obtained from
the HR and Smith approximations and from a numerical fit to the exact
solution. It is evident that the Smith approximation is closer than
HR to the numerically fitted values, especially for $\tau\gtrsim0$.
For $\tau=\tau_{H}$, the Smith approximation coincides with the exact
solution (and the fit). For $\tau=0$ the HR and Smith approximations
coincide. For $\tau=\tau_{Z}$ both the HR and Smith approximations
fail to reproduce the exact result ($R=1$). From Eq. \eqref{eq:ftag0_HR_SMITH}
we see that the fit solutions have a local maxima at the correct range,
since $R\geq1$ only for $\tau_{c}<\tau\leq\tau_{Z}$, while the HR
profiles have a maxima even at $\tau<\tau_{Z}$ and on the other hand,
the Smith profiles do not have maxima at $\tau>\tau_{Z}$. In addition,
for $\tau<\tau_{Z}$, the HR values of $R\left(\tau\right)$ diverge
faster than the fitted values, but at the correct value $\tau=\tau_{c}$,
while the Smith values do not diverge at $\tau_{c}$ but at the lower,
incorrect value $\tau=-\tau_{H}/\left(1+\epsilon\right)$.

In Fig. \ref{fig:xsi0_err} we compare the exact, HR and Smith results
for the front coordinate $\xi_{0}$ as a function of $\tau$ (the
fit approximation uses the exact numerical $\xi_{0}$, and is not
shown). Some of these results for the front coordinate can also be
seen in the profiles shown in Figs. \ref{fig:origin_forms}-\ref{fig:origin_forms-eps013}.
As discussed by Smith in Ref. \cite{smith2010solutions}, we see that
for $\tau>0$ the HR results underestimate the value of $\xi_{0}$,
while the Smith results are accurate to less than 1\%, which is better
than the HR accuracy by about an order of magnitude. For $\tau<0$,
the HR result overestimates the value $\xi_{0}$ while Smith underestimates
it. In the range $\frac{1}{2}\tau_{Z}\lesssim\tau\lesssim0$, the
HR is better than the Smith approximation, with accuracy of 0.1\%
to \textasciitilde few \%. For $\tau_{c}\lesssim\tau\lesssim\frac{1}{2}\tau_{Z}$,
as the exact $\xi_{0}$ diverges, both approximations become increasingly
inaccurate. In Fig. \ref{fig:ftag0_err} we compare the exact results
for $f'\left(0\right)$ with the HR, Smith and fit approximations,
as a function of $\tau$. A similar comparison in made in Fig. \ref{fig:E_err}
for the energy integral. As expected, it is evident that the fit results
are much better than HR and Smith, and are relatively accurate even
for $\tau_{c}<\tau\leq\tau_{Z}$. We note that Hammer and Rosen employ
a different approximation for the total energy (Eq. 29 in Ref. \cite{hammer2003consistent}):
\begin{equation}
\mathcal{E}=\xi_{0}\left(1-\epsilon\right),\label{eq:E_HR_INTEGRAL_apprx}
\end{equation}
which, as seen in Fig. \ref{fig:E_err} is mostly better than the
integrated HR profile (Eq. \eqref{eq:energy_integral_HR_SMITH} using
Eqs. \eqref{eq:xsi0_HR}-\eqref{eq:R_HR}).

Finally, we note that, as expected, it is evident from Figs \ref{fig:origin_forms}-\ref{fig:origin_forms-eps013}
and Figs. \ref{fig:fit_err}-\ref{fig:E_err}, that both the HR, Smith
and fit approximations become increasingly better when $\epsilon$
is decreased.

\section{The optical depth\label{sec:The-optical-depth}}

The optical depth $\mathcal{T}$, is defined as the number of mean
free paths within the heat wave (see i.e. Refs. \cite{rosen2005fundamentals,krief2024self}):
\begin{align}
\mathcal{T}\left(t\right)= & \int_{0}^{\infty}k_{t}\left(T\left(x,t\right)\right)dx.
\end{align}
Using the self-similar solution Eq. \eqref{eq:Tss}, we find{\small{}
\begin{align}
\mathcal{T}\left(t\right) & =t^{\frac{1}{2}\left(\tau\left(4-\alpha-\beta\right)+1\right)}\left(\frac{4aT_{0}^{4-\alpha-\beta}ck_{0}}{3\left(4+\alpha\right)u_{0}}\right)^{\frac{1}{2}}\int_{0}^{\infty}f^{-\alpha}\left(\xi\right)d\xi\nonumber \\
 & =\left[t^{\delta}\left(KT_{0}^{4+\alpha-\beta}\right)^{\frac{1}{2}}\right]k\left(T_{0}t^{\tau}\right)\int_{0}^{\infty}f^{-\alpha}\left(\xi\right)d\xi.\label{eq:Topt_2}
\end{align}
}Evidently, $\mathcal{T}\propto k_{0}^{1/2}$ which shows how the
optical depth increases for opaque problems. The second form, Eq. \eqref{eq:Topt_2},
expresses the optical depth as the ratio between the typical heat
front position (the first term) to the mean free path evaluated at
the current surface temperature (the second term, which is the inverse
of Eq. \eqref{eq:ross_opac_powerlaw}), and corrected by the reduction
of the mean free path due to difference between the surface temperature
and the temperature across the heat wave (the integral term).

For linear conduction, the exact solution {[}Eq. \eqref{eq:f_lin_sol}{]},
results in a diverging integral in Eq. \eqref{eq:Topt_2}. For nonlinear
conduction, the integral is carried from the surface to the heat front,
and Eq. \eqref{eq:Topt_2} can be rewritten as:
\begin{align}
\mathcal{T}\left(t\right) & =\int_{0}^{x_{F}\left(t\right)}k_{t}\left(T\left(x,t\right)\right)dx.\nonumber \\
 & =x_{F}\left(t\right)k\left(T_{0}t^{\tau}\right)\left(\frac{1}{\xi_{0}}\int_{0}^{\xi_{0}}f^{-\alpha}\left(\xi\right)d\xi\right).\label{eq:OPTDEPTH_exact}
\end{align}
The integral in Eq. \eqref{eq:OPTDEPTH_exact} can be calculated analytically
by assuming the similarity profile of the form given in Eq. \eqref{eq:f_hr},
which, as discussed in Sec. \ref{sec:Non-linear-conduction}, represents
the Hammer-Rosen and Smith approximate profiles for general $\tau$,
as well as the exact Henyey and Zel'dovich-Barenblatt profiles for
specific values of $\tau$. As in the calculation of the energy integral
{[}Eq. \eqref{eq:energy_integral_HR_SMITH}{]}, the result is given
in terms of a Gaussian hypergeometric function:

\begin{align}
 & \frac{1}{\xi_{0}}\int_{0}^{\xi_{0}}f^{-\alpha}\left(\xi\right)d\xi=\int_{0}^{1}\left[\left(1-y\right)\left(1+Ry\right)\right]^{\frac{-\alpha}{4+\alpha-\beta}}dy\nonumber \\
= & _{2}F_{1}\left(1,\frac{\alpha}{4+\alpha-\beta};\frac{8+\alpha-2\beta}{4+\alpha-\beta};-R\right)\frac{4+\alpha-\beta}{4-\beta}.
\end{align}
This integral diverges, unless $\frac{\alpha}{4+\alpha-\beta}<1$,
which means that we must have $\beta<4$ in order for the integral
to be finite. For $\beta=4$ the integral diverges logarithmically
(a result which was also found in Refs. \cite{rosen2005fundamentals,krief2024self},
for $\tau=\tau_{H}$).

This divergence of the optical depth integral results from the steep
temperature decrease near the heat front for nonlinear conduction,
and from the infinite extent of the heat wave for linear conduction.
Therefore, as discussed in Ref. \cite{krief2024self}, a more useful
and simple estimate for the optical depth is obtained by ignoring
the temperature variation across the wave, and using the mean-free-path
at the surface temperature. This is equivalent to replacing the integral
$\int_{0}^{\infty}f^{-\alpha}\left(\xi\right)d\xi$ with $\xi_{0}$,
which for nonlinear conduction is the heat front coordinate, while
for linear conduction it is the value of $\xi$ for which the temperature
profile {[}Eq. \eqref{eq:f_lin_sol}{]} has decayed (we take $\xi_{0}=\sqrt{20}$).
Since the average temperature across the profile is usually lower
than the surface temperature, this results in a useful lower bound
for the optical depth

\begin{align}
\mathcal{T}\left(t\right)\gtrsim & x_{F}\left(t\right)k\left(T_{0}t^{\tau}\right)\label{eq:depth_tot}\\
=\xi_{0} & t^{\frac{1}{2}\left(\tau\left(4-\alpha-\beta\right)+1\right)}\left(\frac{4aT_{0}^{4-\alpha-\beta}ck_{0}}{3\left(4+\alpha\right)u_{0}}\right)^{\frac{1}{2}},
\end{align}
which can be used to estimate whether a heat wave is opaque enough
such that the diffusion approximation of radiation transport is applicable. 

\section{Comparison with simulations\label{sec:simulations}}

In this section we define radiation transfer benchmarks based on the
LTE diffusion problem defined in Sec. \ref{sec:Statement-of-the}.
We define setups for gray diffusion and transport calculations, which
are non-LTE calculations that handle the dynamics of the radiation
field and material temperature profiles.

\subsection{Gray diffusion setup}

In the absence of photon scattering, the total opacity $k$ (given
by Eq. \eqref{eq:ross_opac_powerlaw}) and the absorption opacity
are equal, and the gray diffusion equation in slab geometry reads:

\begin{equation}
\frac{\partial E_{r}}{\partial t}=\frac{\partial}{\partial x}\left(\frac{c}{3k}\frac{\partial E_{r}}{\partial x}\right)+ck\left(aT^{4}-E_{r}\right),\label{eq:main_gray}
\end{equation}
\begin{equation}
\frac{\partial u_{m}\left(T\right)}{\partial t}=ck\left(E_{r}-aT^{4}\right),\label{eq:main_mat}
\end{equation}
where $E_{r}$ is the radiation energy density and 
\begin{equation}
u_{m}\left(T\right)=u\left(T\right)-aT^{4}=u_{0}T^{\beta}-aT^{4},\label{eq:u_mat_tr}
\end{equation}
is the material energy density (since Eq. \eqref{eq:utot} represents
the matter+radiation energy at equilibrium). In the LTE limit, which
is reached for optically thick problems (when the optical depth $\mathcal{T\gg}1$,
see Sec. \ref{sec:The-optical-depth}), the material and radiation
temperatures are equal, $E_{r}=aT^{4}$, and the gray diffusion system
\eqref{eq:main_gray}-\eqref{eq:main_mat} is reduced to the LTE diffusion
equation \eqref{eq:main_eq} for the total energy density $u\left(T\right)=u_{m}\left(T\right)+aT^{4}$. 

The boundary condition for gray diffusion is the same as for LTE diffusion,
using either a prescribed surface temperature {[}Eq. \eqref{eq:Tbc}{]}
or bath temperature via the Marshak boundary condition {[}Eqs. \eqref{eq:fick},
\eqref{eq:Tbath_marsh_bc_F}-\eqref{eq:Tbath}{]}.

\subsection{Transport setup\label{subsec:Transport-setup}}

The general one dimensional, one group (gray) radiation transport
equation without scattering and in slab symmetry for the radiation
intensity field $I\left(x,\mu,t\right)$ is given by \cite{su1997analytical,olson2000diffusion,pomraning2005equations,steinberg2022multi,castor2004radiation,mihalas1999foundations,bennett2023benchmark,hu2023rigorous,krief2024self}:

\begin{align}
\left(\frac{1}{c}\frac{\partial}{\partial t}+\mu\frac{\partial}{\partial x}\right)I\left(x,\mu,t\right)+ & k\left(T\right)I\left(x,\mu,t\right)\label{eq:Treq}\\
 & =\frac{ac}{4\pi}k\left(T\right)T^{4}\left(x,t\right),\nonumber 
\end{align}
where $\mu$ is the directional angle cosine, $k\left(T\right)$ the
absorption opacity (given by Eq. \eqref{eq:ross_opac_powerlaw}).
The transport equation for the radiation field is coupled to the material
via the material energy equation:

\begin{align}
\frac{\partial u_{m}\left(T\right)}{\partial t} & =k\left(T\right)\left[2\pi\int_{-1}^{1}d\mu'I\left(x,\mu',t\right)-acT^{4}\left(x,t\right)\right],\label{eq:tr_mat}
\end{align}
where $u_{m}\left(T\right)$ is given in Eq. \eqref{eq:u_mat_tr}.
The radiation energy density is defined by the zeroth angular moment
of the intensity via: 
\begin{equation}
E_{r}\left(x,t\right)=\frac{2\pi}{c}\int_{-1}^{1}d\mu'I\left(x,\mu',t\right).
\end{equation}
The boundary condition for the transport problem is defined by an
incident radiation field for incoming directions $\mu>0$, which is
given by a black body radiation bath: 
\begin{equation}
I\left(x=0,\mu,t\right)=\frac{ac}{4\pi}T_{\text{bath}}^{4}\left(t\right),\label{eq:Ibath_tr}
\end{equation}
where the time dependent bath temperature is given by solution of
the LTE diffusion problem via eq. \eqref{eq:Tbath}, which is obtained
from the Marshak (Milne) boundary condition, as detailed in Sec. \ref{subsec:Marshak-boundary-condition}.

In the absence of photon scattering, the LTE diffusion limit is reached
for optically thick problems ($\mathcal{T\gg}1$), so that $E_{r}=aT^{4}$
and the transport problem \eqref{eq:Treq}-\eqref{eq:Ibath_tr} is
reduced to the LTE diffusion problem defined in Sec. \ref{sec:Statement-of-the}.

\subsection{Test cases}

We define four benchmarks based on the self-similar solutions of the
LTE diffusion equation given in Sec. \ref{sec:Self-Similar-solution}.
We specify in detail the setups for LTE and gray radiation diffusion
as well as radiation transport computer simulations. We have performed
gray diffusion simulations as well as deterministic discrete-ordinates
($S_{N}$) and stochastic implicit Monte-Carlo (IMC) \cite{fleck1971implicit,mcclarren2009modified}
transport simulations. The $S_{N}$ transport calculations were performed
using a numerical method which is detailed in \cite{mcclarren2022data},
while the IMC calculations were performed using a novel numerical
scheme that was recently developed by Steinberg and Heizler in Refs.
\cite{steinberg2022new,steinberg2022multi,steinberg2023frequency}.
All benchmarks are run until the final time $t=1\text{ns}$. The benchmarks
were defined such that the optical depth {[}Eq. \eqref{eq:depth_tot}{]}
is larger than unity after a short transient, so that the LTE diffusion
limit is applicable, and the gray diffusion and transport simulation
results should agree with the self-similar solutions of the LTE diffusion
equation.

We note that unlike LTE diffusion simulations, which use a given total
energy density function $u\left(T\right)$ as the dependent variable, non-LTE simulations
(gray diffusion and transport), in which the radiation energy density
is always present explicitly, solve for the material energy density $u_{m}\left(T\right)$, which, according to Eq. \eqref{eq:u_mat_tr}, must be given
as a difference between a temperature power law $u_{0}T^{\beta}$,
and the radiation energy density $aT^{4}$. This is not an issue if
the material energy density has the form $u_{m}\left(T\right)=aT^{4}/\varepsilon$,
so that the total energy density is $u\left(T\right)=\left(1+\frac{1}{\varepsilon}\right)aT^{4}$
(that is, $\beta=4$, $u_{0}=\left(1+\frac{1}{\varepsilon}\right)a$).
In the more realistic case for which the material energy density is
given by a power law $u_{m}\left(T\right)=u_{0,m}T^{\beta_{m}}$ with
$\beta_{m}\neq4$, the total energy density can be approximated by
a power law only if either the radiation energy is negligible, so
that $u\left(T\right)=u_{0,m}T^{\beta_{m}}+aT^{4}\approx u_{0,m}T^{\beta_{m}}$
or if the material energy is negligible so that $u\left(T\right)\approx aT^{4}$.
The benchmarks below have either a negligible radiation energy density
(tests 1 ,3-5), or $\beta_{m}=4$ (tests 2, 6), so that the total energy
density obeys a power law.

\subsubsection*{TEST 1}

\begin{figure}[t]
\begin{centering}
\includegraphics[scale=0.55]{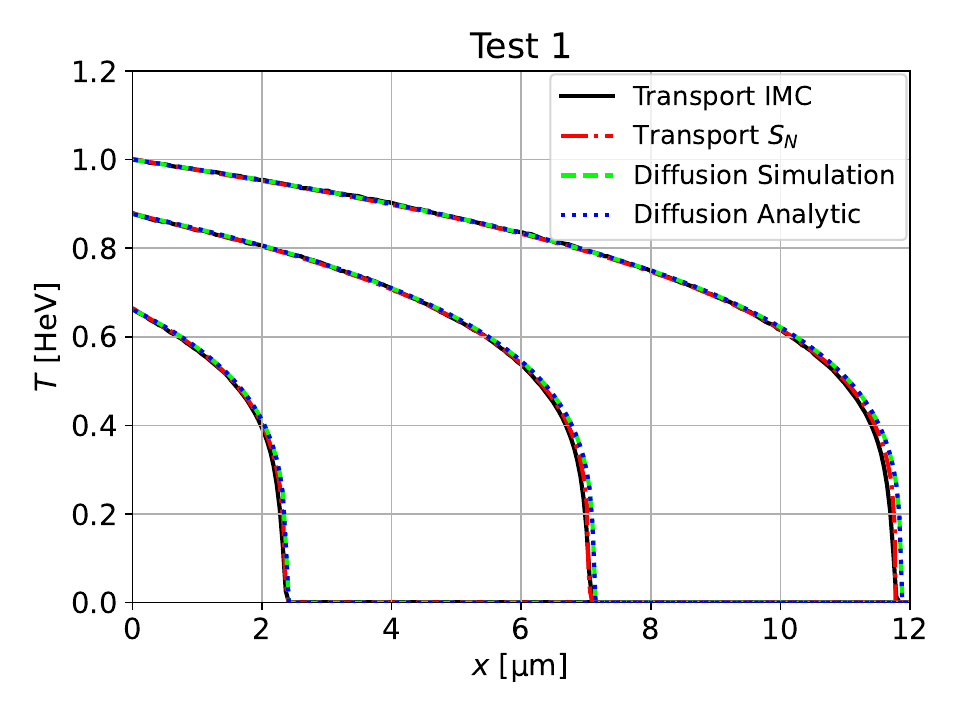} 
\par\end{centering}
\caption{Temperature profiles for test 1. Results are shown at times $t=0.2,\ 0.6$
and 1ns, as obtained from a gray diffusion simulation and from Implicit-Monte-Carlo
(IMC) and discrete ordinates ($S_{N}$) transport simulations, and
are compared to the analytic solution of the diffusion equation {[}Eq.
\eqref{eq:Tprof_anal_1}{]}.\label{fig:simulation_1}}
\end{figure}
\begin{figure}[t]
\begin{centering}
\includegraphics[scale=0.55]{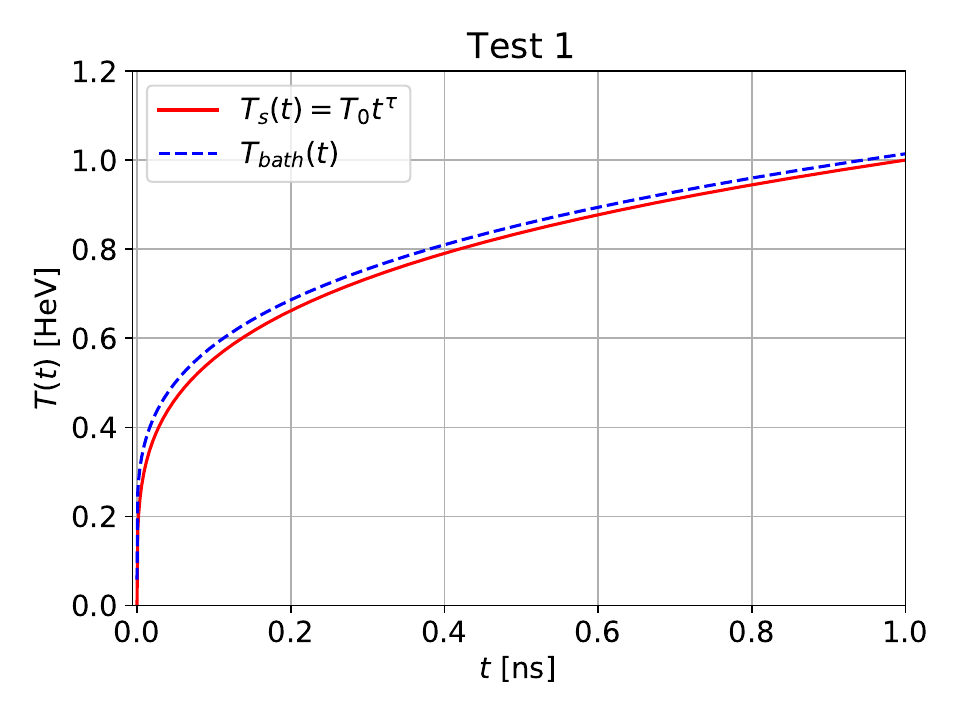}
\par\end{centering}
\caption{A comparison between the surface (red line, Eq. \eqref{eq:Tbound_1})
and bath (blue line, Eq. \eqref{eq:Tbath_1}) driving temperatures,
for test 1.\label{fig:simulation_1_Tbath}}
\end{figure}

For the first test, we take a simple Henyey wave which has an analytical
solution (see Sec. \ref{subsec:The-Henyey-analytic}), for a gold-like
material model used by Hammer and Rosen \cite{hammer2003consistent}, at a temperature scale
of $1\text{HeV}=0.1\text{KeV}\approx1.160452\times10^{6}\text{K}$,
at which the radiation energy can be neglected. We take $\alpha=1.5$,
$\beta=1.6$ with coefficients $k_{0}=10^{4}\text{HeV}^{1.5}/\text{cm}$,
$u_{0}=10^{13}\text{HeV}^{-1.6}\text{erg/cm}^{3}$ so that the opacity
is: 
\begin{equation}
k\left(T\right)=10^{4}\left(\frac{T}{\text{HeV}}\right)^{-1.5}\ \text{cm}^{-1},
\end{equation}
and the total energy density, which is used in LTE diffusion simulations,
is: 
\begin{equation}
u\left(T\right)=10^{13}\left(\frac{T}{\text{HeV}}\right)^{1.6}\ \frac{\text{erg}}{\text{cm}^{3}}.
\end{equation}
We define a Henyey wave, so that $\tau=\tau_{H}=\frac{10}{39}$, and
surface temperature 
\begin{equation}
T_{s}\left(t\right)=\left(\frac{t}{\text{ns}}\right)^{\frac{10}{39}}\ \text{HeV}.\label{eq:Tbound_1}
\end{equation}
At these temperatures the radiation energy density is much smaller
than the material energy density, and can be neglected, so that for
non-LTE simulations (gray diffusion and transport), in which the material
and radiation energies are separated, we can take: 
\begin{equation}
u_{m}\left(T\right)=u\left(T\right)-aT^{4}\approx10^{13}\left(\frac{T}{\text{HeV}}\right)^{1.6}\ \frac{\text{erg}}{\text{cm}^{3}}.
\end{equation}
Eq. \eqref{eq:xsi_0_hy} gives the front coordinate $\xi_{0}=1.187542$,
which via eq. \eqref{eq:xheat} gives the heat front position, which
propagates linearly in time: 
\begin{equation}
x_{F}\left(t\right)=0.00118584\left(\frac{t}{\text{ns}}\right)\ \text{cm}.
\end{equation}
The temperature profile is given analytically by using Eqs. \eqref{eq:f_Hy_Exact}
and \eqref{eq:Tss}:

\begin{equation}
T\left(x,t\right)=\left(\frac{t}{\text{ns}}\right)^{\frac{10}{39}}\left(1-\frac{x}{x_{F}\left(t\right)}\right)^{\frac{10}{39}}\ \text{HeV},\label{eq:Tprof_anal_1}
\end{equation}
which is a familiar Marshak wave with a steep heat front, shown in
Fig. \ref{fig:simulation_1}. From Eqs. \eqref{eq:Tbath}-\eqref{eq:Bbath},
\eqref{eq:FTAG_HY} we find the bath temperature: 
\begin{equation}
T_{\text{bath}}\left(t\right)=\left(1+0.0576603\left(\frac{t}{\text{ns}}\right)^{-\frac{8}{13}}\right)^{\frac{1}{4}}\left(\frac{t}{\text{ns}}\right)^{\frac{10}{39}}\ \text{HeV},\label{eq:Tbath_1}
\end{equation}
which is used in transport simulations via the incoming bath radiation
flux \eqref{eq:Ibath_tr} or in diffusion simulations via the Marshak
boundary condition \eqref{eq:marsh_bc_def}. We note that diffusion
simulations can be run equivalently using the surface temperature
boundary condition \eqref{eq:Tbound_1}. A comparison of the surface
and bath temperatures as a function of time are shown in Figure \ref{fig:simulation_1_Tbath}.
As discussed in Sec. \ref{subsec:Marshak-boundary-condition}, the
bath temperature is slightly larger than the surface temperature,
since $\tau>\tau_{Z}$. The energy integral {[}Eqs. \eqref{eq:Iint},
\eqref{eq:eint_hy_Exact}{]} is $\mathcal{E}=0.842075$, which gives
the total energy as a function of time {[}Eq. \eqref{eq:Etotss}{]}:
\begin{equation}
E\left(t\right)=8.4086998\times10^{9}\left(\frac{t}{\text{ns}}\right)^{\frac{55}{39}}\ \text{erg}.
\end{equation}
According to Eq. \eqref{eq:depth_tot}, the optical depth at the final
time is {\small{}
\[
\mathcal{T}\gtrsim x_{F}\left(t_{\text{final}}\right)k\left(T_{s}\left(t_{\text{final}}\right)\right)=0.00118584\times10^{4}\approx11.86,
\]
}so the heat wave is optically thick and we expect the LTE diffusion
limit to hold. A comparison between the analytical profiles and the
simulation results are presented in Fig. \ref{fig:simulation_1},
showing a good agreement.

\subsubsection*{TEST 2}

\begin{figure}[t]
\begin{centering}
\includegraphics[scale=0.55]{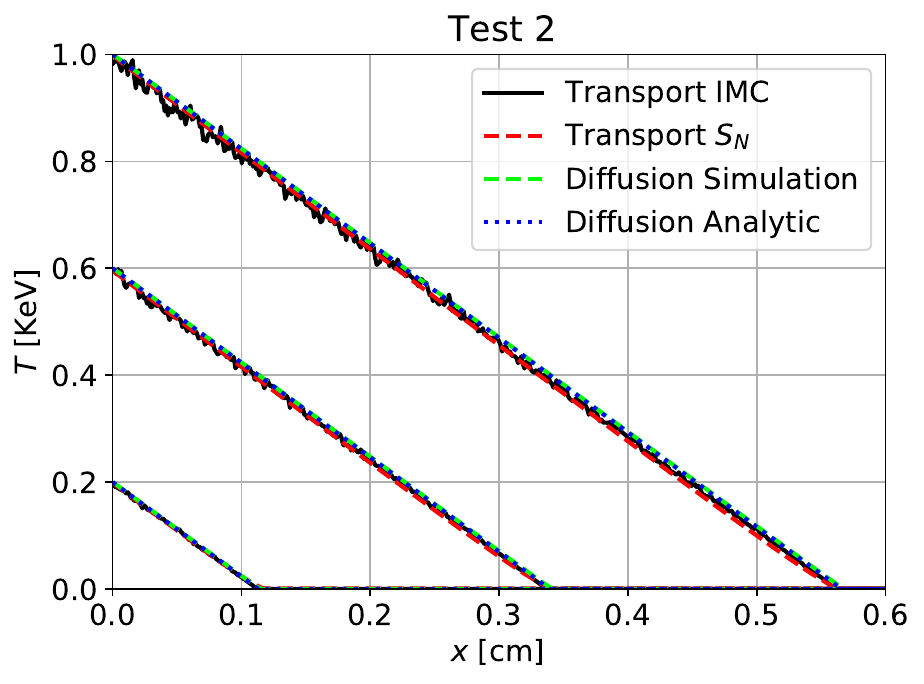}
\par\end{centering}
\caption{Temperature profiles for test 2. Results are shown at times $t=0.2,\ 0.6$
and 1ns, as obtained from a gray diffusion simulation and from Implicit-Monte-Carlo
(IMC) and discrete ordinates ($S_{N}$) transport simulations, and
are compared to the analytic solution of the diffusion equation {[}Eq.
\eqref{eq:Tprof_anal_2}{]}.\label{fig:simulation_2}}
\end{figure}

As in the previous case, we define a Henyey wave, but with a material
model which gives rise to a linear front, and with a higher temperature
scale (KeV), such that the radiation energy density is not negligible.
We take $\alpha=1$, $\beta=4$, so that according to Eq. \eqref{eq:nudef}
it has a linear front, $\nu=1$, with coefficients $k_{0}=10^{2}\text{KeV}/\text{cm}$,
$u_{0}=\frac{5}{4}a$, so that the opacity is: 
\begin{equation}
k\left(T\right)=100\left(\frac{T}{\text{KeV}}\right)^{-1}\ \text{cm}^{-1},
\end{equation}
and the total energy density, which is used in LTE diffusion simulations,
is: 
\begin{equation}
u\left(T\right)=\frac{5}{4}aT^{4}=\frac{5}{4}\times1.372017\times10^{14}\left(\frac{T}{\text{keV}}\right)^{4}\ \frac{\text{erg}}{\text{cm}^{3}}.
\end{equation}
For non-LTE simulations, we have the material energy density: 
\begin{equation}
u_{m}\left(T\right)=u\left(T\right)-aT^{4}=\frac{1}{5}u\left(T\right),
\end{equation}
which is 25\% of the radiation energy density. As in the previous
case, we define a Henyey wave, $\tau=\tau_{H}=1$, and the surface
temperature: 
\begin{equation}
T_{s}\left(t\right)=\left(\frac{t}{\text{ns}}\right)\ \text{KeV}.\label{eq:Tbound-2}
\end{equation}
Eq. \eqref{eq:xsi_0_hy} gives the front coordinate $\xi_{0}=2.236068$
which via eq. \eqref{eq:xheat}, gives the heat front position, which
propagates linearly in time: 
\begin{equation}
x_{F}\left(t\right)=0.56548972\left(\frac{t}{\text{ns}}\right)\ \text{cm}.
\end{equation}
The temperature profile is given analytically by using Eqs. \eqref{eq:f_Hy_Exact}
and \eqref{eq:Tss}:

\begin{equation}
T\left(x,t\right)=\left(\frac{t}{\text{ns}}\right)\left(1-\frac{x}{x_{F}\left(t\right)}\right)\ \text{KeV},\label{eq:Tprof_anal_2}
\end{equation}
which defines a linear heat wave profile, shown in Fig. \ref{fig:simulation_2}.
From Eqs. \eqref{eq:Tbath}-\eqref{eq:Bbath}, \eqref{eq:FTAG_HY}
we find the bath temperature: 
\begin{equation}
T_{\text{bath}}\left(t\right)=1.01158627\left(\frac{t}{\text{ns}}\right)\ \text{KeV},
\end{equation}
The energy integral {[}Eqs. \eqref{eq:Iint}, \eqref{eq:eint_hy_Exact}{]}
is $\mathcal{E}=0.4472136$, which gives the total energy as a function
of time {[}Eq. \eqref{eq:Etotss}{]}: 
\begin{equation}
E\left(t\right)=1.939654\times10^{13}\left(\frac{t}{\text{ns}}\right)^{5}\ \text{erg}.
\end{equation}
According to Eq. \eqref{eq:depth_tot} the optical depth at the final
time is {\small{}
\[
\mathcal{T}\gtrsim x_{F}\left(t_{\text{final}}\right)k\left(T_{s}\left(t_{\text{final}}\right)\right)=0.56549\times100\approx56.5,
\]
}so the heat wave is optically thick and we expect the LTE diffusion
limit to hold. The results are presented in Fig. \ref{fig:simulation_2},
showing a good agreement between the analytic solution and numerical
simulations.

\subsubsection*{TEST 3}

\begin{figure}[t]
\begin{centering}
\includegraphics[scale=0.55]{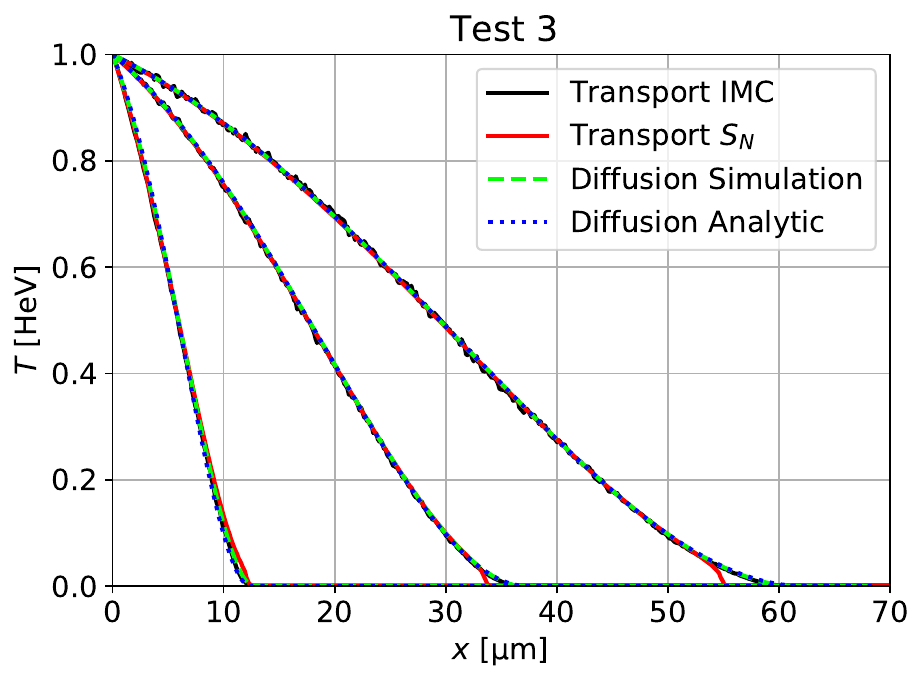}
\par\end{centering}
\caption{Temperature profiles for test 3. Results are shown at times $t=0.04\text{ns},\ 0.36\text{ns}$
and 1ns, as obtained from a gray diffusion simulation and from Implicit-Monte-Carlo
(IMC) and discrete ordinates ($S_{N}$) transport simulations, and
are compared to the analytic solution of the diffusion equation {[}Eq.
\eqref{eq:Tprof_anal_3}{]}. \label{fig:simulation_3}}
\end{figure}
\begin{figure}[t]
\begin{centering}
\includegraphics[scale=0.55]{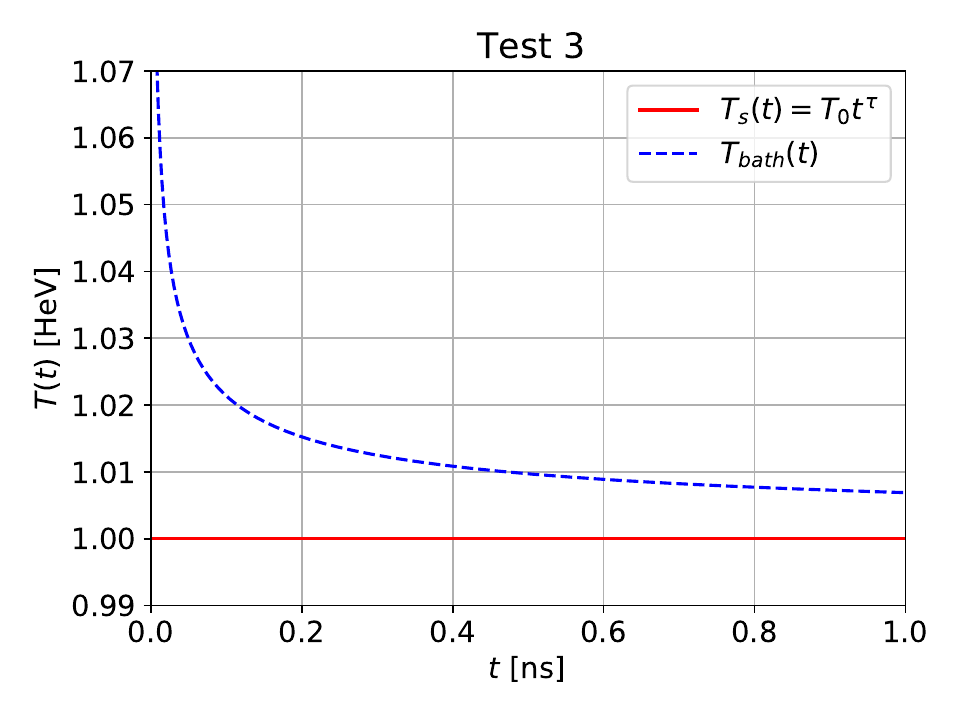}
\par\end{centering}
\caption{A comparison between the surface (red line, Eq. \eqref{eq:Tbound_3})
and bath (blue line, Eq. \eqref{eq:Tbath_3}) driving temperatures,
for test 3.\label{fig:simulation_3_Tbath}}
\end{figure}
\begin{table}[t]
\centering{}%
\begin{tabular}{|c|c|c||c|c|}
\cline{2-5} \cline{3-5} \cline{4-5} \cline{5-5} 
\multicolumn{1}{c|}{} & \multicolumn{2}{c||}{Test 3} & \multicolumn{2}{c|}{Test 5}\tabularnewline
\hline 
$\xi_{0}$ & \multicolumn{2}{c||}{6.2167035} & \multicolumn{2}{c|}{24.826729}\tabularnewline
\hline 
$R$ & \multicolumn{2}{c||}{0.7198876} & \multicolumn{2}{c|}{36.84138}\tabularnewline
\hline 
\hline 
$\xi/\xi_{0}$ & $f\left(\xi\right)$ exact & $f\left(\xi\right)$ fit & $f\left(\xi\right)$ exact & $f\left(\xi\right)$ fit\tabularnewline
\hline 
\hline 
0 & 1 & 1 & 1 & 1\tabularnewline
\hline 
0.015 & 0.99004 & 0.99129 & 1.0802 & 1.0675\tabularnewline
\hline 
0.03 & 0.97966 & 0.98198 & 1.1327 & 1.1161\tabularnewline
\hline 
0.05 & 0.96514 & 0.96864 & 1.1831 & 1.1651\tabularnewline
\hline 
0.1 & 0.92548 & 0.93082 & 1.2643 & 1.2478\tabularnewline
\hline 
0.15 & 0.88107 & 0.88696 & 1.3153 & 1.3016\tabularnewline
\hline 
0.2 & 0.83207 & 0.83756 & 1.3508 & 1.3398\tabularnewline
\hline 
0.25 & 0.77876 & 0.78319 & 1.3762 & 1.3678\tabularnewline
\hline 
0.3 & 0.72153 & 0.7245 & 1.3945 & 1.3883\tabularnewline
\hline 
0.35 & 0.6609 & 0.66223 & 1.4071 & 1.4029\tabularnewline
\hline 
0.4 & 0.59748 & 0.59718 & 1.415 & 1.4126\tabularnewline
\hline 
0.45 & 0.53201 & 0.53023 & 1.4185 & 1.4177\tabularnewline
\hline 
0.5 & 0.46535 & 0.46236 & 1.418 & 1.4187\tabularnewline
\hline 
0.55 & 0.39847 & 0.3946 & 1.4135 & 1.4155\tabularnewline
\hline 
0.6 & 0.33243 & 0.32807 & 1.4048 & 1.4079\tabularnewline
\hline 
0.65 & 0.26842 & 0.26396 & 1.3915 & 1.3957\tabularnewline
\hline 
0.7 & 0.20775 & 0.20356 & 1.373 & 1.378\tabularnewline
\hline 
0.75 & 0.15181 & 0.14821 & 1.3478 & 1.3536\tabularnewline
\hline 
0.8 & 0.10213 & 0.09934 & 1.3141 & 1.3206\tabularnewline
\hline 
0.85 & 0.060326 & 0.05846 & 1.268 & 1.2749\tabularnewline
\hline 
0.9 & 0.028125 & 0.027156 & 1.2009 & 1.208\tabularnewline
\hline 
0.95 & 0.0073687 & 0.0070887 & 1.0876 & 1.0946\tabularnewline
\hline 
0.973 & 0.0021948 & 0.0021079 & 0.99261 & 0.99921\tabularnewline
\hline 
0.99 & 0.0003058 & 0.00029333 & 0.85404 & 0.85984\tabularnewline
\hline 
0.996 & 4.9197E-05 & 4.717E-05 & 0.74238 & 0.74746\tabularnewline
\hline 
0.998 & 1.2322E-05 & 1.1812E-05 & 0.66748 & 0.67206\tabularnewline
\hline 
0.999 & 3.0832E-06 & 2.9555E-06 & 0.60005 & 0.60417\tabularnewline
\hline 
0.9995 & 7.7115E-07 & 7.3919E-07 & 0.53939 & 0.5431\tabularnewline
\hline 
0.9999 & 3.0857E-08 & 2.9578E-08 & 0.42111 & 0.42401\tabularnewline
\hline 
0.99999 & 3.086E-10 & 2.958E-10 & 0.2955 & 0.29753\tabularnewline
\hline 
0.999999 & 3.086E-12 & 2.958E-12 & 0.20735 & 0.20878\tabularnewline
\hline 
\end{tabular}\caption{The exact and fitted similarity temperature profiles as a function
of $\xi/\xi_{0}$, for tests 3 and 5. The exact profiles result from
the numerical solution of the ODE \eqref{eq:ode}. The fitted profile
is given by Eq. \eqref{eq:f_hr}, using exact front coordinate $\xi_{0}$
and the resulting fitted value $R$, which are also given in the table.\label{tab:tests_f}}
\end{table}

In this case with define a non-Henyey wave with a constant surface
temperature, a negligible radiation energy and a material model which
results in a flat parabolic front. We take the same opacity as in
test 1: $\alpha=1.5$, $k_{0}=10^{4}\text{HeV}^{1.5}/\text{cm}$:
\begin{equation}
k\left(T\right)=10^{4}\left(\frac{T}{\text{HeV}}\right)^{-1.5}\ \text{cm}^{-1},
\end{equation}
while for the total energy density we take $\beta=5$, $u_{0}=10^{13}\text{HeV}^{-5}$:
\begin{equation}
u\left(T\right)=10^{13}\left(\frac{T}{\text{HeV}}\right)^{5}\ \frac{\text{erg}}{\text{cm}^{3}}.
\end{equation}
According to Eq. \eqref{eq:nudef} this material model results in
a parabolic front, as $\nu=2$. We take a constant surface temperature,
$\tau=0$: 
\begin{equation}
T_{s}\left(t\right)=1\ \text{HeV},\label{eq:Tbound_3}
\end{equation}
for which, as in test 1, the radiation energy density can be neglected,
so that for non-LTE simulations (gray diffusion and transport), we
can take: 
\begin{equation}
u_{m}\left(T\right)=u\left(T\right)-aT^{4}\approx10^{13}\left(\frac{T}{\text{HeV}}\right)^{5}\ \frac{\text{erg}}{\text{cm}^{3}}.
\end{equation}
Unlike the previous cases, for this value of $\tau$, there is no
analytical solution (see Tab. \ref{tab:origin_behaviou}), and the
ODE \eqref{eq:ode} must be solved numerically, as detailed in Sec.
\ref{subsec:Numerical-solution}. The numerical solution yields the
front coordinate $\xi_{0}=6.2167035$, and via Eq. \eqref{eq:xheat}
we find the heat front position: 
\begin{equation}
x_{F}\left(t\right)=0.00620780785\left(\frac{t}{\text{ns}}\right)^{\frac{1}{2}}\ \text{cm}.
\end{equation}
The numerical similarity profile is tabulated in Tab. \ref{tab:tests_f}.
Using the self-similar solution \eqref{eq:Tss}, the temperature profile
is 
\begin{equation}
T\left(x,t\right)=f\left(\xi_{0}x/x_{F}\left(t\right)\right)\ \text{HeV}.\label{eq:Tprof_anal_3}
\end{equation}
which, as shown in Fig. \ref{fig:simulation_3}, has a flat front.
Alternatively, instead of using the exact tabulated solution, one
can use the fitted closed form solution, which is discussed in Sec.
\ref{subsec:The-Hammer-Rosen-approximate}. A numerical fit of the
exact numerical solution to the from in Eq. \eqref{eq:f_hr}, gives
the parameter $R=0.7198876$, with a maximal error of 4\% and an average
error of 2\%. The fitted solution is compared to the exact solution
in Tab. \ref{tab:tests_f}. By using Eqs. \eqref{eq:f_hr} and \eqref{eq:Tss},
we obtain the approximate fitted temperature profile:

{\small{}
\begin{equation}
T\left(x,t\right)=\left[\left(1-\frac{x}{x_{F}\left(t\right)}\right)\left(1+R\frac{x}{x_{F}\left(t\right)}\right)\right]^{2}\ \text{HeV},
\end{equation}
}which evidently defines a heat wave profile with a flat parabolic
front. The exact numerical solution has $f'\left(0\right)=-0.10447545$,
and using Eqs. \eqref{eq:Tbath}-\eqref{eq:Bbath}, we find the bath
temperature:{\small{}
\begin{equation}
T_{\text{bath}}\left(t\right)=\left(1+0.0279\left(\frac{t}{\text{ns}}\right)^{-\frac{1}{2}}\right)^{\frac{1}{4}}\ \text{HeV},\label{eq:Tbath_3}
\end{equation}
}which is time dependent, while the surface temperature Eq. \ref{eq:Tbound_3}
is constant, as also shown in Fig. \ref{fig:simulation_3_Tbath}.
The energy integral {[}Eq. \eqref{eq:Iint}{]} of the exact numerical
profile is $\mathcal{E}=1.14923$ (the fitted profile has $\mathcal{E}=1.1739$),
which gives the total energy as a function of time {[}Eq. \eqref{eq:Etotss}{]}:
\begin{equation}
E\left(t\right)=11.47585\times10^{9}\left(\frac{t}{\text{ns}}\right)^{\frac{1}{2}}\ \text{erg}.
\end{equation}
According to Eq. \eqref{eq:depth_tot}, the optical depth at the final
time is {\small{}
\[
\mathcal{T}\gtrsim x_{F}\left(t_{\text{final}}\right)k\left(T_{s}\left(t_{\text{final}}\right)\right)=0.0062078\times10^{4}\approx62.1,
\]
}so the heat wave is optically thick and we expect the LTE diffusion
limit to hold. The results are presented in Fig. \ref{fig:simulation_3},
showing a good agreement between the analytic solution and numerical
simulations.

\subsubsection*{TEST 4}

\begin{figure}[t]
\begin{centering}
\includegraphics[scale=0.55]{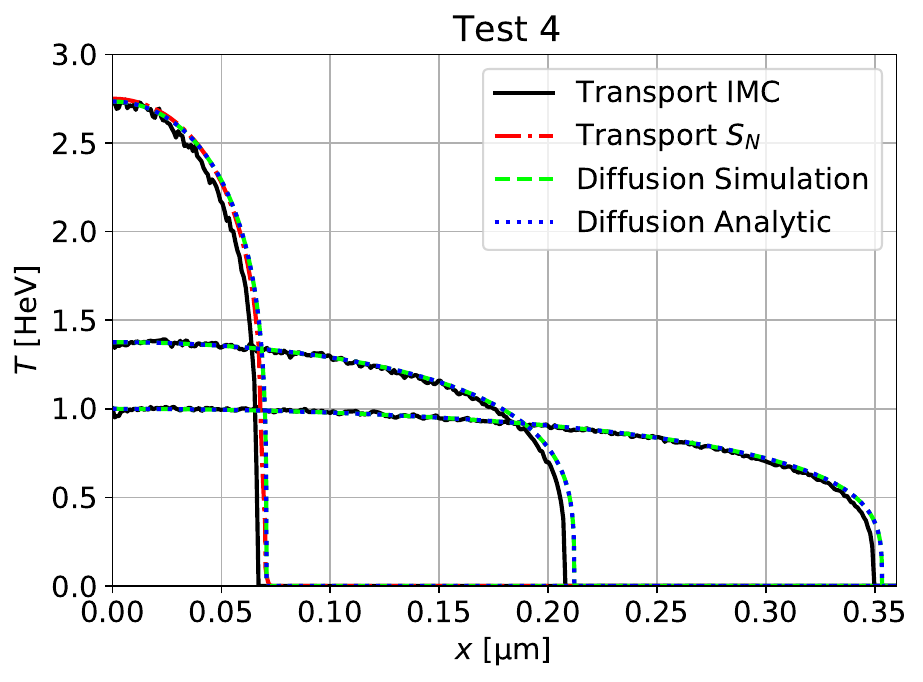} 
\par\end{centering}
\caption{Temperature profiles for test 4. Results are shown at times $t=7.9126254\times10^{-4}\text{ns},\ 0.10364448\text{ns}$
and $1$ns, as obtained from a gray diffusion simulation and from
Implicit-Monte-Carlo (IMC) and discrete ordinates ($S_{N}$) transport
simulations, and are compared to the analytic solution of the diffusion
equation {[}Eq. \eqref{eq:Tprof_anal_4}{]}.\label{fig:simulation_4}}
\end{figure}

In this case define a Zel'dovich-Barenblatt wave (see Sec. \ref{subsec:The-Zel'dovich-Barenblatt-analyt}),
which has a decreasing surface temperature and an analytical solution.
We take $\alpha=1.5$, $\beta=1.6$ with coefficients $k_{0}=10^{8}\text{HeV}^{1.5}/\text{cm}$,
$u_{0}=10^{13}\text{HeV}^{-1.6}\text{erg/cm}^{3}$ so that the opacity
is: 
\begin{equation}
k\left(T\right)=10^{8}\left(\frac{T}{\text{HeV}}\right)^{-1.5}\ \text{cm}^{-1},\label{eq:kt_zeld}
\end{equation}
and the total energy density is: 
\begin{equation}
u\left(T\right)=10^{13}\left(\frac{T}{\text{HeV}}\right)^{1.6}\ \frac{\text{erg}}{\text{cm}^{3}}.
\end{equation}
We define a Zel'dovich-Barenblatt wave for this material model, that
is, $\tau=\tau_{Z}=-\frac{10}{71}$. Since the flux is zero at the
origin for a Zel'dovich-Barenblatt wave, the surface and bath temperatures
are equal {[}see Eqs. \eqref{eq:Tbath}-\eqref{eq:Bbath}{]}: 
\begin{equation}
T_{s}\left(t\right)=T_{\text{bath}}\left(t\right)=\left(\frac{t}{\text{ns}}\right)^{-\frac{10}{71}}\ \text{HeV}.\label{eq:Tbound_4}
\end{equation}
The opacity in Eq. \eqref{eq:kt_zeld} is much higher than in the
previous cases, in order to insure that even at short times when the
temperature is high, the matter and radiation are approximately at
equilibrium. In addition, the radiation energy density can be neglected
after a short period, so that for non-LTE simulations, we take: 
\begin{equation}
u_{m}\left(T\right)=u\left(T\right)-aT^{4}\approx10^{13}\left(\frac{T}{\text{HeV}}\right)^{1.6}\ \frac{\text{erg}}{\text{cm}^{3}}.
\end{equation}
Eq. \eqref{eq:xsi_0_Z} gives front coordinate $\xi_{0}=3.5377995$,
which via Eq. \eqref{eq:xheat} gives the heat front position: 
\begin{equation}
x_{F}\left(t\right)=3.532735\times10^{-5}\left(\frac{t}{\text{ns}}\right)^{\frac{16}{71}}\ \text{cm}.
\end{equation}
The temperature profile is given analytically by using Eqs. \eqref{eq:f_z}
and \eqref{eq:Tss}:

\begin{equation}
T\left(x,t\right)=\left(\frac{t}{\text{ns}}\right)^{-\frac{10}{71}}\left(1-\left(\frac{x}{x_{F}\left(t\right)}\right)^{2}\right)^{\frac{10}{39}}\ \text{HeV}\label{eq:Tprof_anal_4}
\end{equation}
which is a familiar wave for the instantaneous point source problem
with a steep heat front and decreasing surface temperature, as shown
in Fig. \ref{fig:simulation_4}. Using Eq. \eqref{eq:Ezld}, the total
energy is:
\begin{equation}
E=2.8759901\times10^{8}\text{erg}.
\end{equation}
As discussed in Sec. \ref{subsec:The-total-energy}, since the total
energy is time-independent for a Zel'dovich-Barenblatt wave, instead
of using the surface/bath temperature boundary condition \eqref{eq:Tbound_4},
this test can also be run equivalently by depositing this amount of
energy near the origin at $t=0$, and using a reflective boundary
condition at $x=0$. Alternatively, it can be run with twice this
amount of energy deposited at $x=0$, allowing the heat wave to propagate
in both positive and negative directions. We note that in both simulation
modes for this test (surface/bath temperature or energy deposition),
the total energy in the system should remain constant in time.

According to Eq. \eqref{eq:depth_tot}, the optical depth at the final
time is {\small{}
\begin{align*}
\mathcal{T} & \gtrsim x_{F}\left(t_{\text{final}}\right)k\left(T_{s}\left(t_{\text{final}}\right)\right)\\
 & =3.532735\times10^{-5}\times10^{8}\approx3532,
\end{align*}
}so the heat wave is extremely optically thick and we expect the LTE
diffusion limit to hold. The results are presented in Fig. \ref{fig:simulation_4},
showing a good agreement between the analytic solution and numerical
simulations.

\subsubsection*{TEST 5}

\begin{figure}[t]
\begin{centering}
\includegraphics[scale=0.55]{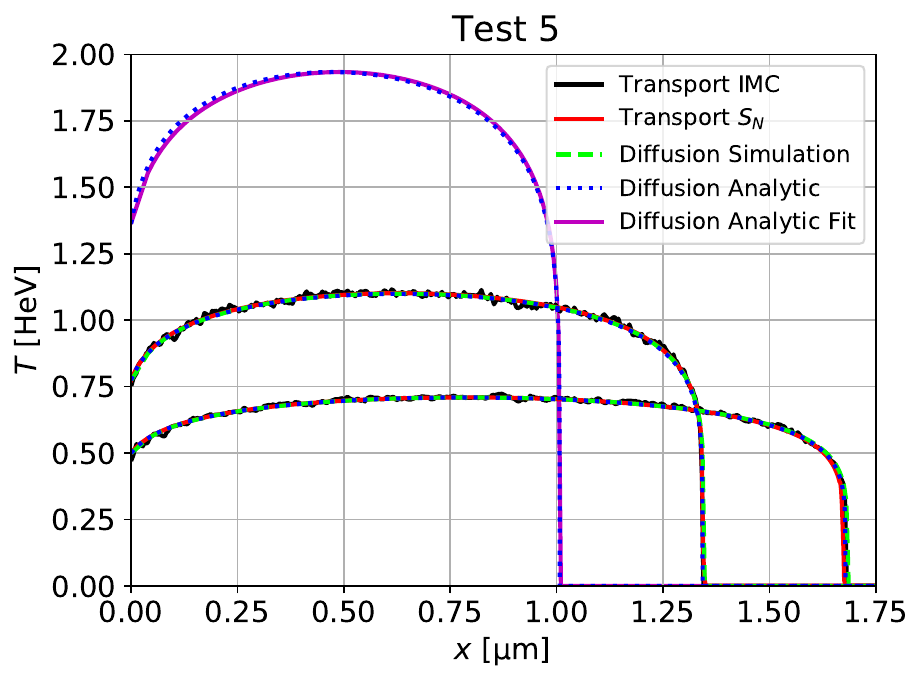} 
\par\end{centering}
\caption{Temperature profiles for test 5. The simulations are initialized with
the analytic temperature profiles {[}Eq. \eqref{eq:Tprof_anal_5}{]}
at the time $t=5.3147117\times10^{-4}\text{ns}$, where the heat front
has reached 60\% of its final distance. The approximated fitted analytic
profile {[}see Eq. \eqref{eq:Tprof_anal_fit_5}{]} is shown for comparison
at the initial time as well. The simulations results are compared
with the exact analytic solution at time $t=0.037118798$ns at which
the front has reached 80\% of the final distance, and at the final
time, $t=1$ns (simulation results at the initialization time are
not shown since they are initialized with the analytic solution).\label{fig:simulation_5}}
\end{figure}

\begin{figure}[t]
\begin{centering}
\includegraphics[scale=0.55]{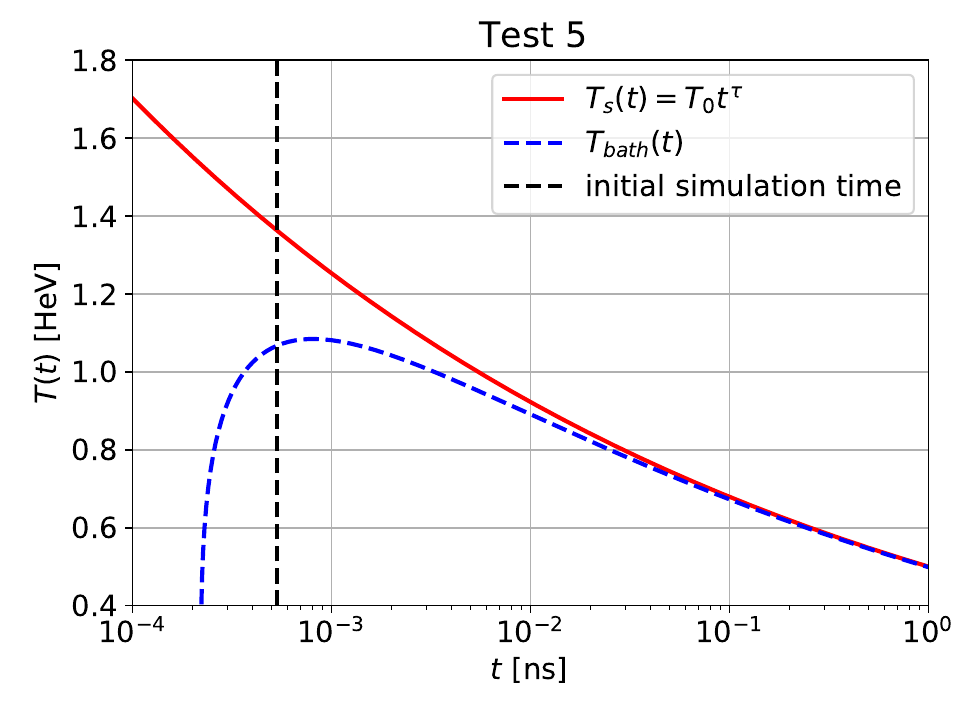}
\par\end{centering}
\caption{A comparison between the surface (red line, Eq. \eqref{eq:Tbound_5})
and bath (blue line, Eq. \eqref{eq:Tbath_5}) driving temperatures,
for test 5. The initial simulation time is also shown.\label{fig:simulation_5_Tbath}}
\end{figure}
\begin{figure}[t]
\begin{centering}
\includegraphics[scale=0.55]{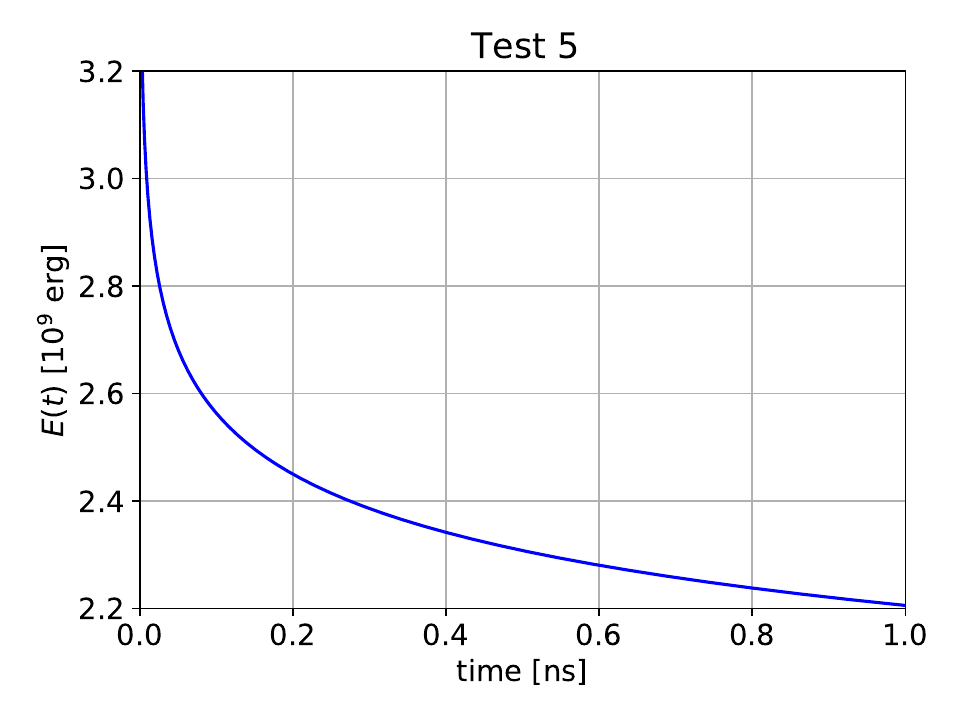} 
\par\end{centering}
\caption{The total energy in the system as a function of time, for test 5.\label{fig:simulation_E_t}}
\end{figure}

We define a wave with $\tau<\tau_{Z}$, for which the surface temperature
decrease faster than the heat wave propagation, and as discussed in
Sec. \ref{subsec:The-total-energy}, results in a local maxima in
the temperature profile and a total energy which decreases over time.
We take $\alpha=3.5$ and $\beta=1$, with the coefficients $k_{0}=8.838835\times10^{5}\text{HeV}^{3.5}/\text{cm}$
and $u_{0}=2\times10^{13}\text{HeV}^{-1}\text{erg/cm}^{3}$ so that
the opacity is: 
\begin{equation}
k\left(T\right)=10^{7}\left(\frac{T}{\frac{1}{2}\text{HeV}}\right)^{-3.5}\ \text{cm}^{-1},\label{eq:opacity_4}
\end{equation}
and the total energy density is: 
\begin{equation}
u\left(T\right)=10^{13}\left(\frac{T}{\frac{1}{2}\text{HeV}}\right)\ \frac{\text{erg}}{\text{cm}^{3}}.
\end{equation}
We take $\tau=-0.133=0.9975\ensuremath{\tau_{c}}<\ensuremath{\tau_{Z}}$
(in this case, $\tau_{c}=-\frac{2}{15}=-0.1333...$ and $\tau_{Z}=-\frac{2}{17}$)
and the surface temperature 
\begin{equation}
T_{s}\left(t\right)=\frac{1}{2}\left(\frac{t}{\text{ns}}\right)^{-0.133}\ \text{HeV}.\label{eq:Tbound_5}
\end{equation}
As in test 4, the opacity in Eq. \eqref{eq:opacity_4} is high enough
to insure that even at short times when the temperature is high, the
matter and radiation are approximately at equilibrium. The radiation
energy density can be neglected after a short period, so that for
non-LTE simulations, we take: 
\begin{equation}
u_{m}\left(T\right)=u\left(T\right)-aT^{4}\approx10^{13}\left(\frac{T}{\text{HeV}}\right)^{1.6}\ \frac{\text{erg}}{\text{cm}^{3}}.
\end{equation}
As in test 3, for this value of $\tau$ there is no analytical solution,
and the ODE \eqref{eq:ode} must be solved numerically. The numerical
integration yields the front coordinate $\xi_{0}=24.826729$, which
gives the heat front position {[}Eq. \eqref{eq:xheat}{]}: 
\begin{equation}
x_{F}\left(t\right)=1.678372\times10^{-4}\left(\frac{t}{\text{ns}}\right)^{0.06775}\ \text{cm}.
\end{equation}
The numerical similarity profile is tabulated in Tab. \ref{tab:tests_f}.
Using the self-similar solution \eqref{eq:Tss}, the temperature profile
is given by: 
\begin{equation}
T\left(x,t\right)=\frac{1}{2}\left(\frac{t}{\text{ns}}\right)^{-0.133}f\left(\xi_{0}x/x_{F}\left(t\right)\right)\ \text{HeV},\label{eq:Tprof_anal_5}
\end{equation}
which as shown in Fig. \ref{fig:simulation_4}, has a local maxima.
As in test 3, instead of using the exact tabulated solution, one can
use the fitted closed form solution. A numerical fit of the exact
numerical solution gives the parameter $R=36.84138$, with a maximal
error of 1.5\% and an average error of 0.5\%. The fitted solution
is compared to the exact solution in Tab. \ref{tab:tests_f}. By using
Eqs. \eqref{eq:f_hr} and \eqref{eq:Tss}, we obtain the approximate
temperature profile:

{\footnotesize{}
\begin{equation}
T\left(x,t\right)=\frac{1}{2}\left(\frac{t}{\text{ns}}\right)^{-0.133}\left[\left(1-\frac{x}{x_{F}\left(t\right)}\right)\left(1+R\frac{x}{x_{F}\left(t\right)}\right)\right]^{\frac{2}{13}}\ \text{HeV},\label{eq:Tprof_anal_fit_5}
\end{equation}
}which is compared to the exact solution in Fig. \ref{fig:simulation_5}. 

The exact numerical solution has $f'\left(0\right)=0.2838745$, and
by using Eqs. \eqref{eq:Tbath}-\eqref{eq:Bbath}, we find the bath
temperature: {\small{}
\begin{equation}
T_{\text{bath}}\left(t\right)=\left(1-0.011198\left(\frac{t}{\text{ns}}\right)^{-0.53325}\right)^{\frac{1}{4}}\frac{1}{2}\left(\frac{t}{\text{ns}}\right)^{-0.133}\ \text{HeV}.\label{eq:Tbath_5}
\end{equation}
}Since the bath constant is negative (as energy flows out of the system),
the bath temperature is undefined at short enough times (negative
incoming flux), as depicted in Fig. \ref{fig:simulation_5_Tbath}.
In order to overcome this issue, which only applies for diffusion
simulations that use the Marshak boundary condition or for transport
simulations (and is irrelevant for diffusion simulations that use
the surface temperature boundary condition), the simulations for this
test are initialized with the analytic profile at the finite time
$t=5.3147117\times10^{-4}\text{ns}$, for which the heat front has
reached 60\% of its final distance (as shown in Fig. \ref{fig:simulation_5}),
and for which the bath temperature Eq. \eqref{eq:Tbath_5} is well
defined. As evident from Fig. \ref{fig:simulation_5_Tbath}, after
a short transient, the bath and surface temperatures become very close,
as this problem as very opaque. In contrast to all previous cases,
here the bath temperature is always lower than the surface temperature,
since $\tau<\tau_{Z}$ (as discussed in Sec. \ref{subsec:Marshak-boundary-condition}).
We also note that due to the relatively high accuracy of the approximate
fitted profile {[}Eq. \eqref{eq:Tprof_anal_fit_5}{]}, the simulations
can equivalently be initialized using it, instead of the exact tabulated
profile.

The energy integral {[}Eq. \eqref{eq:Iint}{]} of the exact profile
is $\mathcal{E}=32.6293$ (the fitted profile has $\mathcal{E}=32.5856$),
which give the total energy as a function of time {[}Eq. \eqref{eq:Etotss}{]}:
\begin{equation}
E\left(t\right)=2.205854\times10^{9}\left(\frac{t}{\text{ns}}\right)^{-0.06525}\ \text{erg},
\end{equation}
which is a decreasing function of time, which we plot in Fig. \ref{fig:simulation_E_t}.
According to Eq. \eqref{eq:depth_tot}, the optical depth at the final
time is {\footnotesize{}
\begin{align*}
\mathcal{T} & \gtrsim x_{F}\left(t_{\text{final}}\right)k\left(T_{s}\left(t_{\text{final}}\right)\right)\\
 & =1.678372\times10^{-4}\times8.838835\times10^{5}\times2^{3.5}\\
 & \approx1678,
\end{align*}
}so the heat wave is extremely optically thick and we expect the LTE
diffusion limit to hold. The results are presented in Fig. \ref{fig:simulation_5},
showing a good agreement between the analytic solution and numerical
simulations.

\subsubsection*{TEST 6}

\begin{figure}[t]
\begin{centering}
\includegraphics[scale=0.55]{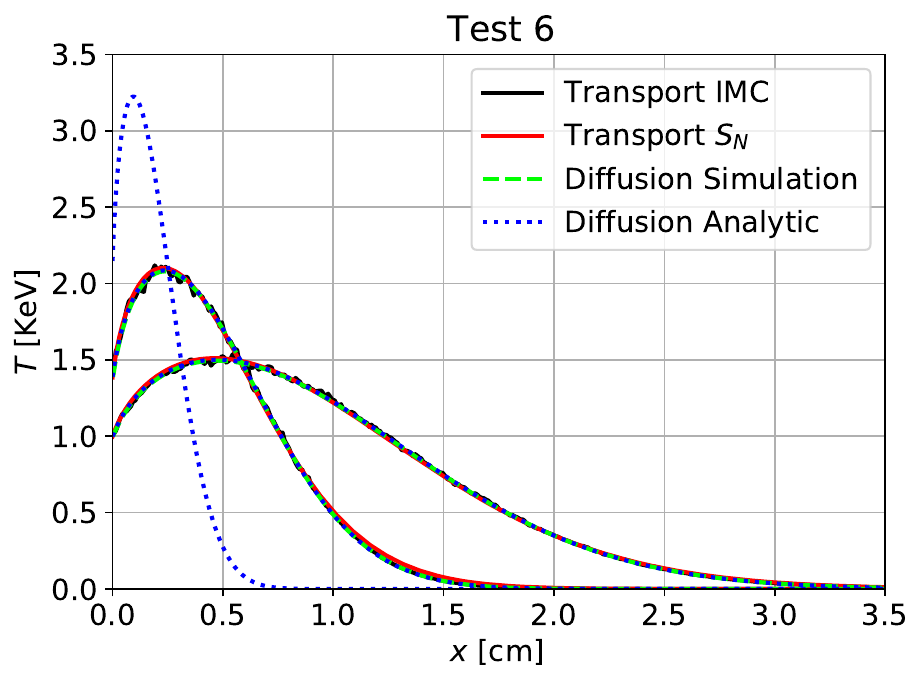} 
\par\end{centering}
\caption{Temperature profiles for test 6. The simulations are initialized with
the analytic temperature profiles (Eq. \eqref{eq:Tprof_anal_6}) at
time $t=0.04\text{ns}$. The simulations results are compared with
the exact analytic solution at time $t=0.25$ns and at the final time,
$t=1$ns (simulation results at the initialization time are not shown
since they are initialized with the analytic solution). \label{fig:simulation_6}}
\end{figure}
\begin{figure}[t]
\begin{centering}
\includegraphics[scale=0.55]{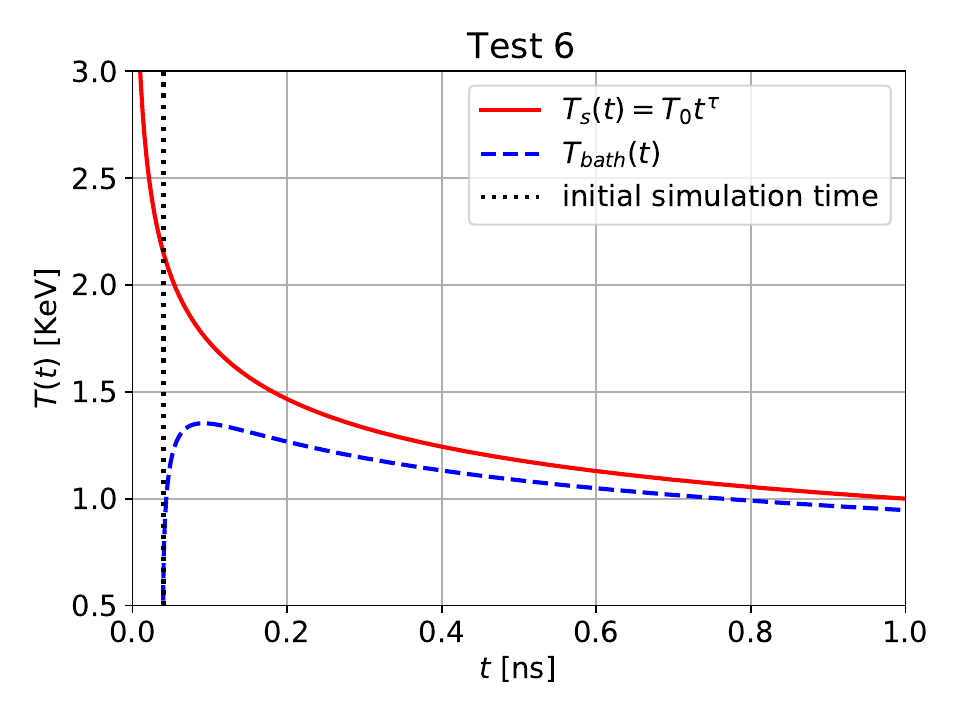}
\par\end{centering}
\caption{A comparison between the surface (red line, Eq. \eqref{eq:Tbound_6})
and bath (blue line, Eq. \eqref{eq:Tbath_6}) driving temperatures,
for test 6. The initial simulation time is also shown.\label{fig:simulation_6_Tbath}}
\end{figure}

As in test 5, we define a wave with $\tau<\tau_{Z}$, but unlike all
previous cases, here we define a material model which results in linear
heat conduction. To that end, we take a constant opacity $\alpha=0$
and $\beta=4$. We take the dimensional coefficients $k_{0}=50\text{cm}^{-1}$
and $u_{0}=\frac{3}{2}a$, so that the opacity is: 
\begin{equation}
k\left(T\right)=50\ \text{cm}^{-1},
\end{equation}
and the total energy density reads: 
\begin{equation}
u\left(T\right)=\frac{3}{2}aT^{4}=\frac{3}{2}\times1.372017\times10^{14}\left(\frac{T}{\text{keV}}\right)^{4}\ \frac{\text{erg}}{\text{cm}^{3}}.
\end{equation}
For non-LTE simulations, we have the material energy density: 
\begin{equation}
u_{m}\left(T\right)=u\left(T\right)-aT^{4}=\frac{1}{2}aT^{4}=\frac{1}{3}u\left(T\right),
\end{equation}
which is always one half of the radiation energy density. We take
$\tau=-0.2375=0.95\ensuremath{\tau_{c}}<\ensuremath{\tau_{Z}}$ (in
this case, $\tau_{c}=-\frac{1}{4}$ and $\tau_{Z}=-\frac{1}{8}$)
and the surface temperature:
\begin{equation}
T_{s}\left(t\right)=\left(\frac{t}{\text{ns}}\right)^{-0.2375}\ \text{KeV}.\label{eq:Tbound_6}
\end{equation}
As the conduction in this case is linear, the heat wave does not have
a well defined front. The heat propagates in space via Eq. \eqref{eq:xsi_def}:
\begin{equation}
\frac{x}{\xi}=0.365022\left(\frac{t}{\text{ns}}\right)^{\frac{1}{2}}\ \text{cm}.\label{eq:xsi_test6}
\end{equation}
The exact analytical linear conduction similarity profile {[}Eq. \eqref{eq:f_lin_sol}{]}
was shown for this case in Fig. \ref{fig:profiles_shapes}. The dimensional
temperature profile is obtained using Eq. \eqref{eq:Tss}:

{\small{}
\begin{equation}
T\left(x,t\right)=\left(\frac{t}{\text{ns}}\right)^{-0.2375}\left[\frac{\Gamma\left(\frac{1}{20}\right)}{\sqrt{\pi}}e^{-\xi^{2}/4}U\left(-\frac{9}{20},\frac{1}{2},\frac{\xi^{2}}{4}\right)\right]^{\frac{1}{4}}\ \text{KeV},\label{eq:Tprof_anal_6}
\end{equation}
}where $\xi=\xi\left(x,t\right)$ is obtained from Eq. \eqref{eq:xsi_test6}.
The profiles are shown in Fig. \ref{fig:simulation_4}, and the existence
of a local maxima is evident. From Eq. \eqref{eq:f'0_linear} we find
$f'\left(0\right)=-\frac{1}{4}\Gamma\left(\frac{1}{20}\right)/\Gamma\left(-\frac{9}{20}\right)\approx1.355332$,
which is positive as energy is leaving the system, and from Eqs. \eqref{eq:Tbath}-\eqref{eq:Bbath},
we find the bath temperature:{\small{} 
\begin{equation}
T_{\text{bath}}\left(t\right)=\left(1-0.19802739\left(\frac{t}{\text{ns}}\right)^{-0.5}\right)^{\frac{1}{4}}\left(\frac{t}{\text{ns}}\right)^{-0.2375}\ \text{KeV},\label{eq:Tbath_6}
\end{equation}
}which, as discussed in test 5, is undefined at short times, and is
always lower than the surface temperature. This is depicted in Fig.
\ref{fig:simulation_6_Tbath}. As for test 5, this issue is overcome
by initializing the simulations with the analytic profile at a finite
time. We take the initialization time $t=0.04\text{ns}$ (as shown
in Fig. \ref{fig:simulation_6}), for which the bath temperature Eq.
\eqref{eq:Tbath_6} is well defined. As opposed to test 5 which is
extremely opaque, here, as evident from Fig. \ref{fig:simulation_6_Tbath},
there is a non-negligible (\textasciitilde 5\%) difference between
the bath and surface temperatures. 

From Eq. \eqref{eq:Eint_linear_anal} we find the energy integral
$\mathcal{E}=\Gamma\left(\frac{1}{20}\right)/\Gamma\left(\frac{11}{20}\right)=12.047394$,
using Eq. \eqref{eq:Etotss} we find the total energy in the system
as a function of time: 
\begin{equation}
E\left(t\right)=9.050299\times10^{14}\left(\frac{t}{\text{ns}}\right)^{-0.45}\ \text{erg}.
\end{equation}
According to Eq. \eqref{eq:depth_tot}, the optical depth at the final
time is {\small{}
\begin{align*}
\mathcal{T} & \gtrsim0.365022\sqrt{20}\times50\approx81.6,
\end{align*}
}so the heat wave is optically thick and we expect the LTE diffusion
limit to hold. The results are presented in Fig. \ref{fig:simulation_6},
showing a good agreement between the analytic solution and numerical
simulations. .

\section{Conclusion}

In this work we have studied in detail the self-similar solutions
of the supersonic LTE Marshak wave problem. We explored the behavior
and characteristics of the solutions in both linear and nonlinear
conduction regimes as a function of the surface temperature drive
exponent $\tau$, the opacity exponent $\alpha$ and energy density
exponent $\beta$, to present a unified theory of these Marshak wave solutions. It was shown that there exists a range of $\tau$
for which the solutions have a local maxima and does not have the
familiar monotonically decreasing character. For those solutions,
the total energy is decreasing in time, due to the very rapid decrease
of the surface temperature. In addition, the values of $\tau$ for
which closed form analytical solutions exist where specified as classical
solutions found by previous authors and were identified as special
cases. The behavior of the solution as a function of $\tau$ is summarized
in Tab. \ref{tab:origin_behaviou}. For nonlinear conduction, we mapped
the values of $\alpha$ and $\beta$ for which the heat front does
not have the familiar sharp character, and demonstrated that the front
can be linear or even flat. The character of the heat front as a function
of $\alpha$ and $\beta$ is summarized in Tab. \ref{tab:front_behaviour}
and Fig. \ref{fig:mesh_ny}. 

We performed a detailed study of the accuracy of the widely used Hammer-Rosen
perturbation theory \cite{hammer2003consistent} and for the series
expansion method of Smith  \cite{smith2010solutions}, by comparing
their results to the exact self-similar Marshak wave solutions in
a wide range of the exponents $\tau,\alpha,\beta$. It is shown that
when the surface temperature does not decrease too fast ($\tau\gtrsim\tau_{Z}$),
the approximation of Smith is more accurate. 
Therefore, we believe that the method of Smith should also be used in the analysis of Marshak wave experiments, which typically employs the Hammer-Rosen method for the general (non power-law) temperature drive (see i.e. \cite{cohen2020key}), as it might result in more accurate predictions.

By using the Hammer-Rosen
and Smith profile, we parameterized the solution by a numerical fit
to the exact solution, which resulted in a very accurate analytical
approximation to the exact numerical solution, in a wide range of
$\tau$, which is given in terms of just two numerical parameters. 
We used the exact and approximate solutions to construct
a set of benchmarks for supersonic LTE radiative heat transfer, including
some which posses the unusual and interesting properties which were
demonstrated, such as local maxima and non sharp fronts. We compared
the solutions to implicit Monte-Carlo and discrete-ordinate transport
simulations as well gray diffusion simulations, showing a good agreement.
This demonstrates the usefulness of these LTE Marshak wave benchmarks,
as a set of simple but non-trivial verification test problems for
radiative transfer simulations. 

\subsection*{Availability of data}

The data that support the findings of this study are available from
the corresponding author upon reasonable request.

\bibliographystyle{unsrt}
\bibliography{datab}

\begin{thebibliography}{10}

\bibitem{lindl2004physics}
John~D Lindl, Peter Amendt, Richard~L Berger, S~Gail Glendinning, Siegfried~H Glenzer, Steven~W Haan, Robert~L Kauffman, Otto~L Landen, and Laurence~J Suter.
\newblock The physics basis for ignition using indirect-drive targets on the national ignition facility.
\newblock {\em Physics of plasmas}, 11(2):339--491, 2004.

\bibitem{robey2001experimental}
Harry~F Robey, JO~Kane, BA~Remington, RP~Drake, OA~Hurricane, H~Louis, RJ~Wallace, J~Knauer, P~Keiter, D~Arnett, et~al.
\newblock An experimental testbed for the study of hydrodynamic issues in supernovae.
\newblock {\em Physics of Plasmas}, 8(5):2446--2453, 2001.

\bibitem{bailey2015higher}
James~E Bailey, Taisuke Nagayama, Guillaume~Pascal Loisel, Gregory~Alan Rochau, C~Blancard, James Colgan, Ph~Cosse, G~Faussurier, CJ~Fontes, F~Gilleron, et~al.
\newblock A higher-than-predicted measurement of iron opacity at solar interior temperatures.
\newblock {\em Nature}, 517(7532):56--59, 2015.

\bibitem{falize2011similarity}
E~Falize, Claire Michaut, and Serge Bouquet.
\newblock Similarity properties and scaling laws of radiation hydrodynamic flows in laboratory astrophysics.
\newblock {\em The Astrophysical Journal}, 730(2):96, 2011.

\bibitem{hurricane2014fuel}
OA~Hurricane, DA~Callahan, DT~Casey, PM~Celliers, Charles Cerjan, EL~Dewald, TR~Dittrich, T~D{\"o}ppner, DE~Hinkel, LF~Berzak Hopkins, et~al.
\newblock Fuel gain exceeding unity in an inertially confined fusion implosion.
\newblock {\em Nature}, 506(7488):343--348, 2014.

\bibitem{cohen2020key}
Avner~P Cohen, Guy Malamud, and Shay~I Heizler.
\newblock Key to understanding supersonic radiative marshak waves using simple models and advanced simulations.
\newblock {\em Physical Review Research}, 2(2):023007, 2020.

\bibitem{heizler2021radiation}
Shay~I Heizler, Tomer Shussman, and Moshe Fraenkel.
\newblock Radiation drive temperature measurements in aluminum via radiation-driven shock waves: Modeling using self-similar solutions.
\newblock {\em Physics of Plasmas}, 28(3):032105, 2021.

\bibitem{back2000diffusive}
CA~Back, JD~Bauer, JH~Hammer, BF~Lasinski, RE~Turner, PW~Rambo, OL~Landen, LJ~Suter, MD~Rosen, and WW~Hsing.
\newblock Diffusive, supersonic x-ray transport in radiatively heated foam cylinders.
\newblock {\em Physics of plasmas}, 7(5):2126--2134, 2000.

\bibitem{sigel1988x}
R~Sigel, R~Pakula, S~Sakabe, and GD~Tsakiris.
\newblock X-ray generation in a cavity heated by 1.3-or 0.44-$\mu$m laser light. iii. comparison of the experimental results with theoretical predictions for x-ray confinement.
\newblock {\em Physical Review A}, 38(11):5779, 1988.

\bibitem{lindl1995development}
John Lindl.
\newblock Development of the indirect-drive approach to inertial confinement fusion and the target physics basis for ignition and gain.
\newblock {\em Physics of plasmas}, 2(11):3933--4024, 1995.

\bibitem{calder2002validating}
Alan~C Calder, Bruce Fryxell, T~Plewa, Robert Rosner, LJ~Dursi, VG~Weirs, T~Dupont, HF~Robey, JO~Kane, BA~Remington, et~al.
\newblock On validating an astrophysical simulation code.
\newblock {\em The Astrophysical Journal Supplement Series}, 143(1):201, 2002.

\bibitem{krumholz2007equations}
Mark~R Krumholz, Richard~I Klein, Christopher~F McKee, and John Bolstad.
\newblock Equations and algorithms for mixed-frame flux-limited diffusion radiation hydrodynamics.
\newblock {\em The Astrophysical Journal}, 667(1):626, 2007.

\bibitem{gittings2008rage}
Michael Gittings, Robert Weaver, Michael Clover, Thomas Betlach, Nelson Byrne, Robert Coker, Edward Dendy, Robert Hueckstaedt, Kim New, W~Rob Oakes, et~al.
\newblock The rage radiation-hydrodynamic code.
\newblock {\em Computational Science \& Discovery}, 1(1):015005, 2008.

\bibitem{lowrie2007radiative}
Robert~B Lowrie and Rick~M Rauenzahn.
\newblock Radiative shock solutions in the equilibrium diffusion limit.
\newblock {\em Shock waves}, 16(6):445--453, 2007.

\bibitem{mcclarren2011benchmarks}
Ryan~G McClarren and Daniel Holladay.
\newblock Benchmarks for verification of hedp/ife codes.
\newblock {\em Fusion Science and Technology}, 60(2):600--604, 2011.

\bibitem{bennett2021self}
William Bennett and Ryan~G McClarren.
\newblock Self-similar solutions for high-energy density radiative transfer with separate ion and electron temperatures.
\newblock {\em Proceedings of the Royal Society A}, 477(2249):20210119, 2021.

\bibitem{mcclarren2021two}
Ryan~G McClarren.
\newblock Two-group radiative transfer benchmarks for the non-equilibrium diffusion model.
\newblock {\em Journal of Computational and Theoretical Transport}, 50(6-7):583--597, 2021.

\bibitem{mcclarren2011solutions}
Ryan~G McClarren and John~G W{\"o}hlbier.
\newblock Solutions for ion--electron--radiation coupling with radiation and electron diffusion.
\newblock {\em Journal of Quantitative Spectroscopy and Radiative Transfer}, 112(1):119--130, 2011.

\bibitem{mcclarren2008analytic}
Ryan~G McClarren, James~Paul Holloway, and Thomas~A Brunner.
\newblock Analytic p1 solutions for time-dependent, thermal radiative transfer in several geometries.
\newblock {\em Journal of Quantitative Spectroscopy and Radiative Transfer}, 109(3):389--403, 2008.

\bibitem{kamm2008enhanced}
James~R Kamm, Jerry~S Brock, Scott~T Brandon, David~L Cotrell, Bryan Johnson, Patrick Knupp, W~Rider, T~Trucano, and V~Gregory Weirs.
\newblock Enhanced verification test suite for physics simulation codes.
\newblock Technical report, Lawrence Livermore National Lab.(LLNL), Livermore, CA (United States), 2008.

\bibitem{rider2016robust}
William Rider, Walt Witkowski, James~R Kamm, and Tim Wildey.
\newblock Robust verification analysis.
\newblock {\em Journal of Computational Physics}, 307:146--163, 2016.

\bibitem{krief2021analytic}
Menahem Krief.
\newblock Analytic solutions of the nonlinear radiation diffusion equation with an instantaneous point source in non-homogeneous media.
\newblock {\em Physics of Fluids}, 33(5):057105, 2021.

\bibitem{giron2021solutions}
Itamar Giron, Shmuel Balberg, and Menahem Krief.
\newblock Solutions of the imploding shock problem in a medium with varying density.
\newblock {\em Physics of Fluids}, 33(6), 2021.

\bibitem{giron2023solutions}
Itamar Giron, Shmuel Balberg, and Menahem Krief.
\newblock Solutions of the converging and diverging shock problem in a medium with varying density.
\newblock {\em Physics of Fluids}, 35(6), 2023.

\bibitem{krief2023piston}
Menahem Krief.
\newblock Piston driven shock waves in non-homogeneous planar media.
\newblock {\em Physics of Fluids}, 35(4), 2023.

\bibitem{marshak1958effect}
RE~Marshak.
\newblock Effect of radiation on shock wave behavior.
\newblock {\em The Physics of Fluids}, 1(1):24--29, 1958.

\bibitem{petschek1960penetration}
Albert~G Petschek, Ralph~E Williamson, and John~K Wooten~Jr.
\newblock The penetration of radiation with constant driving temperature.
\newblock Technical report, Los Alamos National Lab.(LANL), Los Alamos, NM (United States), 1960.

\bibitem{pert1977class}
GJ~Pert.
\newblock A class of similar solutions of the non-linear diffusion equation.
\newblock {\em Journal of Physics A: Mathematical and General}, 10(4):583, 1977.

\bibitem{pakula1985self}
R~Pakula and R~Sigel.
\newblock Self-similar expansion of dense matter due to heat transfer by nonlinear conduction.
\newblock {\em The Physics of fluids}, 28(1):232--244, 1985.

\bibitem{kaiser1989x}
N~Kaiser, J~Meyer-ter Vehn, and R~Sigel.
\newblock The x-ray-driven heating wave.
\newblock {\em Physics of Fluids B: Plasma Physics}, 1(8):1747--1752, 1989.

\bibitem{hammer2003consistent}
James~H Hammer and Mordecai~D Rosen.
\newblock A consistent approach to solving the radiation diffusion equation.
\newblock {\em Physics of Plasmas}, 10(5):1829--1845, 2003.

\bibitem{garnier2006self}
Josselin Garnier, Guy Malini{\'e}, Yves Saillard, and Catherine Cherfils-Cl{\'e}rouin.
\newblock Self-similar solutions for a nonlinear radiation diffusion equation.
\newblock {\em Physics of plasmas}, 13(9):092703, 2006.

\bibitem{smith2010solutions}
CC~Smith.
\newblock Solutions of the radiation diffusion equation.
\newblock {\em High Energy Density Physics}, 6(1):48--56, 2010.

\bibitem{lane2013new}
Taylor~K Lane and Ryan~G McClarren.
\newblock New self-similar radiation-hydrodynamics solutions in the high-energy density, equilibrium diffusion limit.
\newblock {\em New Journal of Physics}, 15(9):095013, 2013.

\bibitem{shussman2015full}
Tomer Shussman and Shay~I Heizler.
\newblock Full self-similar solutions of the subsonic radiative heat equations.
\newblock {\em Physics of Plasmas}, 22(8):082109, 2015.

\bibitem{cohen2018modeling}
Avner~P Cohen and Shay~I Heizler.
\newblock Modeling of supersonic radiative marshak waves using simple models and advanced simulations.
\newblock {\em Journal of Computational and Theoretical Transport}, 47(4-6):378--399, 2018.

\bibitem{hristov2018heat}
Jordan Hristov.
\newblock The heat radiation diffusion equation: Explicit analytical solutions by improved integral-balance method.
\newblock {\em Thermal science}, 22(2):777--788, 2018.

\bibitem{krief2024self}
Menahem Krief and Ryan~G McClarren.
\newblock Self-similar solutions for the non-equilibrium nonlinear supersonic marshak wave problem.
\newblock {\em Physics of Fluids}, 36(1), 2024.

\bibitem{malka2022supersonic}
Elad Malka and Shay~I Heizler.
\newblock Supersonic--subsonic transition region in radiative heat flow via self-similar solutions.
\newblock {\em Physics of Fluids}, 34(6), 2022.

\bibitem{fujita1952exact}
Hiroshi Fujita.
\newblock The exact pattern of a concentration-dependent diffusion in a semi-infinite medium, part i.
\newblock {\em Textile Research Journal}, 22(11):757--760, 1952.

\bibitem{fujita1952exactII}
Hiroshi Fujita.
\newblock The exact pattern of a concentration-dependent diffusion in a semi-infinite medium, part ii.
\newblock {\em Textile Research Journal}, 22(12):823--827, 1952.

\bibitem{fujita1954exact}
Hiroshi Fujita.
\newblock The exact pattern of a concentration-dependent diffusion in a semi-infinite medium, part iii.
\newblock {\em Textile Research Journal}, 24(3):234--240, 1954.

\bibitem{philip1960general}
JR~Philip.
\newblock General method of exact solution of the concentration-dependent diffusion equation.
\newblock {\em Australian Journal of Physics, vol. 13, p. 1}, 13:1, 1960.

\bibitem{boyer1961some}
RH~Boyer.
\newblock On some solutions of a non-linear diffusion equation.
\newblock {\em Journal of Mathematics and Physics}, 40(1-4):41--45, 1961.

\bibitem{tuck1976some}
Brian Tuck.
\newblock Some explicit solutions to the non-linear diffusion equation.
\newblock {\em Journal of Physics D: Applied Physics}, 9(11):1559, 1976.

\bibitem{brutsaert1982some}
Wilfried Brutsaert.
\newblock Some exact solutions for nonlinear desorptive diffusion.
\newblock {\em Zeitschrift f{\"u}r angewandte Mathematik und Physik}, 33:540--546, 1982.

\bibitem{king1990exact}
JR~King.
\newblock Exact similarity solutions to some nonlinear diffusion equations.
\newblock {\em Journal of Physics A: Mathematical and General}, 23(16):3681, 1990.

\bibitem{heaslet1961diffusion}
Max~A Heaslet and Alberta Alksne.
\newblock Diffusion from a fixed surface with a concentration-dependent coefficient.
\newblock {\em Journal of the Society for Industrial and Applied Mathematics}, 9(4):584--596, 1961.

\bibitem{kass1966numerical}
W~Kass and M~O'keeffe.
\newblock Numerical solution of fick's equation with concentration-dependent diffusion coefficients.
\newblock {\em Journal of Applied Physics}, 37(6):2377--2379, 1966.

\bibitem{castor2004radiation}
John~I Castor.
\newblock {\em Radiation hydrodynamics}.
\newblock 2004.

\bibitem{mihalas1999foundations}
Dimitri Mihalas and Barbara Weibel-Mihalas.
\newblock {\em Foundations of radiation hydrodynamics}.
\newblock Courier Corporation, 1999.

\bibitem{heizler2016self}
Shay~I Heizler, Tomer Shussman, and Elad Malka.
\newblock Self-similar solution of the subsonic radiative heat equations using a binary equation of state.
\newblock {\em Journal of Computational and Theoretical Transport}, 45(4):256--267, 2016.

\bibitem{pattle1959diffusion}
R.~E. Pattle.
\newblock Diffusion from an instantaneous point source with a concentration-dependent coefficient.
\newblock {\em The Quarterly Journal of Mechanics and Applied Mathematics}, 12(4):407--409, 1959.

\bibitem{ames1965similarity}
W~Fc Ames.
\newblock Similarity for nonlinear diffusion equation.
\newblock {\em Industrial \& Engineering Chemistry Fundamentals}, 4(1):72--76, 1965.

\bibitem{zeldovich1967physics}
Ya~B Zeldovich, Yuri~Petrovich Raizer, WD~Hayes, and RF~Probstein.
\newblock {\em Physics of shock waves and high-temperature hydrodynamic phenomena. Vol. 2}.
\newblock Academic Press New York, 1967.

\bibitem{buckingham1914physically}
Edgar Buckingham.
\newblock On physically similar systems; illustrations of the use of dimensional equations.
\newblock {\em Physical review}, 4(4):345, 1914.

\bibitem{barenblatt1996scaling}
Grigory~Isaakovich Barenblatt.
\newblock {\em Scaling, self-similarity, and intermediate asymptotics: dimensional analysis and intermediate asymptotics}.
\newblock Number~14. Cambridge University Press, 1996.

\bibitem{zel1959propagation}
B~Zeldovich~Ya and AS~Kompaneets.
\newblock On the propagation of heat for nonlinear heat conduction.
\newblock {\em Collection dedicated to the seventieth Birthday of Academician AF Ioffe (PI Lukirskii, ed.) Izdat. Acad. Nauk SSSR, Moskow}, 1959.

\bibitem{lonngren1974field}
KE~Lonngren, WF~Ames, A~Hirose, and J~Thomas.
\newblock Field penetration into a plasma with nonlinear conductivity.
\newblock {\em The Physics of Fluids}, 17(10):1919--1920, 1974.

\bibitem{pomraning1979non}
GC~Pomraning.
\newblock The non-equilibrium marshak wave problem.
\newblock {\em Journal of Quantitative Spectroscopy and Radiative Transfer}, 21(3):249--261, 1979.

\bibitem{bingjing1996benchmark}
Su~Bingjing and Gordon~L Olson.
\newblock Benchmark results for the non-equilibrium marshak diffusion problem.
\newblock {\em Journal of Quantitative Spectroscopy and Radiative Transfer}, 56(3):337--351, 1996.

\bibitem{fleck1971implicit}
Joseph~A Fleck~Jr and JD~Cummings~Jr.
\newblock An implicit monte carlo scheme for calculating time and frequency dependent nonlinear radiation transport.
\newblock {\em Journal of Computational Physics}, 8(3):313--342, 1971.

\bibitem{larsen1988grey}
Edward Larsen.
\newblock A grey transport acceleration method far time-dependent radiative transfer problems.
\newblock {\em Journal of Computational Physics}, 78(2):459--480, 1988.

\bibitem{olson2000diffusion}
Gordon~L Olson, Lawrence~H Auer, and Michael~L Hall.
\newblock Diffusion, p1, and other approximate forms of radiation transport.
\newblock {\em Journal of Quantitative Spectroscopy and Radiative Transfer}, 64(6):619--634, 2000.

\bibitem{densmore2012hybrid}
Jeffery~D Densmore, Kelly~G Thompson, and Todd~J Urbatsch.
\newblock A hybrid transport-diffusion monte carlo method for frequency-dependent radiative-transfer simulations.
\newblock {\em Journal of Computational Physics}, 231(20):6924--6934, 2012.

\bibitem{yee2017stable}
Ben~Chung Yee, Allan~Benton Wollaber, Terry~Scot Haut, and H~Park.
\newblock A stable 1d multigroup high-order low-order method.
\newblock {\em Journal of Computational and Theoretical Transport}, 46(1):46--76, 2017.

\bibitem{mclean2022multi}
KW~McLean and SJ~Rose.
\newblock Multi-group radiation diffusion convergence in low-density foam experiments.
\newblock {\em Journal of Quantitative Spectroscopy and Radiative Transfer}, 280:108070, 2022.

\bibitem{steinberg2022multi}
Elad Steinberg and Shay~I Heizler.
\newblock Multi-frequency implicit semi-analog monte-carlo (ismc) radiative transfer solver in two-dimensions (without teleportation).
\newblock {\em Journal of Computational Physics}, 450:110806, 2022.

\bibitem{steinberg2023frequency}
Elad Steinberg and Shay~I Heizler.
\newblock Frequency-dependent discrete implicit monte carlo scheme for the radiative transfer equation.
\newblock {\em Nuclear Science and Engineering}, pages 1--13, 2023.

\bibitem{zhang2023fully}
Xiaojiang Zhang, Peng Song, Yi~Shi, and Min Tang.
\newblock A fully asymptotic preserving decomposed multi-group method for the frequency-dependent radiative transfer equations.
\newblock {\em Journal of Computational Physics}, 491:112368, 2023.

\bibitem{liu2023implicit}
Chang Liu, Weiming Li, Yanli Wang, Peng Song, and Kun Xu.
\newblock An implicit unified gas-kinetic wave--particle method for radiative transport process.
\newblock {\em Physics of Fluids}, 35(11), 2023.

\bibitem{li2024unified}
Weiming Li, Chang Liu, and Peng Song.
\newblock Unified gas-kinetic particle method for frequency-dependent radiation transport.
\newblock {\em Journal of Computational Physics}, 498:112663, 2024.

\bibitem{rosen2005fundamentals}
Mordecai~D Rosen.
\newblock Fundamentals of icf hohlraums.
\newblock Technical report, Lawrence Livermore National Lab.(LLNL), Livermore, CA (United States). Also appears as: "Fundamentals of ICF Hohlraums", Mordecai D. Rosen, "Lectures in the Scottish Universities Summer School in Physics, 2005, on High Energy Laser Matter Interactions", D. A. Jaroszynski, R. Bingham, and R. A. Cairns, editors, CRC Press Boca Raton. Pgs 325-353, (2009)., 2005.

\bibitem{virtanen2020scipy}
Pauli Virtanen, Ralf Gommers, Travis~E Oliphant, Matt Haberland, Tyler Reddy, David Cournapeau, Evgeni Burovski, Pearu Peterson, Warren Weckesser, Jonathan Bright, et~al.
\newblock Scipy 1.0: fundamental algorithms for scientific computing in python.
\newblock {\em Nature methods}, 17(3):261--272, 2020.

\bibitem{carslaw_jaeger1959}
H.~S. Carslaw and J.~C. Jaeger.
\newblock {\em Conduction of heat in solids / H.S. Carslaw and J.C. Jaeger.}
\newblock Clarendon Press, Oxford, second edition. edition, 1959.

\bibitem{kot2019parabolic}
VA~Kot.
\newblock Parabolic profile in heat-conduction problems. 2. semi-bounded space with a time-varying surface temperature.
\newblock {\em Journal of Engineering Physics and Thermophysics}, 92:333--354, 2019.

\bibitem{jost1952diffusion}
Wilhelm Jost.
\newblock Diffusion in solids, liquids, gases.
\newblock {\em Zeitschrift f{\"u}r Physikalische Chemie}, 201(1-2):319--320, 1952.

\bibitem{kot2018parabolic}
VA~Kot.
\newblock Parabolic profile in heat-conduction problems. 1. semi-bounded space with a surface of constant temperature.
\newblock {\em Journal of Engineering Physics and Thermophysics}, 91:1391--1412, 2018.

\bibitem{hristov2016integral}
Jordan Hristov.
\newblock Integral solutions to transient nonlinear heat (mass) diffusion with a power-law diffusivity: a semi-infinite medium with fixed boundary conditions.
\newblock {\em Heat and Mass Transfer}, 52:635--655, 2016.

\bibitem{nelson2009semi}
Eric~M Nelson and James Reynolds.
\newblock Semi-analytic solution for a marshak wave via numerical integration in mathematica.
\newblock Technical Report LA-UR-09-04551, Los Alamos National Laboratory, 2009.

\bibitem{polyanin2003handbook}
Andrei~D Polyanin and Valentin~F Zaitsev.
\newblock {\em Handbook of nonlinear partial differential equations: exact solutions, methods, and problems}.
\newblock Chapman and Hall/CRC, 2003.

\bibitem{su1997analytical}
Bingjing Su and Gordon~L Olson.
\newblock An analytical benchmark for non-equilibrium radiative transfer in an isotropically scattering medium.
\newblock {\em Annals of Nuclear Energy}, 24(13):1035--1055, 1997.

\bibitem{pomraning2005equations}
Gerald~C Pomraning.
\newblock {\em The equations of radiation hydrodynamics}.
\newblock Courier Corporation, 2005.

\bibitem{bennett2023benchmark}
William Bennett and Ryan~G McClarren.
\newblock Benchmark solutions for radiative transfer with a moving mesh and exact uncollided source treatments.
\newblock {\em Nuclear Science and Engineering}, pages 1--31, 2023.

\bibitem{hu2023rigorous}
Yuan Hu, Chang Liu, Huayun Shen, Gang Xiao, and Jinghong Li.
\newblock A rigorous model reduction for the anisotropic-scattering transport process.
\newblock {\em Physics of Fluids}, 35(12), 2023.

\bibitem{mcclarren2009modified}
Ryan~G McClarren and Todd~J Urbatsch.
\newblock A modified implicit monte carlo method for time-dependent radiative transfer with adaptive material coupling.
\newblock {\em Journal of Computational Physics}, 228(16):5669--5686, 2009.

\bibitem{mcclarren2022data}
Ryan~G McClarren and Terry~S Haut.
\newblock Data-driven acceleration of thermal radiation transfer calculations with the dynamic mode decomposition and a sequential singular value decomposition.
\newblock {\em Journal of Computational Physics}, 448:110756, 2022.

\bibitem{steinberg2022new}
Elad Steinberg and Shay~I Heizler.
\newblock A new discrete implicit monte carlo scheme for simulating radiative transfer problems.
\newblock {\em The Astrophysical Journal Supplement Series}, 258(1):14, 2022.

\end{thebibliography}

\pagebreak{}

\appendix

\section{Dimensional analysis\label{sec:Dimensional-analysis}}

\begin{table}[H]
\centering{}%
\begin{tabular}{|c|c|c|c|c|}
\hline 
$T$  & $T_{0}$  & $K$  & $x$  & $t$\tabularnewline
\hline 
\hline 
$\left[T\right]$  & $\left[T\right]\left[\text{time}\right]^{-\tau}$  & $\frac{\left[\text{length}\right]^{2}}{\left[T\right]^{4+\alpha-\beta}\left[\text{time}\right]}$  & $\left[\text{length}\right]$  & $\left[\text{time}\right]$\tabularnewline
\hline 
\end{tabular}\caption{The dimensional quantities in the problem (upper row) and their dimensions
(lower row).\label{tab:The-dimensional-quantities}}
\end{table}

In this appendix, the method of dimensional analysis will be employed
in order to find a self-similar ansatz for the solution of the problem
defined by Eqs. \eqref{eq:Tpde}-\eqref{eq:Tbc}. The dimensional
quantities which define the problem are listed in table \ref{tab:The-dimensional-quantities}.
The problem is defined by $M=5$ dimensional quantities which are
composed of $N=3$ different units: time, length and temperature.
Therefore, from the theory of dimensional analysis \cite{buckingham1914physically,zeldovich1967physics,barenblatt1996scaling},
the problem can be solved using $M-N=2$ dimensionless variables,
which are given in terms of power laws of the dimensional quantities.
Since the boundary temperature $T_{0}t^{\tau}$ has units of temperature,
we can write the dimensionless temperature similarity profile directly
as: 
\begin{equation}
f\left(\xi\right)=\frac{T\left(x,t\right)}{T_{0}t^{\tau}}.\label{eq:fapp}
\end{equation}
The dimensionless independent coordinate is written as: 
\begin{equation}
\xi=xt^{-\delta}K^{\eta}T_{0}^{\theta}.\label{eq:xsiapp}
\end{equation}
The requirement that $\xi$ is dimensionless result in the following
system of linear equations:

\begin{align*}
 & -\delta-\eta-\theta\tau=0\\
 & \eta\left(\beta-\alpha-4\right)+\theta=0\\
 & 1+2\eta=0
\end{align*}
which has the solution: 
\begin{equation}
\delta=\frac{1}{2}\left(1+\tau\left(4+\alpha-\beta\right)\right)
\end{equation}
\begin{equation}
\eta=-\frac{1}{2}
\end{equation}
\begin{equation}
\theta=-\frac{4+\alpha-\beta}{2}
\end{equation}
Hence, it is seen that the resulting dimensionless quantities in Eqs.
\eqref{eq:fapp}-\eqref{eq:xsiapp} give the self-similar ansatz in
Eqs. \eqref{eq:xsi_def}-\eqref{eq:Tss}. 
\end{document}